\documentclass [prb,preprint,superscriptaddress,floatfix,amsmath,amssymb] {revtex4-2}
\usepackage{amsmath}
\usepackage{graphicx}
\usepackage{hyperref}
\usepackage{color}
\usepackage{amssymb}
\usepackage{bm}

\newcommand{\beq} {\begin{equation}}
\newcommand{\eeq} {\end{equation}}
\newcommand{\bea} {\begin{eqnarray}}
\newcommand{\eea} {\end{eqnarray}}
\newcommand{\be} {\begin{equation}}
\newcommand{\ee} {\end{equation}}

\renewcommand{\[}{\left[}
\renewcommand{\]}{\right]}

\DeclareMathOperator{\Tr}{\text{Tr}}
\DeclareMathOperator{\Ree}{\text{Re}}
\DeclareMathOperator{\Imm}{\text{Im}}

\definecolor{darkgreen}{RGB}{0,170,0}

\begin{document}
\title { Superconductivity out of a non-Fermi liquid. Free energy analysis.}
\author{Shang-Shun Zhang}
\affiliation{School of Physics and Astronomy and William I. Fine Theoretical Physics Institute,
University of Minnesota, Minneapolis, MN 55455, USA}
\author{Yi-Ming Wu}
\affiliation{Institute for Advanced Study, Tsinghua University, Beijing 100084, China}
\author{Artem Abanov}
\affiliation{Department of Physics, Texas A\&M University, College Station,  USA}
\author{Andrey V. Chubukov}
\affiliation{School of Physics and Astronomy and William I. Fine Theoretical Physics Institute,
University of Minnesota, Minneapolis, MN 55455, USA}

\date{\today}
\begin{abstract}
  In this paper, we present in-depth analysis of the condensation energy $E_c$  for a
   superconductor in a situation when superconductivity  emerges out of a non-Fermi liquid due to pairing mediated by a massless boson.
    This is the case for electronic-mediated pairing near a quantum-critical point in a metal, for pairing in SYK-type models, and
   for phonon-mediated pairing in the properly defined limit, when the dressed Debye frequency vanishes.
    We consider a subset of these quantum-critical models, in which the pairing in a channel with a proper spatial symmetry is described by an effective $0+1$ dimensional model with the effective dynamical interaction
   $V(\Omega_m) = {\bar g}^\gamma/|\Omega_m|^\gamma$, where $\gamma$ is model-specific (the $\gamma$-model). In previous papers,  we argued that the pairing in the $\gamma$ model is qualitatively different from that in a Fermi liquid, and the
      gap equation at $T=0$ has an infinite number  of topologically distinct solutions, $\Delta_n  (\omega_m)$, where an integer $n$, running between $0$  and infinity, is the
   number of  zeros of $\Delta_n  (\omega_m)$ on the positive Matsubara axis. This gives rise to the set of extrema of $E_c$ at
   $E_{c,n}$, of which $E_{c,0}$  is the global minimum.
   The spectrum  $E_{c,n}$ is  discrete for a generic $\gamma <2$,  but becomes continuous at $\gamma = 2-0$. Here, we discuss in more detail the profile of the condensation energy near each $E_{c,n}$ and the  transformation from a discrete to a continuous spectrum at  $\gamma \to 2$.
   We also discuss the free energy and the specific heat of the $\gamma$-model in the
    normal state.
         \end{abstract}
\maketitle

\section{ Introduction.}

 This work
  extends
  our previous analysis~\cite{paper_1,paper_2,paper_3,paper_4,paper_5,paper_6} of the interplay between non-Fermi liquid (NFL) physics and  superconductivity  for quantum-critical (QC) itinerant fermionic systems, whose low-energy dynamics can be described  by an effective model of dispersion-full electrons, interacting by exchanging fluctuations of the critical order parameter.  Viewed in  the particle-hole channel, this interaction gives rise to singular self-energy $\Sigma (k, \omega)$ and NFL physics. Viewed in the particle-particle channel, it gives rise to a
  strong attraction in at least one pairing channel and to a spontaneous  appearance of an anomalous  pairing vertex $\Phi (k, \omega)$  at an elevated $T_c$. The two phenomena compete with each other: a NFL self-energy destroys Cooper logarithm in the particle-particle channel, while the opening of a pairing gap transfers the spectral weight out of low energies, rendering  a coherent Fermi liquid behavior of  low-energy fermions.
    Because both NFL and pairing come from the same source, the characteristic scales for the two phenomena are comparable. This makes the analysis of the competition quite challenging.

  We analyze the interplay between pairing and NFL for a subset of quantum-critical systems, in which
  critical order parameter fluctuations are slow modes compared to fermions for one reason or another. Examples include
    systems at the verge of spin-density-wave  order ~\cite{Millis1992,Altshuler1995a,Sachdev1995,acf,*acs,*acs2,*finger_2001,*acn,Subir,Subir2,nick_b,wang,max2}, 
    charge-density wave order~\cite{vojta,efetov,efetov2,efetov3,tsvelik,Bauer2015,ital,ital2,ital3,wang23,wang_22}, ferromagnetic and Ising-nematic order~\cite{triplet,triplet2,triplet3,sslee2,steve_sam} in 3D and 2D systems, fermions at a half-filled Landau level~\cite{PALee1989,Monien1993,Nayak1994,Altshuler1994,Kim_1994}, electron-phonon systems~\cite{Allen_1975,Karakozov_1976,Marsiglio_91,combescot,Chubukov_2020b} in the properly defined limit when the dressed Debye frequency vanishes~ 
    \footnote{At small dressed Debye frequency the physics that we study can be preempted by the development of 
     a bi-polaronic superconductivity~\cite{kolya}. We assume without proof that this does not happen.}
  In all these cases,  fermionic self-energy, excluding the thermal piece, can be treated as momentum-independent $\Sigma (\omega)$
(barring potential logarithms~\cite{metlitski2010quantum1,metlitski2010quantum2}),
  while the pairing vertex $\Phi (k, \omega)$, is either $k-$independent or can be factorized as $\Phi (k, \omega)= f_k \Phi (\omega)$, where $f_k$ is the spatial gap structure for the most strongly divergent  channel ($s-$wave, $d-$wave, etc). The theory then becomes effectively $0+1$ dimensional and reduces to the  set of two coupled equations for  ${\tilde \Sigma} (\omega) = \omega + \Sigma (\omega)$ and $\Phi (\omega)$  with an effective dynamical 4-fermion interaction $V(\Omega)$.  These
  can be obtained diagrammatically by restricting to
rainbow approximation
for the self-energy and ladder approximation for the pairing vertex.  On the Matsubara axis the two equations are
   \bea \label{eq:gapeq}
    \Phi (\omega_m) &=& \pi T \sum_{m'} \frac{\Phi (\omega_{m'})}{\sqrt{{\tilde \Sigma}^2 (\omega_{m'}) +\Phi^2 (\omega_{m'})}} V\left(\omega_m - \omega_{m'}\right), \nonumber \\
     {\tilde \Sigma} (\omega_m) &=& \omega_{m}
   +  \pi T \sum_{m'}  \frac{{\tilde \Sigma}(\omega_{m'})}{\sqrt{{\tilde \Sigma}^2 (\omega_{m'})
     +\Phi^2 (\omega_{m'})}}  V\left(\omega_m - \omega_{m'}\right)
   \eea
 At $T=0$, $\pi T \sum_{m'} = (1/2) \int d\omega'_{m}$.
The interaction $V(\Omega_m)$ is real and positive (attractive in our sign convention) and
 at a small but finite
bosonic mass
 $\omega_D$
  has the form
   \beq
  V(\Omega_m) = \frac{{\bar g}^\gamma}{(\Omega^2_m + \omega^2_D)^{\gamma/2}}
 \label{eq:1}
  \eeq
 The $0+1$ dimensional model  with $V(\Omega_m)$ as in Eq. (\ref{eq:1}) has been nicknamed the $\gamma-$model. Different  $\gamma$ correspond to different physical realizations (see Ref. \cite{paper_1}
 for details).  At a critical point, where
 $\omega_D =0$,
 $V(\Omega_m)$ becomes a singular function of frequency,  $V(\Omega_m) = ({\bar g}/|\Omega_m|)^\gamma$.
   For electronic pairing, the full set contains an additional equation describing  the feedback from the pairing on the bosonic propagator, but we will neglect it as the feedback preserves Eqs. (\ref{eq:gapeq}) and only changes $\gamma$ into $\gamma_{eff}$, which depends on $T$. The physics of the feedback can then
   be
   analyzed by  moving  along the line $\gamma = \gamma_{eff} (T)$ in  the $(\gamma, T)$ plane.

A rather similar, although not identical set of equations holds for dispersion-less fermions randomly coupled to each other~\cite{Sachdev_22,Chowdhury_2020}, or to phonons~\cite{Schmalian_19,Wang_19,Schmalian_19a,Classen_21}.
  The equations for  $\Sigma (\omega_m)$ and $\Phi (\omega_m)$  can be partly decoupled by  introducing $\Delta (\omega_m) = \Phi (\omega_m)/Z(\omega_m)$, often called the gap function,  and  $Z(\omega_m) = 1 + \Sigma (\omega_m)/\omega_m$, which is the inverse quasiparticle residue.
 The equations for $\Delta (\omega_m)$ and $Z(\omega_m)$ are straightforwardly obtained from (\ref{eq:gapeq}):
  \beq
   \Delta (\omega_m) = \pi T \sum_{m'} \frac{\Delta (\omega_{m'}) - \Delta (\omega_m) \frac{\omega_{m'}}{\omega_m}}{\sqrt{(\omega_{m'})^2 +\Delta^2 (\omega_{m'})}} V(\omega_m-\omega_{m'})
     \label{ss_11_0}
  \eeq
  \beq
    Z(\omega_m) = 1 +  \frac{\pi T}{\omega_m}  \sum_{m'} \frac{\omega_{m'}}
    {\sqrt{(\omega_{m'})^2 +\Delta^2 (\omega_{m'})}} V(\omega_m-\omega_{m'})
     \label{ss_11_01}
  \eeq
Eqs. (\ref{ss_11_0}) and (\ref{ss_11_01})  have the same form as Eliashberg equations for electron-phonon problem with a finite $V(0)$, and it is customary to keep calling them
  Eliashberg equations even when $V(0)$ diverges.

Observe that Eq.  (\ref{ss_11_0}) is an integral equation that does not contain $Z(\omega_m)$, and
   Eq.  (\ref{ss_11_01}) expresses $Z(\omega_m)$ via
   $\Delta (\omega_{m'})$. Note also that potentially dangerous self-action term with $m=m'$ can be safely eliminated in (\ref{ss_11_0}) because the numerator there vanishes at $m=m'$. It should, however, be kept in (\ref{ss_11_01}).

 In writing Eqs. (\ref{ss_11_0}) (\ref{ss_11_01}) we assumed that $Z(\omega_m)$ is positive for all $\omega_m$.
  The authors of Ref.~\cite{yuz} analyzed potential solutions  with sign-changing $Z(\omega_m)$.
  We do not consider such solutions here as they only exist at a finite $T$, while our
  primary
  interest is to understand system behavior at zero temperature. We comment on the solutions with sign-changing $Z(\omega_m)$ in Appendix \ref{sec:causality}. We argue that (i) at $T=0$ such solutions do not exist  and (ii)
   a finite $T$ solutions exist, but do not obey  the causality principle.

For large enough bosonic mass  $\omega_D$, the normal state is a Fermi liquid. Solving  the gap equation, one obtains  a pairing instability at some $T_{p,0}$. The corresponding  gap function  $\Delta_0 (\omega_m)$ is a sign-preserving function of frequency.  As $\omega_D$ decreases, new pairing instabilities  emerge at $T_{p,n} < T_{p,0}$ (Ref. \cite{paper_2}).
  The number of solutions increases with decreasing $\omega_D$ and becomes infinite at $\omega_D =0$, when a pairing boson becomes massless. In this limit,
   there exists an infinite, discrete set of solutions of the non-linear gap equation at $T=0$. One end of the set is the sign-preserving solution  $\Delta_0 (\omega_m)$. The other  end is
  $\Delta_{\infty} (\omega_m)$ with infinitesimally small amplitude. For the  latter we found the  exact analytical expression by solving exactly the linearized gap equation at $T=0$ (Refs. \cite{paper_1,paper_4}).
   All gap functions from
   the
   set are analytic in the upper half-plane of complex  frequency $z = \omega' + i \omega^{''}$, i.e., the corresponding Green's functions are
   causal.

\begin{figure}
  \includegraphics[width=8.5cm]{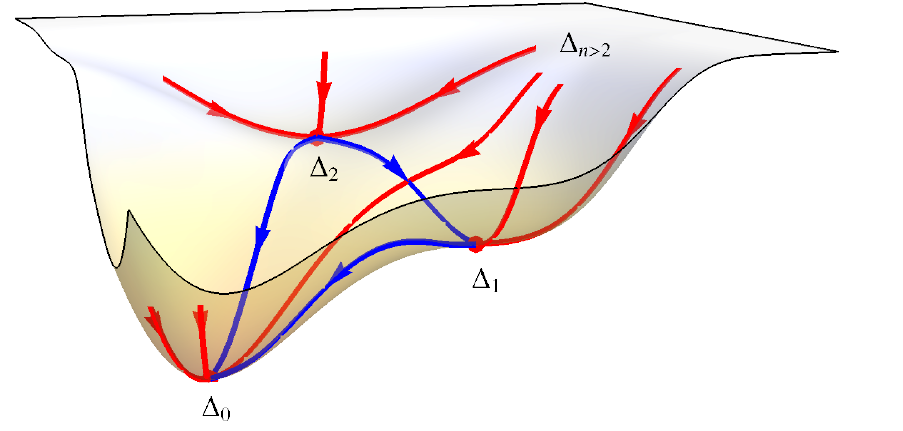}
  \caption{A schematic plot of the free energy configuration in the functional space and the saddle points corresponding to the gap function $\Delta_n(\omega_m)$.
  The $n>0$ solution is a saddle point of the free energy, which has $n$ unstable principle axes connected with the $n'<n$ solutions, as illustrated by the flows in the plot.
  The $n=0$ solution is a stable minimum. }
  \label{fig:intro}
\end{figure}

In this communication we address two issues. One is the  profile of the
condensation energy around $\Delta_n (\omega_m)$. The condensation energy $E_c  = F_{sc}  - F_n$ is the difference between the free energy of a superconductor, $F_{sc}$, and that
of a would be normal state at the same temperature.
The condensation energy is a functional of $\Delta (\omega_m)$ and $Z(\omega_m)$,  and  the Eliashberg equations are stationary conditions $\delta E_c/\delta \Delta =0$, $\delta E_c/\delta Z$ =0.  We
      expand $E_c$  to second order in deviations from $E_{c,n} $ for stationary gap function ${\Delta_n (\omega_m)}$ with different $n$ and analyze the eigenvalues.  For  $n=0$,  all eigenvalues are positive, i.e.,
     $\Delta_0 (\omega_m)$ is a minimum of $E_c$.  All other $\Delta_n (\omega_m)$ are saddle points, but rather specific ones (see
     Fig.\ref{fig:intro}).
      Namely,
     for $n=1$, we show that there is a single negative eigenvalue, and the corresponding eigenfunction shifts $\Delta_1 (\omega_m)$ towards $\Delta_0 (\omega_m)$.  For $n=2$, there are two negative eigenvalues, which  shift  $\Delta_2 (\omega_m)$ towards $\Delta_0 (\omega_m)$ and towards $\Delta_1 (\omega_m)$. For a generic $\Delta_n (\omega_m)$, there are $n$ negative eigenvalues. One can find a ``direction'' in the space of $\Delta (\omega_m)$, along which  all $E_{c,n}$  are connected by a single curve. Along this direction, each $E_{c,n}$ with $n >0$  is  an inflection point (see Fig.~\ref{fig:freeE}).

The other issue is the evolution of  $\Delta_n (\omega_m)$ and $E_{c,n}$  near $\gamma =2$. We argued in \cite{paper_4,paper_5} that at $T=0$ and $\omega_D =0$, the set of $\Delta_n (\omega_m)$ becomes continuous at $\gamma \to 2$ in the particular double limit, which we  discuss below.
Specifically, we argued  that all $\Delta_n (\omega_m)$ with
     finite $n$ become equivalent to $\Delta_0 (\omega_m)$ down to $\omega_m \to 0$,
      while $\Delta_n (\omega_m)$ with $n \to \infty$
      form a one-parameter continuous set
      $\Delta (\epsilon,\omega_m)$, in which
       $\epsilon$ is a function of $(n^*/n)$, where $n^* \sim |\log (2-\gamma)|/(2-\gamma)$.
       By construction, $\epsilon$ runs between $0+$, for which  $\Delta (0+,\omega_m)$ is infinitesimally small and $\epsilon_{max}$, for which $\Delta (\epsilon_{max}, \omega_m) = \Delta_0 (\omega_m)$.
        We further argued that
        the condensation energy $E_{c,n}$ also
        becomes continuous
        $E_{c}(\epsilon)$
        at $\gamma \to 2$.
        We emphasize that this happens only  at $T=\omega_D =0$.
          At any finite  $\omega_D$,
           or $T$,  $\Delta_n (\omega_m)$ and $E_{c,n}$  remain  discrete at $\gamma =2$.

          Here we discuss the transformation from a discrete spectrum  to a continuous one
          for both $\Delta_n (\omega_m)$ and $E_{c,n}$
         in more detail. We show that the limit $\gamma \to 2$  has to be taken with special care because $E_{c,n}$ formally diverges as $1/(2-\gamma)$  for all values of $n$.
         Specifically, we show that for large but finite $n$,
         \beq
          E_{c,n} = -N_F
          \bar{g}^2
          \left(\frac{1}{2-\gamma} -
          c
          n + ...\right)
        \label{nnn_1}
        \eeq
         where
         $c = O(1)$
         and
         $N_F \propto N/E_F$ is the total density of states at the Fermi surface ($N$ is the number of electrons and $E_F$ is the Fermi energy).
         The
         prefactor $b$ in (\ref{nnn_1}) tends to a  finite value at $\gamma \to 2$, yet, because the first term diverges, the condensation energy can be
          re-expressed as
           \beq
          E_{c,n}
          = -  \frac{
           \bar{g}^2
          N_F}{2-\gamma}  \left(1 -
          c
          n (2-\gamma) + ...\right)
        \eeq
       At $\gamma \to 2$, the term proportional to $b$ vanishes for all finite $n$. We computed this term at
        $n \to \infty$ and found
         \beq
          E_{c} (\epsilon) = - \frac{
           \bar{g}^2
          N_F}{2-\gamma}
          g(\epsilon)
        \eeq
        where $\epsilon$ is the same as for the set of the gap functions, and
        $g(0) =0$, $g(\epsilon_{max}) =1$.

       The divergence in the overall factor in $E_c (\epsilon)$ can be avoided  if we treat $\gamma \to 2$ as the double limit,
       in which
        $N_F$ scales as
         $2-\gamma$, and the product $N_F/(2-\gamma) $ remains finite. In this double limit one obtains a non-divergent, continuous
          spectrum of $E_c (\epsilon)$ at $\gamma \to 2$.

           The emergence of a dispersion-full  $E_c (\epsilon)$ may sound surprising
            because all gap functions
            $\Delta (\epsilon,\omega_m)$
            satisfy
            $\delta E_c [\Delta (\epsilon, \omega_m)]/\delta \Delta (\epsilon,\omega_m) =0$,
            hence
             $d E_c (\epsilon)/d \epsilon =0$ and $E_c (\epsilon)$ is apparently flat.
               We show here that while
               $d E_c (\epsilon)/d \epsilon$ vanishes,  $d E^2_c (\epsilon)/d \epsilon^2$ and all higher-order derivatives diverge at $\gamma \to 2$. In this case, Taylor expansion around any  $\epsilon$  has to be taken to an infinite order, and the summation of Taylor series yields the dispersion-full $E_c (\epsilon)$.

            As a part of our analysis  we discuss two other issues.  One is the relation between Luttinger-Ward (LW) variational free energy and the actual one, obtained using Hubbard-Stratonovich transformation.
              We argue that the variational
             free energy yields  the correct Eliashberg equations as stationary conditions,  but in general should not be used to expand around stationary points as it does not correctly describe fluctuations of
           inverse quasiparticle residue
           $Z(\omega)$.
            However, when stationary
            $Z(\omega)$
            is infinite, which  at
            $\omega_D =0$
            holds for $\gamma >1$ at $T=0$ and for all $\gamma$ at a finite $T$, both stationary solutions and fluctuations around it are properly described by the variational (spin chain) free energy.  For $\gamma =2$ this result  has been obtained in Ref. \cite{yuz}.

             The second issue is the specific heat in the normal state. It was argued in
            Ref. \cite{yuz_1} that $C(T)$ is negative for $\gamma \leq 2$ in some $T$ range above the onset of pairing, i.e., the $\gamma-$model is unstable.  We argue that this comes about because the authors of Ref.~\cite{yuz_1} subtracted a particular  temperature-dependent piece
              from
            the free energy.
              We analyze the free energy without the subtraction  and find that $C(T)$ is positive at all $T$.

             The structure of the paper is the following. In the next section we present the expressions for the
               variational LW
                free energy $F^{\text{var}}_{sc}$ and the actual, more complex free energy $F_{sc}$,  obtained by introducing normal and anomalous self-energies $\Sigma (\omega_m)$ and $\Phi (\omega_m)$  as auxiliary fields and integrating out fermions using Hubbard-Stratonovich transformation~ \cite{yuz_private,boyack,yuz}.
                We show that the condition that the free energy is stationary with
               respect to variations of $\Sigma$ and $\Phi$ yields the correct Eliashberg  equations for
                both $F^{\text{var}}_{sc}$ and $F_{sc}$.  We argue that to account for fluctuation corrections one generally  has to use $F_{sc}$, but when stationary
                $Z(\omega_m) = 1 + \Sigma(\omega_m)/\omega_m$
                is infinite, one can use more simple
               $F^{\text{var}}_{sc}$.
               We discuss the free energy in the normal state and the specific
               heat
               in Sec.~\ref{spec_heat}.
               In Sec.~\ref{sec:generic} we analyze the free energy expansion around the discrete set of the gap functions $\Delta_n (\omega_m)$
                for $\gamma <2$.We discuss the discrete set in Sec.~\ref{sec:generic1},
             obtain the free energy expansion around $\Delta_n (\omega_m)$  with $n=0,1$, and $2$ in Sec.~\ref{sec:generic2}, and obtain the condensation energy for different $n$
              in Sec. ~\ref{sec:generic3}.
              In Sec.~\ref{sec:limit} we analyze the transformation from a discrete spectrum of $\Delta_n (\omega_m)$  to a continuous one-parameter set
              $\Delta (\epsilon,\omega_m)$
              at $\gamma \to 2$ and discuss the corresponding transformation for the condensation energy and the need to take the double limit $\gamma \to 2$, $N_F \to 0$,
              $N_F/(2-\gamma) \to \text{const}$.
             We present our conclusions in Sec.~\ref{sec:sumary}.

 \section{Eliashberg equations as stationary points of the free energy}

 As we said, the Eliashberg equations, Eqs. (\ref{eq:gapeq}), or, equivalently, Eqs. (\ref{ss_11_0}) and (\ref{ss_11_01}), can be obtained diagrammatically in one-loop/ladder approximation.
 By general arguments, these equations also emerge as conditions on stationary points of the variational
   free energy. This has been demonstrated by Eliashberg, who extended to a superconductor the variational approach, pioneered by Luttinger and Ward (Ref. \cite{lw})
  The LW functional is expressed via  variational functions $\Sigma_k$  and $\Phi_k$, where $k = ({\bf k}, \omega_m)$ and $\omega_m = (2m+1)\pi T$ is fermionic Matsubara frequency. It has the form
   \bea
    F^{\text{var}}_{sc}&=& -T V \sum_{k} \left[ \log (- \det \hat{G}^{-1}_k) - i \Tr  ( \hat{\Sigma}_k \hat{G}_k ) \right]  \nonumber \\
    && +
     V T^2
     \sum_{k,k'} \left[G_k V(k-k') G_{k'} - F^*_k V(k-k') F_{k'}\right] + ...,
   \label{eq:var_energy}
   \eea
where $V$ is the system's volume, $\sum_k =
 \int d^d {\bf k}/(2\pi)^d \sum_{m}$ ($d$ is the spatial dimensionality), ${\tilde \Sigma}_k = \omega_m + \Sigma_k$, $\epsilon_{\bf k}$ is the fermionic dispersion, and $V(k-k')$ is a  4-fermion interaction, which in general depends on both frequency and momentum transfer.

  The matrices ${\hat G}_k$ and $i{\hat \Sigma}_k$ are  the single-particle Green's function and the self-energy in Nambu representation of electron operators, $\Psi_{\bm k} = (\psi_{{\bm k}, \uparrow},\psi^{\dagger}_{-{\bm k}, \downarrow})^\text{T}$.   The dots in (\ref{eq:var_energy}) stand for higher-order terms in $G_k$ and $F_k$.

  In explicit form
\beq
{\hat G}_k =
\[ \begin{array}{cc}
G_k & F_k \\
F_k^* & -G_{-k}
\end{array}\],
i{\hat \Sigma}_k =
\[ \begin{array}{cc}
i\Sigma_k & i\Phi_k \\
-i\Phi_k^* & -i\Sigma_{-k}
\end{array}\],
\eeq
where the diagonal (off-diagonal) elements are the normal (anomalous) components. The elements of
${\hat G}_k$ and ${\hat \Sigma}_k$
are related by the Dyson equation:
   \bea
   G_k &=& - \frac{\epsilon_{-\bf k} - i {\tilde \Sigma}_{-k}}{(\epsilon_{\bf k} - i {\tilde \Sigma}_k)(\epsilon_{-\bf k} - i {\tilde \Sigma}_{-k}) + |\Phi_k|^2}, \nonumber \\
   F_k &=&  i \frac{\Phi_k}{(\epsilon_{\bf k} - i {\tilde \Sigma}_k)(\epsilon_{-\bf k} - i {\tilde \Sigma}_{-k}) + |\Phi_k|^2}.
   \eea
The stationary solutions for $\Sigma_k$ and $\Phi_k$ are obtained from $\delta F^{\text{var}}_{sc}/\delta \Sigma_k =0$, $\delta F^{\text{var}}_{sc}/\delta \Phi_k =0$. This leads to
  \bea
  \[ \begin{array}{ccc}
2{\delta F_k \over \delta \Sigma_k} & - {\delta G_k \over \delta \Sigma_k} & - {\delta G_{-k} \over \delta \Sigma_k} \\
2{\delta F_k \over \delta \Sigma_{-k}} & - {\delta G_k \over \delta \Sigma_{-k}}  & - {\delta G_{-k} \over \delta \Sigma_{-k}} \\
2{\delta F_k \over \delta \Phi_k} & - {\delta G_k \over \delta \Phi_k}  & - {\delta G_{-k} \over \delta \Phi_k}
\end{array}\]
\[ \begin{array}{c}
I_{1,k} \\
I_{2,k} \\
I_{2,-k}
\end{array}\]
=0,
\label{eq:2}
 \eea
  where
  \bea
  && I_{1,k} = \Phi^*_k -
   i T\sum_{k'} V(k-k') F_{k'}^*, \nonumber \\
   && I_{2,k} =  {\Sigma}_k -
    i T\sum_{k'} V(k-k') G_{k'}.
   \label{eq:3}
   \eea
   The determinant of the $3 \times 3$ matrix in Eq.~(\ref{eq:2}) equals to $2i(\epsilon_{\bm k}-i\tilde{\Sigma}_k)(\epsilon_{-\bm k}-i\tilde{\Sigma}_{-k})$ and is generally non-zero. The three equations are then independent, which implies that  $I_{1,k}=I_{2,k} =0$, i.e.,
  \bea
  && \Phi^*_k  =  i T \sum_{k'} V(k-k') F_{k'}^*, \nonumber \\
   && {\Sigma}_k = i T \sum_{k'} V(k-k') G_{k'}.
   \label{eq:3_1}
   \eea
  Assuming further that a soft boson is a slow excitation compared to a fermion, one can factorize the integration over ${\bf k}$  along and transverse to the Fermi surface.  The integration transverse to the Fermi surface is over fermionic dispersion. When the Fermi energy $E_F$ is much larger that fermion-boson coupling ${\bar g}$ in Eq.~(\ref{eq:1}), one can convert the momentum integration into that over $\epsilon_{\bf k}$ in infinite limits, keeping the density of states at its value at the Fermi surface.  The
  integration
  along the Fermi surface
  involves $V(k-k')$ with both ${\bf k}$ and ${\bf k}'$ on the Fermi surface.
  Restricting to hot regions on the Fermi surface, when the interaction is peaked at a finite momentum transfer or projecting into the most attractive pairing component, when the interaction is peaked at zero momentum transfer,
    we obtain the two Eliashberg equations (\ref{eq:gapeq}).
    We refer a reader to Ref. \cite{paper_1} for more detailed discussion of
    how the $\gamma-$model description emerges from particular lattice models.

Derivation of  Eliashberg equations from variational $F^{\text{var}}_{sc}$  has been reported several times
in the literature~\cite{,haslinger,secchi2020phonon,benlagra2011luttinger}.
There is one caveat here, which will be relevant to our consideration below. Namely, under the assumptions used to derive Eq.~(\ref{eq:gapeq}),
$\Sigma_k = \Sigma (\omega_m)$ and $\Phi_k = \Phi (\omega_m)$. Then
one can integrate over momentum right in Eq.~(\ref{eq:var_energy}). The integration is straightforward and yields~\cite{haslinger,paper_1}
  \bea
  F^{\text{var}}_{sc} &=& -2\pi T N_F  \sum_{m} \frac{\omega_m {\tilde \Sigma}_m}{\sqrt{{\tilde \Sigma}^2_m + |\Phi_m|^2}} \nonumber \\
   && - \pi^2 T^2 N_F \sum_{m,m'} V(m-m') \frac{{\tilde \Sigma}_m {\tilde \Sigma}_{m'} + \frac{1}{2} \left(\Phi_m \Phi^*_{m'} + \Phi^{*}_m \Phi_{m'}\right)}{\sqrt{{\tilde \Sigma}^2_m + |\Phi_m|^2} \sqrt{{\tilde \Sigma}^2_{m'} + |\Phi_{m'}|^2}}
   \label{eq:var_energy_1}
   \eea
where
$N_F \sim N/E_F$ is the total density of states ($N$ is the number of particles in the system), and
 we introduced the shorthand notations $\Phi_m = \Phi (\omega_m), ~{\tilde \Sigma}_m = {\tilde \Sigma} (\omega_m)$, $V(m-m') = V(\omega_m-\omega_{m'})$.  From
 $\delta F^{\text{var}}_{sc}/\delta \Sigma_m =0$
 and
 $\delta F^{\text{var}}_{sc}/\delta \Phi_m =0$.
  We then obtain, instead of Eq.~(\ref{eq:2}),
 \bea
 && {\Phi_m} {\tilde \Sigma}_m I_1 - |\Phi_m|^2 I_2 =0 \nonumber \\
 &&  {\tilde \Sigma}^2_m I_1  - \Phi^*_m {\tilde \Sigma}_m I_2 =0
 \label{eq:2_1}
 \eea

 One can immediately verify that the determinant of the $2 \times 2$ matrix  vanishes, hence there is only one independent equation.
  Introducing $Z_m$ and $\Delta_m$ as $Z_m = 1 + \Sigma_m/\omega_m$, $\Delta_m = \Phi_m/Z_m$, we find
   that this is Eq. (\ref{ss_11_0})  for $\Delta_m$.  There is no  equation (\ref{ss_11_01}) on $Z_m$.  This can be seen already from Eq. (\ref{eq:var_energy_1}). Indeed, expressing
$\Phi_m$ and $\Sigma_m$ in terms of $\Delta_m$ and $Z_m$, we immediately find that $F^{\text{var}}_{sc}$ depends only on $\Delta_m$:
\bea
F^{\text{var}}_{sc} &=& -2\pi T N_F  \sum_{m} \frac{\omega^2_m}{\sqrt{\omega^2_m + |\Delta_m|^2}} \nonumber \\
&& - \pi^2 T^2 N_F \sum_{m,m'} V(m-m') \frac{\omega_m \omega_{m'} + \frac{1}{2} \left(\Delta_m \Delta^*_{m'} + \Delta^{*}_m \Delta_{m'}\right)}{\sqrt{\omega^2_m + |\Delta_m|^2} \sqrt{\omega^2_{m'} + |\Delta_{m'}|^2}}
\label{eq:var_energy_2}
\eea
 The independence of  $F^{\text{var}}_{sc}$ on $Z_m$ seem to imply  that  fluctuations of $Z_m$ decouple from fluctuations of $\Delta_m$ for any $\omega_D$.  However, variational $F^{\text{var}}_{sc}$ is {\it not} the
 free energy, which appears in the partition function ${\cal Z} = \int {\cal D}[\Delta,Z] e^{-F_{sc} (\Delta, Z)/T}$. To obtain $F_{sc}$,
 one has to follow a standard procedure: introduce auxiliary fields $\Sigma_m$ and $\Phi_m$ and integrate out fermions using Hubbard-Stratonovich transformation.  This has been implemented in  Refs. \cite{yuz_private,boyack,yuz}, and the result is
 \bea
   F_{sc} &=& -2\pi T N_F \sum_m Z_m \sqrt{\omega^2_m + |\Delta_m|^2}  + T^2 N_F \sum_{m,m'} V^{-1} (m-m') \nonumber \\
   &&  \times \left[(Z_m-1) (Z_{m'}-1)\omega_m \omega_{m'} + {1\over 2} Z_m Z_{m'} \left(\Delta_m \Delta^*_{m'} + \Delta^*_m \Delta_{m'}\right)\right],
 \label{eq:var_energy_actual}
   \eea
where the inverse matrix $V^{-1} (m-m')$ satisfies $T^2 \sum_{m''} V(m-m^{''}) V^{-1} (m^{''}-m') = \delta_{m,m'}$
 (see Appendix \ref{app:B} for detail).
 The stationary conditions $\delta F_{sc}/\delta \Delta_m =0$, $\delta F_{sc}/\delta Z_m =0$ yield Eqs. (\ref{ss_11_0}) and (\ref{ss_11_01}), as one can straightforwardly verify.

 The free energy (\ref{eq:var_energy_actual}) depends on both $\Delta_m$ and $Z_m$, and the quadratic form in deviations from a stationary solution generally contains terms with variations of $\Delta_m$ and  $Z_m$. The authors of Ref.~\cite{yuz} argued that for $\gamma =2$ (the case they considered) fluctuations of $Z_m$ get steeper when the bosonic mass gets smaller, and the free energy $F_{sc} (\Delta_m, Z_m)$  with arbitrary $\Delta_m$ and  stationary $Z_m$, given by Eq.~(\ref{ss_11_01}), coincides with $F^{\text{var}}_{sc} (\Delta_m)$ up to regular corrections in powers of $\omega_D/{\bar g}$.
 We extended the analysis of Ref. \cite{yuz} to $\gamma <2$ and found that this holds when the stationary $Z_m$ diverges at $\omega_D \to 0$. This is the case for $\gamma >1$ at $T=0$ and for all $\gamma >0$ at a finite $T$.
We show the details in Appendix \ref{app:B}.

For the rest of the paper we consider the limit of vanishing $\omega_D$ and use $F^{\text{var}}_{sc} (\Delta_m)$ for the free energy both at
and
away
from
stationary points.  At $T=0$, we will be chiefly interested in the system behavior for $\gamma$ close to $2$, when the equivalence between $F^{\text{var}}_{sc}$ and $F_{sc}$ holds. We will also present some numerical results for smaller $\gamma$. These results are obtained at a small, but finite $T$, when the equivalence between $F^{\text{var}}_{sc}$ and $F_{sc}$  again holds.

\subsection{Condensation energy}
\label{sec:generic3_a}

Condensation energy $E_c$ is the difference between the free energy of a superconductor with a stationary $\Delta_m$ and that of a would be normal state at the same $T$.  For an $s-$wave  BCS superconductor $E_c = - N_F \Delta^2/2$.  In our case, stationary
$F_{sc}$ is given by (\ref{eq:var_energy_2}), and the
free energy in the normal state is
 \beq
 F_{\text{norm}} = -2\pi T N_F \sum_m |\omega_m| - \pi^2 T^2 \sum_{m,m'} V(\omega_m-\omega_{m'}) {\text{sign}} (\omega_m \omega_{m'})
\label{eq:n1}
\eeq
The condensation energy
\bea
E_{c} &=& 2\pi T N_F \sum_m \left(|\omega_m|-\frac{\omega^2_m}{\sqrt{\omega^2_m + \Delta^2_m}}\right) -  \pi^2 T^2 N_F \sum_{m, m'} V(\omega_m-\omega'_m)   \nonumber \\
 && \times \left(  \frac{\omega_m \omega'_m + \Delta_m \Delta_{m'} }{\sqrt{\omega^2_m + \Delta^2_m } \sqrt{(\omega'_m)^2 + \Delta^2_{m'}}}  - {\text{sign}} (\omega_m \omega_{m'}) \right).
\label{eq:n2}
\eea
In writing (\ref{eq:n2}) we assumed for simplicity  that $\Delta_m$ is real, i.e., we set its $U(1)$ phase to be zero.

We note that both $F_{sc}$ and $F_{\text{norm}}$ contain thermal contribution from $m=m'$.
 It is proportional to $V(0) = ({\bar g}/\omega_D)^\gamma$ and diverges at  $\omega_D =0$ for any $\gamma >0$.
 \footnote{Similarly, at $T=0$,  both $F^{\text{var}}_{sc}$ and $F^{\text{var}}_{\text{norm}}$  contain
  $\int V(\Omega) d \Omega$. This integral is regular in the infra-red for $\gamma <1$, but becomes singular
   at $\gamma >1$ and scales as $({\bar g}/\omega_D)^{\gamma-1}$.}
This term gives rise to singular temperature-independent entropy at $T > \omega_D$, which we discuss in Sec. \ref{spec_heat} below.   At the same time, the thermal pieces in (\ref{eq:var_energy_2}) and (\ref{eq:n1}) are identical
 and cancel out in the expression for the condensation energy, Eq. (\ref{eq:n2}), which therefore remains non-singular at $\omega_D =0$.

Using the gap equation (\ref{ss_11_0}) one can re-express $E_{c}$  as
\bea
E_{c} &=& -
\pi T N_F
\sum_{m}
|\omega_m| \frac{\left(\sqrt{1 + D^2 (\omega_m)}-1\right)^2}{\sqrt{1 + D^2 (\omega_m)}} -
\frac{1}{2}\pi^2 T^2 N_F
\sum_{m, m'} V(\omega_m-\omega'_m)  \nonumber \\
 && \times \frac{{\text{sign}} (\omega_m \omega_{m'})}{\sqrt{1 + D^2 (\omega_m)} \sqrt{1 + D^2 (\omega_{m'})}}
  \left(\sqrt{1 + D^2 (\omega_m)} -\sqrt{1 + D^2 (\omega_{m'})}\right)^2
 \label{eq:n3}
\eea
where we remind that $D (\omega_m) = \Delta (\omega_m)/\omega_m$.  The advantage of using Eq. (\ref{eq:n3})
 is in that it explicitly shows that $E_{c} <0$, i.e., that there is an energy gain from pairing
  To see this explicitly one should convert the sum into the one over $m, m' \geq 0$ and use $V(\omega_m - \omega_{m'}) > V(\omega_m + \omega_{m'})$.

At $T=0, 2\pi T\sum_{m \geq 0} \to \int_0^\infty d \omega_m$.
At  $\gamma<2$,
the two integrals in (\ref{eq:n3}) are convergent both in the infra-red and ultra-violet limits
 even when  $\omega_D = 0$.
Convergence in the ultra-violet limit follows from the fact that
at large $\omega_m$, $D (\omega_m)$ falls off as
$1/\rvert \omega_m \rvert^{\gamma+1}$.
In the infrared  limit $\Delta (\omega_m)$ approaches a finite value $\Delta$, and $D(\omega_m)$ becomes large. Keeping only the leading terms in (\ref{eq:n3}) we obtain
\beq
E_c \propto \int_0^\Delta d \omega_m\int_0^\Delta d \omega_{m'} \frac{(\omega_m -\omega_{m'})^2}{\omega_m \omega_{m'}} \left(\frac{1}{|\omega_m- \omega_{m'}|^\gamma} - \frac{1}{|\omega_m + \omega_{m'}|^\gamma}\right)
\eeq
This integral is convergent for all $\gamma <2$.

\subsection{Specific heat}
\label{spec_heat}
 The specific heat
 is $C(T) = T dS(T)/dT$, where the entropy
 $S(T) = - dF/dT$ and $F = F_{\text{norm}} + E_c$.

 Computation of the temperature dependence of the free energy for any  $\gamma >0$  requires care for two reasons, both related to $F_{\text{norm}}$.  First,
  in the $\gamma-$model,
  $F_{\text{norm}}$
   contains a singular  thermal contribution from $m=m'$.  Second, the frequency summation in both terms in (\ref{eq:n1})
     does not converge, raising
     a possibility that
      there is a contribution to the specific heat that depends on the upper energy cutoff of the model.
         The
         frequency sums in Eq. (\ref{eq:n3}) for $E_c$ are ultraviolet-convergent, so the jump of $C(T)$ at $T = T_c -0$ is cutoff-independent.

     Because it is $F_{\text{norm}}$
     that is
      potentially problematic for the entropy and the specific heat, we
     consider the normal state.  The first term in (\ref{eq:n1}) is the free energy of the free Fermi gas,
      \beq
      F^{\text{free}}_{\text{norm}} = -2\pi T N_F \sum_m |\omega_m|
      \eeq
       The summation
      over Matsubara frequencies can be performed using Euler-Maclaurin formula in the form presented in~\cite{landau2013statistical}:
      \beq
       F^{\text{free}}_{\text{norm}} = -8 \pi^2 T^2 N_F \sum_{m=0}^{\infty} (m+1/2) \approx -2N_F \int_0^\infty x dx - \frac{\pi^2 T^2 N_F}{3}
      \label{nnn_2}
      \eeq
      The upper limit of the integral is actually the upper energy cutoff in the theory, $\Lambda$, then
        $ F^{\text{free}}_{\text{norm}} = -N_F (\Lambda^2 + \pi^2 T^2/3)$.
        The $\Lambda^2$ term is $T$ independent and therefore is irrelevant for $S(T)$ and $C(T)$.  Keeping the second term, we reproduce the known result for a  Fermi gas: $S(T) = C(T) =
        (2/3) \pi^2 T N_F$.

    Alternatively,  we can evaluate
      $T \sum_m |\omega_m|$ directly, by restricting to $M$ Matsubara numbers for positive and for negative $\omega_m$, i.e., keeping the particle-hole symmetry.  An elementary calculation then sets the
      relation between $M$ and  $\Lambda$:
      \beq
      M = \frac{\Lambda}{2\pi T} + \frac{\pi T}{12 \Lambda} + O\left(\frac{1}{\Lambda^2}\right)
      \label{nnn_3}
        \eeq
        It is essential that there is no
         $\Lambda$-independent term in this relation.
  We verified the absence of the $\Lambda$-independent term in the complimentary calculation, in which we
  integrated
  over fermionic dispersion $\epsilon_k$ over a finite range between $-\Lambda$ and $\Lambda$,
  where $\Lambda \sim E_F \gg {\bar g}$,
and then
summed over
all Matsubara frequencies, as in this calculation the Matsubara sum converges at $|\omega_m| >\Lambda$.

 We now use Eq.~(\ref{nnn_3}) for the calculation of the second
 (interacting)
 term in Eq.~(\ref{eq:n1}),
 \beq
 F^{\text{int}}_{\text{norm}} = - \pi^2 T^2 N_F {\bar g}^\gamma \sum_{m,m'} \frac{{\text{sign}} (\omega_m \omega_{m'})}{
 \left(|\omega_m - \omega_{m'}|^2 + \omega^2_D\right)^{\gamma/2}}
 \eeq
  Summing up over $M$ positive and $M$ negative Matsubara frequencies, we obtain
  after some straightforward algebra that for $T \gg \omega_D$ the second term in the free energy of Eq. (\ref{eq:n1}) is
  \bea
  &&F^{\text{int}}_{\text{norm}}  = -N_F  \pi T \Lambda  \left( \frac{{\bar g}}{{\omega_D}}  \right) ^\gamma \nonumber \\
 && + N_F {\bar g}^\gamma \Lambda^{2-\gamma} \frac{2(2^{-\gamma}-1)}{(1-\gamma)(2-\gamma)} \nonumber \\
  &&-N_F {\bar g}^\gamma  (2\pi T)^{1-\gamma} \Lambda \zeta (\gamma) \nonumber \\
   && + \frac{3}{2} N_F {\bar g}^\gamma (2\pi T)^{2-\gamma} \zeta (\gamma-1) + O(1/\Lambda)
   \label{nnn_4}
  \eea
  This result is valid for $\gamma \neq 1$ (we discuss $\gamma =1$ below).
  In obtaining this expression we used the fact that $\sum_1^m 1/k^\gamma$ is a Harmonic number $H_\gamma (m)$,
   used
   the summation formula~\cite{kronenburg2011some}
  \beq
  \sum_{m=1}^n H_\gamma (m) = (n+1) H_\gamma (n) - H_{\gamma-1} (n)
\eeq
 and the asymptotic expansion of $H_\gamma (m)$
  at large $m$
 \beq
 H_\gamma (m) = \frac{m^{1-\gamma}}{1-\gamma} + \zeta (\gamma-1) + O\left(\frac{1}{m^\gamma}\right).
  \eeq
  The $T^{1-\gamma}$ term in
  Eq.~(\ref{nnn_4})
  comes from fermions with $|\omega_m - \omega_{m'}| = O(T)$, while $\omega_m, \omega_{m'} = O(\Lambda)$. The subleading $T^{2-\gamma}$ term  with $\Lambda$-independent prefactor  comes from fermions with frequencies
    $|\omega_m|, |\omega_{m'}| = O(T)$.

Differentiating twice over temperature, we obtain for the specific heat
\beq
C (T) \approx 2\pi \Lambda N_F \left(\frac{\bar g}{2\pi T}\right)^\gamma Q_\gamma \left(1 +  O\left(\frac{T}{\Lambda}\right)\right)
\label{nnn_5}
\eeq
where
\beq
Q_\gamma = \gamma (\gamma -1) \zeta (\gamma)
\eeq
This $Q_\gamma$ is positive for all $\gamma >0$ and evolves smoothly through $\gamma =1$.  At $\gamma=1$,
the evaluation of
$F_{\text{norm}}^{\text{int}}$
 has to be done with some extra care. The outcome is that the term in $F^{\text{int}}_{\text{norm}}$, which yields $C(T) \propto 1/T$, scales as $\log T$.

We see that $C(T) \propto 1/T^\gamma$, and the prefactor is positive and scales with the upper energy cutoff $\Lambda$.
The latter is generally
a fraction of the Fermi energy $E_F$.
Using that
$N_F \sim N/E_F$, where $N$ is the number of electrons, we find that  $C(T)$ per particle is of order
$({\bar g}/T)^\gamma$.

On a more careful look we found that the  $1/T^\gamma$ term in $C(T)$ originates from
 the non-analytic  $T^{1-\gamma}$ term in the fermionic self-energy at large $\omega_m$.
Indeed, the self-energy is
\beq
\Sigma (\omega_m)
= {\bar g}^\gamma \left( \frac{\pi T}{\omega^\gamma_D} + H_\gamma (m)(2\pi T)^{1-\gamma}\right)
\label{nnn_19}
\eeq
 up to corrections of order
 $T/\Lambda^{\gamma}$
 (we set $\omega_m >0$ for definiteness).
    At large $m>0$, $H_\gamma (m)(2\pi T)^{1-\gamma} \approx  (\omega_m)^{1-\gamma}/(1-\gamma) + \zeta (\gamma) (2\pi T)^{1-\gamma}$,
     hence at  $\omega_m \gg T$~
\footnote{ The prefactor for the
$T^{1-\gamma}$
is negative for $\gamma <1$, hence for these $\gamma$, $\Sigma (\omega_m)$ has a negative temperature-dependent offset. This is consistent with Ref. \cite{avi}, where a negative offset has been found for $\gamma=1/3$ and $1/2$.}
 \beq
\Sigma (\omega_m) \approx {\bar g}^\gamma \left( \frac{\pi T}{\omega^\gamma_D} +
\frac{\omega^{1-\gamma}_m}{1-\gamma} + \zeta (\gamma) (2\pi T)^{1-\gamma} \right)
\label{nnn_19_1}
\eeq
The first two terms in the self-energy do not contribute to $C(T)$, while the last  $T^{1-\gamma}$
term
  gives rise to $1/T^{\gamma}$ scaling of $C(T)$,
   as one can straightforwardly verify by substituting this asymptotic form into
     (\ref{nnn_19}) and then into $F^{\text{int}}_{\text{norm}}  = - 2 \pi T N_F \sum_{m \geq 0}
     \Sigma(\omega_m)$

Our result for $C(T)$ differs from the one in Ref. \cite{yuz_1}. The authors of that work
regularized the free energy with the aim to eliminate the
 singular contribution from thermal fluctuations. For this they subtracted from the free energy
 the term
\beq
- \pi^2 T^2 N_F
\sum_{m,m'} V(\omega_m - \omega_{m'})
\label{nnn_7}
\eeq
This substraction  changes $F^{\text{int}}_{\text{norm}}$ in (\ref{eq:n1}) to
\beq
F^{\text{int}}_{\text{norm, reg}} = - \pi^2 T^2 N_F
\sum_{m,m'}  V\left(\omega_m - \omega_{m'}\right)
\left({\text{sign}} (\omega_m \omega_{m'})-1\right)
 \label{nnn_6}
  \eeq
Because the numerator is nonzero only when $\omega_m$ and
$\omega_{m'}$
have opposite sign, thermal contribution from $m=m'$ vanishes, hence
 (\ref{nnn_6}) can be evaluated right at $\omega_D =0$.
   This eliminates
   the $\Lambda$-dependent terms
   in the free energy and the entropy.
 However, the added term also introduces an additional temperature dependence, not present in the original free energy, and hence affects the specific heat.
  Specifically,
  $F^{\text{int}}_{\text{norm, reg}} = 4 \pi^2 N_F \zeta (\gamma-1)({\bar g}/2\pi)^\gamma  T^{2-\gamma}$,
   which is similar to the last term in (\ref{nnn_4}). The $\Lambda T^{1-\gamma}$ term gets cancelled out.
    Differentiating $F^{\text{int}}_{\text{norm, reg}}$ over $T$, the authors of \cite{yuz_1} found
     the  interaction contribution to specific heat $C(T) = A_\gamma T^{1-\gamma}$, where $A_\gamma$ changes sign at $\gamma =1$ and is negative for $1<\gamma <2$.  They conjectured that this may imply that the normal state of the $\gamma$-model is unstable for such $\gamma$.
  In our view, a positive  $C(T)$, which we obtained
from the original free energy with no added term,
   is the actual specific heat in the $\gamma$-model.
   \footnote{We also note in passing that the regularization, aimed to eliminate the singular thermal piece, is not unique. In particular, one can subtract $- \pi^2 T^2 N_F {\bar g}^\gamma \sum_{m,m'} (1+b - b {\text{sgn}} (\omega_m \omega_{m'}))/(|\omega_m - \omega_{m'}|^\gamma + \omega^\gamma_D)$ with arbitrary $b$. One can then choose $b$ such that even after regularization $C(T)$ remains positive down to any chosen $T$. }

 The specific heat for the underlying fermionic model with $\gamma =1/3$ has been recently analyzed using Quantum Monte Carlo technique~\cite{Erez_c}.  In the  regime of vanishing bosonic mass the specific heat increases with decreasing $T$ at strong enough coupling.  This is consistent with $C(T) \propto 1/T^{1/3}$.  It is however, possible that the observed increase is due to superconducting fluctuations above $T_c$, as the authors of \cite{Erez_c} suggested.  To address this issue in full, one has to compute $C(T)$ for the underlying Ising-nematic model in 2D.

\subsection{Applicability of the $\gamma-$model at finite $T$ and $m \to 0$}

Eq. (\ref{nnn_4}), taken at a face value, implies that
 the entropy $S(T) \approx \pi N_F \Lambda  \left( \frac{{\bar g}}{{\omega_D}}  \right) ^\gamma$ is infinite at $\omega_D =0$ at {\it any} $T$. This result is clearly an
 artifact,
 as we show below.  To understand this, we analyze the
  applicability  of the normal state analysis.

\begin{figure}
\centering
\includegraphics[scale=1]{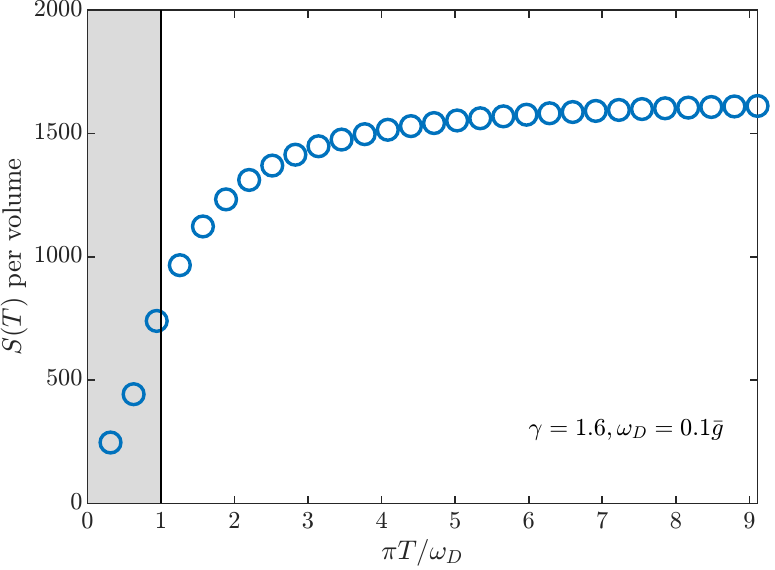}\caption{Entropy per volume $S(T)$ as a function of $\pi T/\omega_D$, where we take the parameters $\omega_D = 0.1 {\bar g}$ and $\gamma=1.6$. A cutoff $\Lambda \sim 15 {\bar g}$ is used to regularize the ultra-violet divergence. The grey region highlights the Fermi liquid regime where $\pi T \ll \omega_D$.}
\label{fig:entropy}
\end{figure}

  First,
  as
  we already stated, the analysis leading to Eq. (\ref{nnn_4}) is valid only at $T > \omega_D$. At smaller $T$,
  the system displays a Fermi liquid behavior and the entropy  and the specific heat per particle scale as $A T$, where
 $A \sim \Lambda {\bar g}^\gamma
 N_F /\omega^{\gamma +1}_D$.
  We show the numerical result for the entropy in Fig.~\ref{fig:entropy}.

Second, the Eliashberg equation for the self-energy is valid as long as vertex corrections are small.  A straightforward
 computation of vertex corrections at a finite
 $T \gg \omega_D$
when thermal fluctuations are the strongest,  shows that
  they are small as long as
  \beq
  \left(\frac{\bar g}{\omega_D}\right)^\gamma \ll \frac{\Lambda}{T}
  \eeq
 or, equivalently, as long as $T \ll T_{max}$,
where $T_{max} \sim \Lambda (\omega_D/{\bar g})^\gamma$.
 The temperature window, where Eq. (\ref{nnn_4}) and the corresponding expressions for the entropy and the specific heat are valid, is then
 \beq
  \omega_D \ll T \ll T_{max}
  \eeq
  This window formally collapses when $\omega_D \to 0$.  A way to keep it finite is to consider the double  limit $\omega_D \to 0$, $\Lambda \to \infty$ (i.e., $E_F \to \infty$) such that
  $T_{max}$
  remains finite.

  Third, normal state description of the $\gamma-$model is valid only above the pairing instability temperature $T_p$. The latter is  of order ${\bar g}$ for a generic $\gamma$.  The normal state analysis is then applicable when
   $T_{max} > {\bar g}$.
  This strengthen the requirement on the double limit: it should be $\Lambda \omega^{\gamma}_D >
 {\bar g}^{\gamma +1}$.
   At the lower boundary  $T = T_p \sim {\bar g}$, we then have
   \beq
   S (T_p) \sim \frac{\Lambda^2 N_F}{\bar g}, ~~~ C(T_p) \sim (\Lambda N_F) \leq N
   \eeq
   This implies that the specific heat per particle at $T_p < T < T_{max}$ is at most $O(1)$.
 Still, this $C(T)$ is much larger than the jump of $C(T)$ at $T_p$, which scales as
 ${\bar g} N_F/N \sim {\bar g}/E_F$ per particle.
   The analysis of the free energy at $T > T_{max}$ requires one to include series of vertex corrections and is beyond the scope of this paper.  We conjecture that at such $T$,  $C(T) \propto T^{1-\gamma}$ with a positive coefficient.
We also emphasize that both thermal self-energy and the singular part of the $T=0$ self-energy at $\gamma >1$
  cancel out in the Eliashberg equation for the superconducting gap.  To avoid vertex corrections at $\omega_D \to 0$, one still need to employ the double limit $\omega_D \to 0$, $E_F \to \infty$,
   \footnote{Vertex corrections at $T=0$  become singular at $\gamma >1$, and to keep them small one
   needs
   to take the double limit $\omega_D \to 0, E_F \to \infty$ such that
   $E_F \omega^{\gamma-1}_D \ll {\bar g}^\gamma$.},
 however the
Eliashberg
gap equation does not contain $E_F$, so the need for the double limit does not affect the pairing problem.

  There is an additional condition, related to derivation of the Eliashberg
  equation
  at a given $\gamma$ from the
    corresponding microscopic model with momentum and frequency dependent $V(k-k')$.  The derivation assumes that bosons are slow modes compared to fermions, in which case the momentum integration can be factorized such that
     the integrations transverse and along the Fermi surfaces are performed separately in fermionic and bosonic propagators, respectively.  This holds at $T=0$, but may or may not
     hold at a finite $T$, when thermal fluctuations are present~\cite{avi}.
      As a result,  the thermal  self-energy in the $\gamma$-model and in the underlying microscopic model
      may differ even without vertex corrections.
 Finally, the free energy of fermions interacting with a near-critical boson contains additional contribution from bosonic propagator:
 \beq
 F_{\text{bos}} = \frac{1}{2} T \sum_{q} \left[\log{D^{-1} (q) + \Pi (q) D(q)}\right]
 \eeq
  where $q = ({\vec q}, \Omega_m)$,  $D(q)$ is the bosonic propagator, and $\Pi (q)$ is the bosonic polarization
   ($D^{-1} (q) = D^{-1}_0 (q) - \Pi (q)$,
   where $D_0$ is a bare bosonic propagator).
  This contribution again has to be calculated within the original microscopic model with momentum and frequency
   dependent $D(q)$.

\section{A generic $\gamma <2$. A discrete set of solutions of the gap equation at $T=\omega_D=0$, and free energy expansion around them}
\label{sec:generic}

In this section, we analyze the profile of the free energy in the superconducting state for a generic $\gamma<2$.

To set the stage for our analysis, we first briefly review the results for
 $\Delta (\omega_m)$ at stationary points at $T=0$.
We consider even-frequency gap functions,
 for which $\Delta (-\omega_m) = \Delta (\omega_m)$. For the analysis of odd-frequency gap functions, see Refs.~\cite{Wu2022,Balatsky_2020}.

\subsection{Discrete set of solutions, $\Delta_n (\omega_m)$}
\label{sec:generic1}

For large enough $\omega_D \geq {\bar g}$, the normal state is a Fermi liquid with
$\Sigma (\omega_m) \approx \lambda \omega_m$, where $\lambda = ({\bar g}/\omega_D)^\gamma (\gamma/2)$.
  The gap equation has a single sign-preserving solution $\Delta_0 (\omega_m)$. It has a finite
     value at $\omega_m=0$ and decreases as $1/|\omega_m|^\gamma$ at large frequencies.

When $\omega_D/{\bar g}$ gets smaller, new solutions of the gap equation appear one-by-one. We label these solutions as $\Delta_n (\omega_m)$, where $n=1,2...$.  A function $\Delta_n (\omega_m)$ changes sign $n$ times along the positive Matsubara axis. These solutions are topologically distinct as each zero of $\Delta_n (\omega_m)$ is the center of a  dynamical vortex on the complex frequency plane $z=\omega^{\prime}+i \omega^{\prime\prime}$.

At  $\omega_D =0$, the  number of solutions become infinite, and they
form a discrete set, in which $n$ ranges from zero to infinity. The overall magnitude of $\Delta_n (\omega_m)$ decreases exponentially with $n$, and the ``end point'' of the set, $\Delta_{\infty} (\omega_m)$, is the solution of the linearized gap equation
 \beq
   \Delta_\infty (\omega_m) = \frac{{\bar g}^{\gamma}}{2} \int d \omega'_m \frac{\Delta_\infty (\omega'_{m}) - \Delta_\infty (\omega_m) \frac{\omega'_{m}}{\omega_m}}{|\omega'_{m}|} \frac{1}{|\omega'_m-\omega_m|^\gamma}
     \label{ss_11_02}
  \eeq
As a proof that such set does exist, we obtained the exact analytical solution of this equation (Refs. \cite{paper_1,paper_4,paper_5}). At small $\omega_m \ll {\bar g}$, $\Delta_\infty (\omega_m)$ oscillates as a function of $\log {({\bar g}/|\omega_m|)^\gamma}$ as
\beq
\Delta_{\infty} (\omega_m) =2 \epsilon {\bar g}^{1-\gamma/2} |\omega_m|^{\gamma/2}  \cos{\left(\beta(\gamma) \log{({\bar g}/|\omega_m|)^\gamma} + \phi (\gamma)\right)},
 \label{eq:7}
 \eeq
where $\beta(\gamma)$
and $\phi (\gamma)$ are $\gamma$-dependent parameters and $\epsilon$ is an arbitrarily small dimensionless overall factor.  At larger $\omega_m$,  $\Delta_{\infty} (\omega_m) \propto \epsilon/|\omega_m|^\gamma$.
 In simple terms, at small $\omega_m$, $\Delta_\infty (\omega_m)$  in the l.h.s. of (\ref{ss_11_02}) can be neglected, which is equivalent to neglecting the bare $\omega_m$ in the fermionic propagator  compared to the self-energy, as one can easily verify.  The solution of (\ref{ss_11_02}) without the l.h.s. is the sum of the two power-laws $\Delta_\infty (\omega_m) \propto |\omega_m|^a$  with complex exponents $a = \gamma/2 \pm i \beta_\gamma$. This is  Eq. (\ref{eq:7}).  A relative phase $\phi$ is a free parameter in this approximation. At large $\omega_m \gg {\bar g}$, the dependence $\Delta_{\infty} (\omega_m) \propto 1/|\omega_m|^\gamma$ follows from a'posteriori verified assumption that typical $\omega'_m$ in (\ref{ss_11_02}) are of order ${\bar g}$, in which case one can approximate  $1/|\omega'_m-\omega_m|^\gamma$ in the r.h.s. of (\ref{ss_11_02}) by $1/|\omega_m|^\gamma$.  The  exact solution shows that there exists a particular $\phi$, for which  $\Delta_\infty (\omega_m)$  gradually transforms between the limits of small and large $|\omega_m|/{\bar g}$.

Qualitative understanding of the appearance of a  discrete set of solutions of Eq.~(\ref{ss_11_0})
can be obtained by expanding the r.h.s.  of the non-linear gap equation (\ref{ss_11_0}) in powers of $D(\omega_m)=\Delta(\omega_m)/\omega_m$ and analyzing the structure of perturbation series in the gap amplitude $\epsilon$, which we now treat as finite.   The analysis is tedious but straightforward.  At small $\omega_m \ll {\bar g}$, the expansion holds in powers of
  $\epsilon^* = \epsilon ({\bar g}/|\omega_m|)^{1-\gamma/2}$ and yields
 \bea
\Delta (\omega_m) &=& 2 \epsilon^* |\omega_m|  \left[ Q_1 \cos({\psi}) + (\epsilon^*)^2 Q_3 \cos({3\psi + \phi_3}) \right. \nonumber \\
&& \left. + (\epsilon^*)^4 Q_5 \cos({5\psi + \phi_5}) + ...\right]
 \label{eq:8}
 \eea
 where
 \bea
 \psi &=& \beta(\gamma)
 \log{({\bar g}/|\omega_m|)^\gamma}  + \phi_{\epsilon^*}
   \label{eq:8a} \\
   \phi_{\epsilon^*} &=& \phi (\gamma) +  s_1 (\epsilon^*)^2 + s_2 (\epsilon^*)^4 + ..., \label{eq:8aa}\\
  \phi_{2k+1} &=& t_{0,2k+1}+t_{1,2k+1} (\epsilon^*)^2 + t_{2,2k+1} (\epsilon^*)^4 + ..., \label{eq:8b}\\
 Q_{2k+1} &=& r_{0,2k+1}+r_{1,2k+1} (\epsilon^*)^2 + r_{2,2k+1}(\epsilon^*)^4 + .... \label{eq:8c}
 \eea
The coefficients
$s_i$,
$t_{l, 2k+1}$ and $r_{l,2k+1}$ are $\gamma$-dependent functions, except for $r_{0,1} =1$.  We present the exact expressions in Appendix~\ref{sec:perturb}. For a generic $\gamma <2$, they all are of order one.

\begin{figure}
\centering
\includegraphics[scale=1]{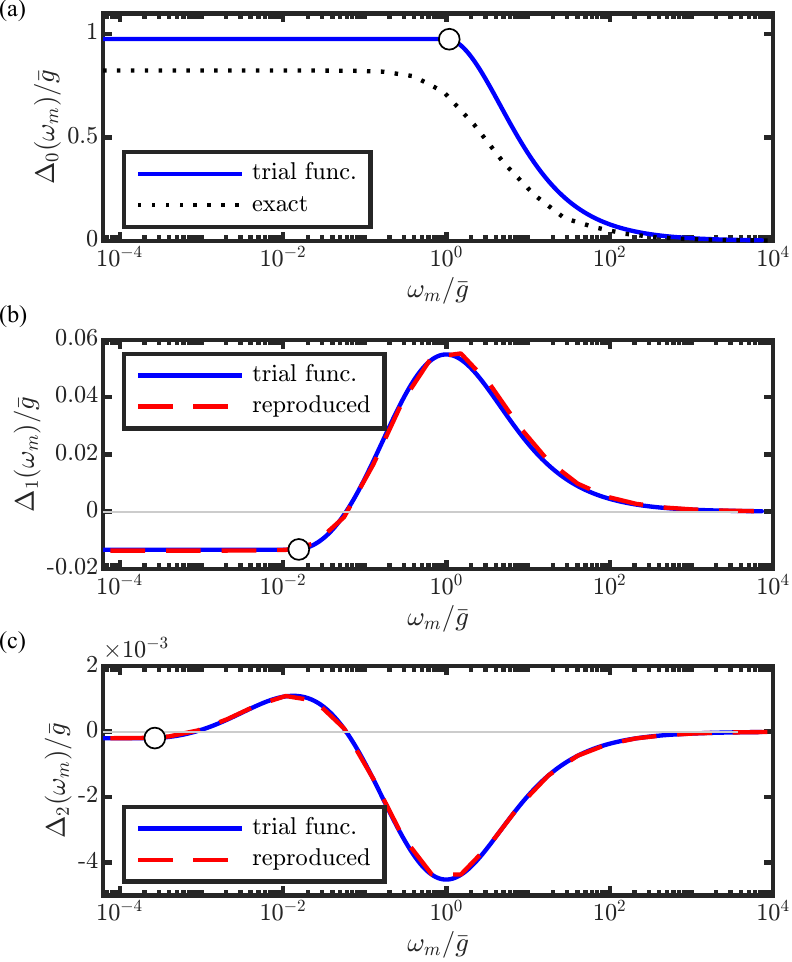}\caption{Trial gap function (blue solid line) of (a) $n=0$, (b) $n=1$, and (c) $n=2$ solutions, where $\gamma=0.8$. Exact gap function of $n=0$ solution is shown by the black dotted line.
In panels (b),(c), the reproduced gap function by substituting the trial gap function back to the gap equation is shown by red dashed lines. Circles
indicate
$\omega_c$,
below which the trial gap function $\Delta_n (\omega_m)$  becomes a constant.}
\label{fig:gap_func_lt1}
\end{figure}

We see that the expansion
 generates higher-order harmonics with multiples of the argument $\psi$ and phases $\phi_{2k+1}$.
 These extra phase factors
 are most relevant to our reasoning. Without them, oscillations from
  $\cos{\psi}, \cos{3\psi}, \cos{5\psi}$...
  would all be in phase,
  and $\Delta (\omega_m)$ would oscillate as a function of $\log ({\omega_m}/{\bar g})$ down to $\omega_m =0$. Because of $\phi_{2k+1}$, the positions of the nodal points in different harmonics are shifted by different amounts, and when the corrections from the non-linear terms become of order one, oscillations get destroyed.  At smaller frequencies, it is natural to assume that $\Delta (\omega_m)$ approaches a constant value $\Delta (0)$.
  For a generic  $\gamma <2$, this transformation occurs  when $\epsilon^*$ becomes of order one, i.e., at a critical Matsubara frequency $|\omega_m| \sim \omega_c = {\bar g} \epsilon^{2/(2-\gamma)}$.
   To first approximation, the gap function then follows $\Delta_\infty (\omega_m)$ from (\ref{eq:7}) down to
   $\omega_c$  and saturates at $\Delta (\omega_m) \sim \omega_c$ at smaller frequencies.  Such a gap function is smooth
    (to discontinuity of a derivative) if $\omega_m = \omega_c$ is one of the extrema of $\Delta_\infty (\omega_m)$.
   The largest $\omega_c = O({\bar g})$ is above the highest frequency where
    $\Delta_\infty (\omega_m)$ changes sign. The corresponding  solution of the non-linear gap equation is sign-preserving $\Delta_0 (\omega_m)$, and the corresponding $\epsilon_0 = O(1)$. However, this is not the only option --
    because $\Delta_\infty (\omega_m)$ oscillates,  $\omega_c$ can be chosen as  an extremum  of  $\Delta_\infty (\omega_m)$ after it changes sign
    $n$ times.
      In this case,
      $\omega_c \sim {\bar g} e^{-n \pi/(\beta({\gamma}) \gamma)}$.
     The corresponding solution is $\Delta_n (\omega_m)$ that changes sign $n$ times at positive $\omega_m$,
     the corresponding
     $\epsilon_n \sim e^{-\pi n(2-\gamma)/(2\beta({\gamma}) \gamma)}$,
     and the corresponding $\Delta_n (0)$ is
\beq
 \Delta_n (0) \sim
 \omega_c
 \sim {\bar g}  e^{-\pi n/(\beta({\gamma})\gamma)},
 \label{eq:9}
 \eeq
At large $n$, both $\epsilon_n$ and $\Delta_n (0)$
 decrease exponentially with $n$.
 In Fig.~\ref{fig:gap_func_lt1}  we show the trial functions $\Delta_n (\omega_m)$, constructed this way, for  a representative $\gamma =0.8$ and for $n=0,1,2$.
We see that this reasoning works quite well. Namely, upon substituting
 the trial
 functions into the r.h.s. of Eq.~(\ref{ss_11_0}), one recovers the same function
  in the l.h.s. with reasonable accuracy
   (more on this below).

These results present our most direct evidence of a discrete set of the gap functions $\Delta_n (\omega_m)$, which satisfy the non-linear gap equation.  The other evidences are (i) the exact analytical result for $\Delta_\infty (\omega_m)$,
 (ii) strong numerical evidence for the existence of a discrete set of the solutions of the non-linear differential equation, which well approximates Eq.~(\ref{ss_11_0}) at small $\gamma$ (Ref.~\cite{paper_1}), and (iii) a highly accurate numerical solution of the linearized gap equation at a finite $T$, yielding a discrete set of critical pairing temperatures  $T_{p,n}$ for $n$ up to $17$~\cite{paper_2}. These critical $T_{p,n}$ decrease exponentially with $n$, consistent with $\Delta_n (0)$ in Eq.~(\ref{eq:9}), and the corresponding eigenfunction changes sign $n$ times along the positive Matsubara axis.

 We emphasize that these results hold for  $\gamma$, which are not particularly close to $2$.  For $\gamma$ close to 2, the analysis
  of $\Delta_n (\omega_m)$ has to be modified, as we show in Sec. \ref{sec:limit}.

We also emphasize that all functions $\Delta_n (\omega_m)$ from the discrete set can be analytically continued into the upper half-plane of complex
frequency. We verified this analytically for $n = \infty$ by replacing $i\omega_m$ by $z = \omega' +i \omega^{''}$ in the analytical formula for $\Delta_{\infty} (\omega_m)$, and we checked this numerically for $n=0,1,2$ by extending $\Delta_n (\omega_m)$ that we just obtained,  to the upper frequency half-plane using Pade approximants.

\subsection{Free energy expansion near $\Delta_n (\omega_m)$}
\label{sec:generic2}

We now analyze the form of the free energy $F_{sc}$ near the stationary solutions $\Delta_n (\omega_m)$. Our goal here is to understand whether these solutions correspond to local minima or saddle points.  To address this issue we
 expand $F_{sc}$ to second order in  $\delta \Delta (\omega_m) = \Delta (\omega_m) - \Delta_n (\omega_m)$ and analyze
 the  quadratic form in $\delta \Delta (\omega_m)$.

 We discuss the expansion of the free energy in detail in Appendix \ref{app:B} and here just present the result.  At $\omega_D =0$, the free energy is $F_{sc} = F_{sc,n} + \delta F_{sc,n}$, where
   \bea
   F^{(0)}_{sc,n} &=& - N_F 2\pi T \sum_{m} \frac{\omega^2_m}{\sqrt{\omega^2_m + \Delta^2_n (\omega_m)}}\nonumber \\
  && -  \pi^2 T^2 N_F \sum_{m, m'} V(\omega_m-\omega'_m) \frac{\omega_m \omega'_m + \Delta_n (\omega_m) \Delta_n (\omega'_m)}{\sqrt{\omega^2_m + \Delta^2_n (\omega_m)} \sqrt{(\omega'_m)^2 + \Delta^2_n (\omega'_m)}}
  \label{eq:10_a}
  \eea
 is the free energy for stationary $\Delta_n (\omega_m)$, and
 \beq
\delta F_{sc,n} =  N_F \pi^2 T^2
\sum_{m, m' \geq 0} \frac{1}{D_n (\omega_m)D_n (\omega_{m'})}
   \frac{1}{(1 + D^2_n (\omega_{m}))^{3/2}
  (1 + D^2_n (\omega_{m'}))^{3/2}} Q_{m,m'}
\label{eq:23_a}
\eeq
with
\bea
&& Q_{m,m'} = {\bar g}^\gamma \times \nonumber \\
  && \left[
  \left(\delta D (\omega_m) D_n (\omega_m) - \delta D (\omega_{m'}) D_n (\omega_{m'})\right)^2 \left(\frac{1}{|\omega_m-\omega_{m'}|^\gamma} + \frac{1}{|\omega_m+\omega_{m'}|^\gamma}\right) \nonumber
  \right.
  \\
  && \left.
  + \frac{\left(\delta D (\omega_m) D^2_n (\omega_m) - \delta D (\omega_{m'}) D^2_n (\omega_{m'})\right)^2}{|\omega_m-\omega_{m'}|^\gamma} +
\frac{\left(\delta D (\omega_m) D^2_n (\omega_m) + \delta D (\omega_{m'}) D^2_n (\omega_{m'})\right)^2}{|\omega_m+\omega_{m'}|^\gamma}
\right]
\label{eq:23_aa}
 \eea
 is the variation of the free energy to
 the
second order in $\delta \Delta (\omega_m)$. In (\ref{eq:23_aa}) we
 introduced $D_n (\omega_m) = \Delta_n (\omega_m)/\omega_m$ and $\delta D_m (\omega_m) = \delta \Delta (\omega_m)/\omega_m$ and restricted to even-frequency variations $\delta \Delta (\omega_m) = \delta \Delta (-\omega_m)$.

Because $Q_{m,m'}$ is positive (the sum of full squares with positive prefactors),  $\delta F_{sc,0}$ is  definitely positive for the sign-preserving $D_0 (\omega_m)$.
For $n >0$ this is not guaranteed, however,  as $D_n (\omega_m)$ changes sign $n$ times along the positive Matsubara axis, and the product $D_n (\omega_m)D_n (\omega_{m'})$ in the denominator of (\ref{eq:23_aa}) can be of either sign.

\begin{figure}
\centering
\includegraphics[scale=1]{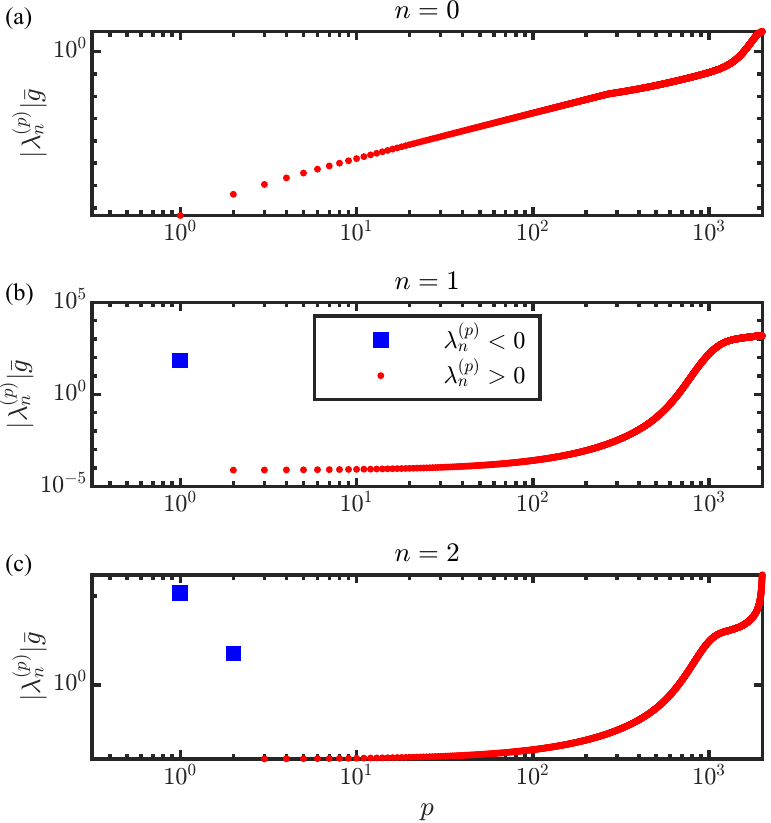}\caption{Eigenvalues of the kernel matrix ${\cal K}_{\omega,\omega^{\prime}}$ for (a) $n=0$, (b) $n=1$, and (c) $n=2$ solutions, where $\gamma=0.8$.}
\label{fig:eigs_lt1}
\end{figure}

It is convenient to re-express $\delta F_{sc,n}$  as
\beq
\delta F_{sc,n} = (\pi T)^2 N_F \sum_{m,m'} {\cal K}_{n}(\omega_m, \omega_{m'}) \delta \Delta(\omega_m) \delta \Delta(\omega_{m'}),
\label{eq:n4}
\eeq
where
\bea
{\cal K}_{n}(\omega_m,\omega_{m'})& = &  {\delta_{m m'}\over 2\pi T} {\omega_m^2 Z_n (\omega_m) \over (\omega_m^2+ \Delta_n(\omega_m)^2)^{3/2} } -  V(\omega_m-\omega_{m'}) \nonumber  \\
& & \times {\omega_m \omega_{m'} (\omega_m \omega_{m'}+\Delta_n(\omega_m)\Delta_{n}(\omega_{m'})) \over  (\omega_m^2+ \Delta_n(\omega_m)^2)^{3/2}( \omega_{m'}^2+ \Delta_n(\omega_{m'})^2)^{3/2} },
\eea
and $Z_n (\omega_m) = 1 + (\pi T/\omega_m) \sum_{m^{''}} V(\omega_m-\omega_{m^{''}}) \omega_{m^{''}}/\sqrt{\omega^2_{m^{''}} +\Delta^2_n (\omega_{m^{''}})}$ is the
inverse
 quasiparticle residue at the stationary point.

The symmetric matrix ${\cal K}_{n}(\omega_m,\omega_{m^{\prime}})$ can be rewritten in the diagonal form as
\beq
 {\cal K}_{n}(\omega_m,\omega_{m^{\prime}}) = \sum_{p} \lambda^{(p)}_n \Delta^{(p)}_n (\omega_m) \Delta^{(p)}_n (\omega_{m'}),
 \eeq
  where $\lambda^{(p)}_n$ ($p=1,2...$) are the eigenvalues, and  $\Delta^{(p)}_n (\omega_m)$ are the eigenfunctions, which we will have to find.
    The eigenfunctions form a complete and orthogonal set and obey the normalization condition
  \beq
  T \sum_m \Delta^{(p)}_n (\omega_m) \Delta^{(p')}_n (\omega_m)  = \delta_{p,p'}
\eeq
Expanding $ \delta \Delta (\omega_m) = \sum_p C^{p}_n \Delta^{(p)}_n (\omega_m)$ and substituting into Eq.~(\ref{eq:n4}), we obtain $\delta F_{sc,n} = \sum_p \lambda^{p}_n (C^{p}_n)^2$.
Obviously, if there is a negative $\lambda^{p}_n$ for at least one value of $p$,  the free energy gets reduced by deviations from $\Delta_n (\omega_m)$ along this direction. In this situation, $\Delta_n (\omega_m)$ is a saddle point of the free energy. If  $\lambda^{p}_n >0$ for all $p$, $\Delta_n (\omega_m)$ is a local minimum.

We obtained the eigenvalues and eigenfunctions of ${\cal K}_{n}(\omega_m,\omega_{m^{\prime}})$ for $n=0,1$ and $2$ numerically
  on a non-uniform mesh of $2000 \times 2000$ Matsubara frequencies.
   A non-uniform grid was chosen to  reach extremely low temperatures $\sim 10^{-10}{\bar g}$ (see
    Ref.~\cite{paper_2} for detail).  For stationary $\Delta_n (\omega_m)$
    we used the trial functions,  constructed in the previous section (Fig.~\ref{fig:gap_func_lt1}).

\begin{figure}
\centering
\includegraphics[scale=1]{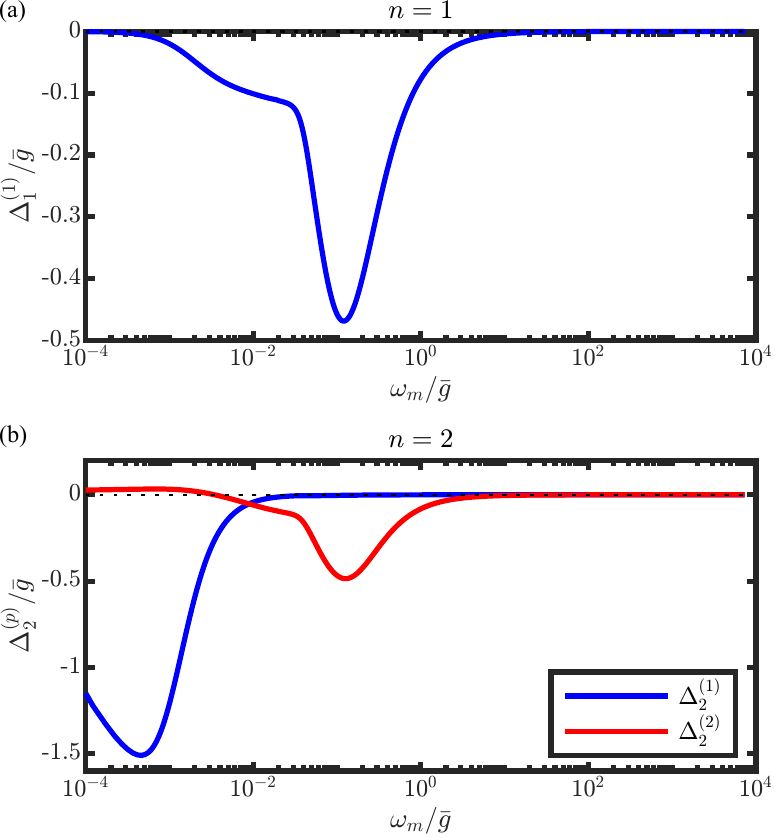}\caption{Eigenfunctions corresponding to the negative eigenvalues of the kernel matrix ${\cal K}_{\omega,\omega^{\prime}}$ for (a) $n=1$ and (b) $n=2$.
  We set $\gamma=0.8$.}
\label{fig:func_lt1}
\end{figure}

We used two  computational procedures to obtain $\lambda^{(p)}_n$ and $\Delta^{(p)}_n (\omega_m)$. In the first, we
  set $\delta \Delta_n (\omega_m)$ to be  real and even under $\omega_m \to - \omega_m$, but did not require that it is
   analytic in the upper half-plane of frequency.  Like we just said, our $\Delta_n (z)$ are analytic functions of $z = \omega' +i \omega^{''}$ at $\omega^{''} >0$. However, fluctuating $\delta \Delta_n (\omega_m)$ do not have to be analytic.
 In the second procedure, we restricted $\delta \Delta_n (\omega_m)$ to analytic functions  in the upper half-plane
  (see Appendix \ref{app:C} for details).
 We found the same structure of eigenvalues and eigenfunctions in the two cases. This equivalence implies that for physically relevant fluctuations,   $\delta \Delta_n (z)$ is an analytic function of $z$.

\begin{figure}
\centering
\includegraphics[scale=1]{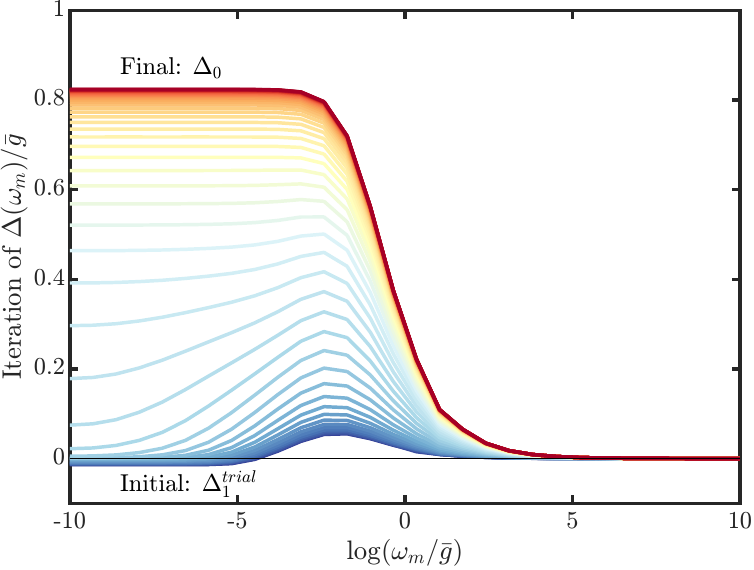}
\caption{Iteration process that solves the stationary points of the non-linear gap equation, Eq.~(\ref{ss_11_0}), which starts from the input $\Delta^{\text{trial}}_1(\omega_m)$ and saturates at $\Delta_0(\omega_m)$.}
\label{fig:iteration}
\end{figure}

 Below we present the results obtained using the first computational procedure.  We show the eigenvalues $\lambda^{(p)}_n$ for a representative $\gamma =0.8$ in Fig. \ref{fig:eigs_lt1}.  We see that for $n=0$,
 all $\lambda^{(p)}_0$ are positive, as expected, i.e.  $\Delta_0 (\omega_m)$  is a minimum of the free
  energy functional (a global minimum as we will see momentarily).
For $n=1$, there is one \textit{negative} eigenvalue $\lambda^{(1)}_1$. The corresponding eigenfunction $\delta \Delta^{(1)}_1 (\omega_m)$ sets the ``direction'', along which the free energy is reduced upon deviations from $\Delta_1 (\omega_m)$. We show $\delta \Delta^{(1)}_1 (\omega_m)$ in Fig.~\ref{fig:func_lt1} (a). We see that it is sign-preserving and shifts the gap function towards the one with  $n=0$.
 This can also be seen by analyzing how our trial gap function $\Delta_1 (\omega_m)$  evolves under iterations.
 As we showed in Fig.~\ref{fig:gap_func_lt1},
 this function satisfies the non-linear gap equation quite well.  Still, it is not exactly
  $\Delta_1 (\omega_m)$, and likely contains some amount of  $\delta \Delta^{(1)}_1 (\omega_m)$.
Then it is natural to expect that after a number of iterations the trial $\Delta_1 (\omega_m)$ will start flowing towards $\Delta_0 (\omega_m)$. Fig. \ref{fig:iteration}
 shows that this is exactly the case: after the large number of iterations the trial $\Delta_1 (\omega_m)$ approaches $\Delta_0 (\omega_m)$.

\begin{figure}
\centering
\includegraphics[scale=1]{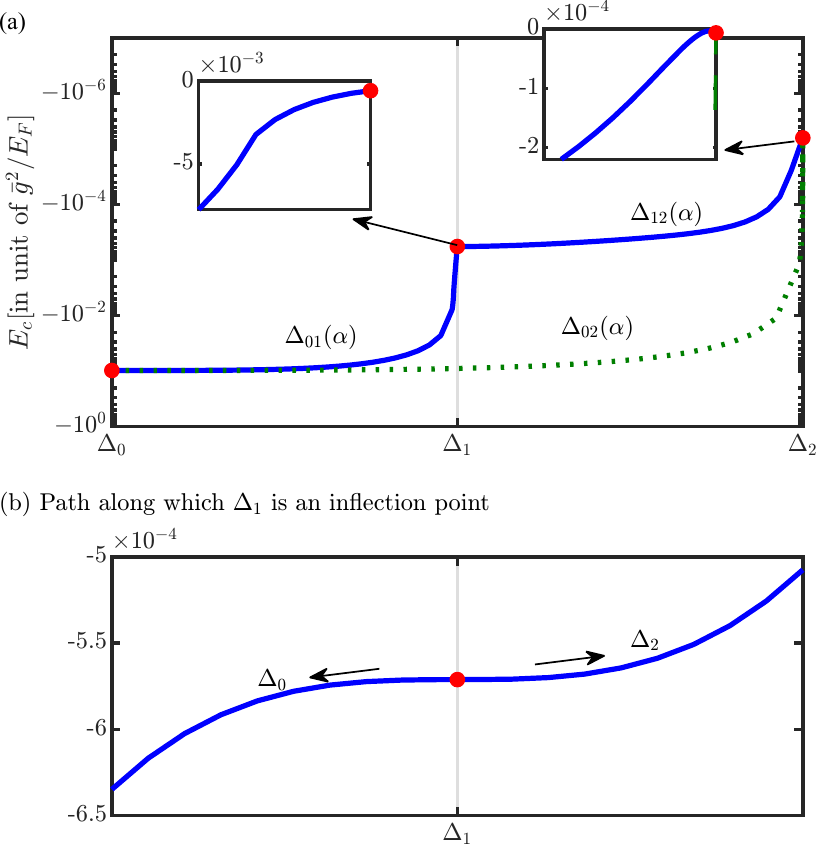}
\caption{(a) Free energy relative to the normal state value $F_{n}$ (i.e., the condensation energy in Eq.~(\ref{eq:n2})) along different paths that smoothly connect $\Delta_0 (\omega_m)$, $\Delta_1 (\omega_m)$, and $\Delta_2 (\omega_m)$ (red dots).
These paths correspond to the trial functions $\Delta_{01}(\omega_m; \alpha)$, $\Delta_{12}(\omega_m; \alpha)$ and
$\Delta_{0 z2}(\omega_m; \alpha)$.
We set $\gamma=0.8$.
 Insets show zoom-ins around $\Delta_1(\omega_m)$ and $\Delta_2(\omega_m)$.
(b) The two-parameter trial gap function, smoothly interplolating between $\Delta_0$, $\Delta_1$, and $\Delta_2$.
 For this function, $\Delta_1$ is an inflection point.}
\label{fig:freeE}
\end{figure}

Another way to see this is to construct a one-parameter  trial gap function
\beq
\Delta_{01}(\omega_m;\alpha) = \cos{\alpha}  \Delta_1(\omega_m) + \sin{\alpha} \Delta_0(\omega_m),
\eeq
which interpolates between $\Delta_1 (\omega_m)$ at $\alpha =0$ and $\Delta_0 (\omega_m)$ at $\alpha = \pi/2$,
and compute the free energy $F_{sc}  (\alpha)$  as a function of $\alpha$. The result is shown in  Fig.~\ref{fig:freeE}.
 We see that $F_{sc} (\alpha)$ indeed gets smaller when $\alpha$ increases.
  We verified that $F_{sc} (\alpha)$ is quadratic in $\alpha$  in the vicinity of $\Delta_1$, however the numerical prefactor is very small.

For $n=2$ the same calculation yields two \textit{negative} eigenvalues, $\lambda^{(1)}_2$ and $\lambda^{(2)}_2$. We show the result in Fig.~\ref{fig:eigs_lt1} (c). The eigenfunction $\Delta^{(1)}_2$  corresponding to $\lambda^{(1)}_2$ is sign-preserving and the eigenfunction $ \Delta^{(2)}_2$, corresponding to $\lambda^{(2)}_2$ has one nodal point (see Fig. ~\ref{fig:func_lt1} (b)).  The free energy gets reduced upon deviations from $\Delta_2 (\omega_m)$  along both directions. To show this more explicitly, we construct the trial functions
\beq
\Delta_{12}(\omega_m;\alpha) = \cos(\alpha) \Delta_2(\omega_m) + \sin (\alpha) \Delta_1(\omega_m),
\eeq
which interpolates between $\Delta_2(\omega_m)$ and $\Delta_1(\omega_m)$, and
\beq
\Delta_{02}(\omega_m;\alpha) = \cos(\alpha) \Delta_2(\omega_m) + \sin (\alpha) \Delta_0(\omega_m),
\eeq
which interpolates between $\Delta_2(\omega_m)$ and $\Delta_0(\omega_m)$. For both functions, we  compute $F_{sc}$ as a function of $\alpha$. We show the results in Fig.~\ref{fig:freeE}. We see that $F_{sc}$ monotonically decreases when the trial gap function moves from $\Delta_2$ to $\Delta_1$ or to $\Delta_0$. This incidentally also shows that $F_{sc}$ increases upon deviation from $\Delta_1 (\omega_m)$ in the direction of $\Delta_2 (\omega_m)$, consistent with the result that there is only one negative eigenvalue for $n=1$.

One can also construct a two-parameter trial gap function that smoothly interpolates between $\Delta_0 (\omega_m)$, $\Delta_1 (\omega_m)$, and $\Delta_2 (\omega_m)$, see Fig.~\ref{fig:freeE} (b). For this trial function,  $F_{sc}$ for $n=0$ is a minimum, $F_{sc}$ for $n=2$ is a maximum, and $F_{sc}$  for $n=1$ is an inflection point.

\subsection{A discrete set of condensation energies $E_{c,n}$}
\label{sec:generic3}

The condensation energy for the stationary $\Delta_n (\omega_m)$ is given by Eq.~(\ref{eq:n3}).
For $\gamma <2$, $\Delta_n (\omega_m)$ form a discrete set, and accordingly $E_{c,n}$  also form a discrete set.
In Fig.~\ref{fig:freeE} (a),
we present $E_{c,n}$ for representative $\gamma =0.8$ for $n=0,1$ and $2$, using the gap functions from Fig.~\ref{fig:gap_func_lt1}.  We see that $E_{c,0}$ is indeed the largest by magnitude.
At large $n$,  $E_{c,n}$ is exponentially small in $n$, as expected.

\section{Limit $\gamma \to 2$}
\label{sec:limit}

In this section we discuss how this set $\Delta_n (\omega_m)$ and the condensation energy $E_{c,n}$
 evolve as $\gamma \to 2$.  The analysis below is for $T=0$.

\subsection{$\Delta_n (\omega_m)$ at $\gamma =2-0$}

We argued in the previous Section that at $\gamma <2$ the non-linear gap equation (\ref{ss_11_0}) has
an
infinite number of solutions $\Delta_n(\omega_m)$, whose amplitudes decrease exponentially with $n$.
 We obtained this result by using  the exact solution of the linearized gap equation,  $\Delta_{\infty}(\omega_m)$, as the starting point and analyzing the structure of the expansion in the gap amplitude $\epsilon$.
  We found that $\epsilon$ is quantized into a discrete set of  $\epsilon_n$.  The quantization follows from the fact that at $\omega_m \ll {\bar g}$,
   each term in the expansion oscillates as a function of $\log{|\omega_m|}$ with its own phase.
   The explicit result for the expansion is given by Eq. (\ref{eq:8}),   which we reproduce here for convenience of a reader:
   \bea
\Delta (\omega_m) &=& 2 \epsilon^* |\omega_m|  \left[ Q_1 \cos({\psi}) + (\epsilon^*)^2 Q_3 \cos({3\psi + \phi_3}) \right. \nonumber \\
&& \left. + (\epsilon^*)^4 Q_5 \cos({5\psi + \phi_5}) + ...\right]
 \label{eq:8_1}
 \eea
 where $\epsilon^* = \epsilon ({\bar g}/|\omega_m|)^{1-\gamma/2}$,
  $\psi = \beta ({\gamma}) \log{({\bar g}/|\omega_m|)^\gamma} + \phi$,
 and $Q_{2k+1}$, $\phi_{2k+1}$, and $\phi$ are functions of $(\epsilon^*)^2$.
 The quantization sets $\epsilon$ to be $\epsilon_n \sim e^{-b n (2-\gamma)/2}$, where $b = O(1)$.

We now analyze Eq. (\ref{eq:8_1}) at $\gamma \to 2$. Here
$\epsilon^* \approx \epsilon$ down to smallest $\omega_m$. The series  $Q_{2k+1}$ in Eq.~(\ref{eq:8_1}) remain non-singular and evolve continuously at  $\gamma \to 2$ towards
\begin{align}
Q_{1} & =1+1.3049\epsilon^{2} +2.99051\epsilon^{4}+...,\label {eq:q1} \\
Q_{3} & =0.222548 + 0.618278 \epsilon^{2}+...,\label {eq:q3}\\
Q_{5}& = 0.0426647 + ....\label {eq:q5}
\end{align}
 The series for $\psi$ do become singular and transform $\beta (\gamma =2) = 0.38187$  (Ref. \cite{paper_5})
  into $\epsilon-$dependent $\beta_\epsilon$.
Explicitly,
\bea
\psi &=& 2 \beta (2) \log \left( {{\bar g}\over \rvert \omega_m \rvert}\right) \left(1 - 0.5 \epsilon^2 -0.889 \epsilon^4 + ...\right) + \phi_\epsilon \nonumber \\
&& = 2\beta_\epsilon  \log \left( {{\bar g}\over \rvert \omega_m \rvert}\right) ^2 + \phi_\epsilon
\label{nnn_8}
\eea
where
\beq
\beta_\epsilon = \beta (2) \left(1 - 0.5 \epsilon^2 -0.889 \epsilon^4 + ...\right).
\label{nnn_9}
\eeq
The transformation $\beta (2) \to \beta_\epsilon$ shifts the frequencies, at which $\psi = \pi/2 + n \pi$, to
$\omega_m \propto e^{-n\pi/2\beta_{\epsilon}}$.
The corrections to  phase $\phi (2)$, defined in Eq.\eqref{eq:8aa},
are also singular and transform it into $\phi_\epsilon = \phi (2) + (\beta_2 \epsilon^2 /(2-\gamma)) (1 + 0.889 \epsilon^2 +...)$. This transformation is, however, irrelevant as the phase is defined up to an integer number  of $2\pi$.

Finally, the series for $\phi_3$, $\phi_5$, etc, also  remain regular, but each term contains $2-\gamma$ as a prefactor.
 In explicit form,
 \begin{align}
\phi_{3} & = (2-\gamma) \left[1/8 - 0.577535 (\epsilon^*)^{2} + ...\right] , \label{eq:phi3} \\
\phi_{5} & = (2-\gamma)  \left[0.452288+....\right] \label{eq:phi5}.
\end{align}
We keep $\epsilon^*$ here as it will be relevant for the understanding of the behavior at small but finite $2-\gamma$.
 We see that all $\phi_{2k+1}$ vanish at $\gamma =2$. In this case,
the scale, at which oscillations would be destroyed  due to phase randomization, vanishes.  This eliminates the argument for the discretization of $\epsilon$.

\subsection{Crossover from a discrete to a continuous set at $\gamma \to 2$}

It is instructive to analyze in more detail how the continuous set of $\Delta_\epsilon$ emerges.  At $\gamma \leq 2$ we expect by continuity that
the phases $\phi_{2k+1}$  become of order
one  at some $\epsilon^*$.  This happens when
 the rest of
 $\phi_{2k+1}$  compensates the overall $2-\gamma$.  This
 happens either because prefactors become singular at some finite $\epsilon^*$, as we suggested in ~\cite{paper_5}, or because they diverge at $\epsilon^* \to \infty$ as, e.g.,
$(\epsilon^*)^{1/a}$.
In the last case, which we analyze here, oscillations survive down to $\omega_m \sim {\bar g} (\epsilon (2-\gamma)^{a})^{2/(2-\gamma)}$.  Matching
this scale with the position of one of the extrema of $\cos{\psi}$ at $\omega_m \propto {\bar g} e^{-n \pi/2\beta_\epsilon}$, we obtain
 self-consistent equation on discretized $\epsilon_n$
\beq
\epsilon_n \sim  \frac{\exp[-{\pi \over 4\beta_{\epsilon_n}} n (2-\gamma)]}{(2-\gamma)^a}.
\label{nnn_15}
\eeq
We see that at $\gamma \to 2$, all $\epsilon_n$  with finite $n \ll n^* =|\log{(2-\gamma)}| /(2-\gamma)$ become  equal to $\epsilon_{n=0} = \epsilon_{max}  \sim {1\over (2-\gamma)^{a}}$.
 As we anticipated, $\epsilon_{max} = \infty$ at  $\gamma =2$.
\footnote{If the phases $\phi_{2k+1}$ become $O(1)$ at a finite $\epsilon^*$, the results are the same, only $a=0$, hence $\epsilon_{max}$ is finite.}
Simultaneously, $\epsilon_n$ with
$n \geq n^* \to \infty$  form a continuous spectrum, and at $\gamma =2-0$,
$\Delta_n(\omega_m)$
form a continuous  one-parameter set
\begin{align}
\Delta(\epsilon,\omega_{m})
& =2\epsilon\rvert\omega_{m}\rvert\left(Q_{1}(\epsilon)\cos(\psi (\epsilon))+Q_{3}(\epsilon)\epsilon^{2}\cos(3\psi (\epsilon))+...\right),
\label{eq:delta_2}
\end{align}
   As  expected, all
   $\Delta(\epsilon,\omega_{m})$ with finite $\epsilon$
    change sign infinite number of times at  $\omega_m \propto e^{-\pi k/(2\beta_\epsilon)}$, i.e., in our classification they all are  $n=\infty$ solutions.  The lower point of the set is $\epsilon =0+$, which corresponds to the solution  of the linearized gap equation, $\Delta_{\infty} (\omega_m)$.   It indeed changes sign infinite number of times.

     The other end point is $\epsilon \to \infty$.  We argue below that this limit is continuous, i.e., this end point is the sign-preserving $\Delta_{n=0} (\omega_m)$.  This is the case if $\beta_{\infty} =0$ as then
     $\psi$
      becomes independent on frequency and $\Delta(\infty,\omega_{m})$ reduces to a constant  (up to corrections in powers of $\omega_m/{\bar g}$, which we  neglected in (\ref{eq:delta_2})).
      The vanishing of $\beta_{\infty}$ in turn implies that the
   frequencies $\omega_m \propto e^{-\pi k/(2\beta_{\epsilon_n})}$, at which $\Delta_n (\omega_m)$ with a finite $n$ changes sign, all vanish at $\gamma \to 2$, as they should.

\begin{figure}
\centering
\includegraphics{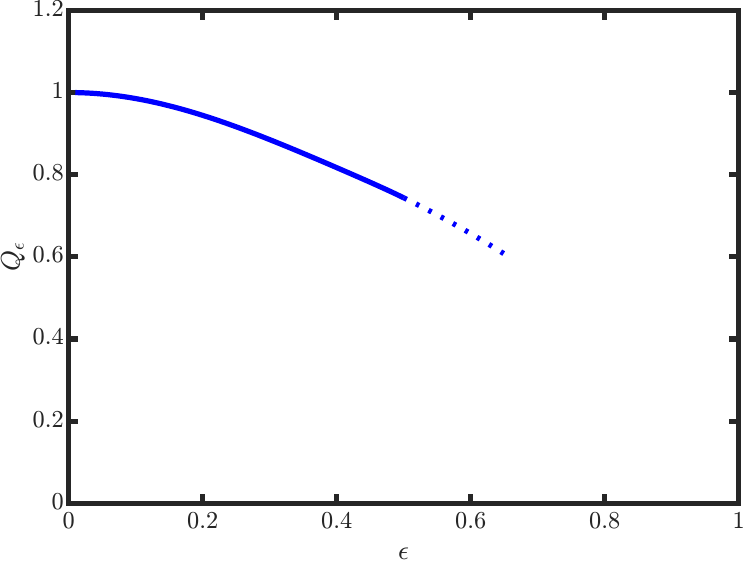}\caption{The series for $Q_{\epsilon}$ given by Eq. (\ref{eq:critical_eps}).
The series converge for $\epsilon \leq 0.5$.
The dashed line is an exrapolation to somewhat larger $\epsilon$.}
\label{fig:critical_eps}
\end{figure}

   Below we  present several arguments in favor of our assertion that $\beta_{\infty} =0$.
   First, we look at series expansion in $\epsilon$.   Eq. (\ref{nnn_9}) already shows that
    $\beta_\epsilon$ decreases with $\epsilon$. To obtain more precise result, we redo the expansion in $\epsilon$ self-consistently, by evaluating each term assuming that $\beta_\epsilon$ is vanishingly small.  We obtain $\beta^2_\epsilon \propto Q_\epsilon$, where
   \begin{equation}
Q_{\epsilon}\equiv 1- \frac{3}{2}\epsilon^{2}+3\epsilon^{4}-\frac{103}{16}\epsilon^{6}+\frac{915}{64}\epsilon^{8}-\frac{4149}{128}\epsilon^{10}+\frac{19075}{256}\epsilon^{12}-\frac{354279}{2048}\epsilon^{14}+.... \label{eq:critical_eps}
\end{equation}
  Our assertion is valid if $Q_{\infty} =0$.
We plot $Q_\epsilon$ in  Fig. \ref{fig:critical_eps}. The series converge well up to $\epsilon \sim 0.5$. Within this range, $Q_\epsilon$ smoothly decreases with increasing $\epsilon$. The behavior is at least consistent
 with the vanishing of $Q_\epsilon$ and hence $\beta_\epsilon$ at $\epsilon = \infty$.

Second, we re-evaluate $\Delta_\epsilon (\omega_m)$  along the same lines as before, by redoing the expansion in $\epsilon$  keeping $\beta_\epsilon$ vanishingly small.  The result is, up to  order $\epsilon^{15}$,
\begin{align}
\Delta(\epsilon,\omega_m)=
&  \frac{\epsilon |\omega_m|}{1024}\Bigg[2048\epsilon\cos(\psi)\nonumber \\
+ & \epsilon^{3}\left(1024-768\epsilon^{2}+768\epsilon^{4}-960\epsilon^{6}+1392\epsilon^{8}-2226\epsilon^{10}+3810\epsilon^{12}+...\right)\cos(3\psi)\nonumber \\
- & \epsilon^{5}\left(-768+1280\epsilon^{2}-1920\epsilon^{4}+3000\epsilon^{6}-4970\epsilon^{8}+8670\epsilon^{10}+...\right)\cos(5\psi)\nonumber \\
+ & \epsilon^{7}\left(640-1680\epsilon^{2}+3360\epsilon^{4}-6328\epsilon^{6}+11886\epsilon^{8}+...\right)\cos(7\psi)\nonumber \\
- & \epsilon^{9}\left(-560+2016\epsilon^{2}-5040\epsilon^{4}+11130\epsilon^{6}+...\right)\cos(9\psi)\nonumber \\
+ & \epsilon^{11}\left(504-2310\epsilon^{2}+6930\epsilon^{4}+...\right)\cos(11\psi)\nonumber \\
- & \epsilon^{13}\left(-462+2574\epsilon^{2}+...\right)\cos(13\psi)\nonumber \\
+ & \epsilon^{15}\left(429+...\right)\cos(15\psi)\nonumber \\
+ & ...\Bigg].
\label{nnn_10}
\end{align}
We expect that at $\epsilon \to \infty$,
$\Delta (\epsilon, \omega_m)$
becomes independent on $\omega_m$
  at $\omega_m \to 0$.  This holds when the expression in square brackets in (\ref{nnn_10})
   compensates the overall factor  $|\omega_m|$ in the r.h.s. of (\ref{nnn_10}).  A simple experimentation shows that this can happen if $\phi_{\infty} = \pi/2$, in which case
    $\cos{[(2k+1) \psi]} = (-1)^{k+1} [(2k+1)2\beta_{\epsilon} \log{({\bar g}/|\omega_m|)} + ....]$, where dots stand for higher order of the logarithm.
  The expression in square brackets in (\ref{nnn_10}) then becomes the series in $\log{({\bar g}/|\omega_m|)}$ with
  $\epsilon-$dependent prefactors.  If the series are exponential, they can cancel the overall $|\omega_m|$ in (\ref{nnn_10}), however, for this the prefactors  must tend to finite values at $\epsilon =\infty$

  We found that the prefactor $C_\epsilon$ for $\log{({\bar g}/|\omega_m|)}$ is proportional to $Q_\epsilon$ from  (\ref{eq:critical_eps}):
   \begin{align}
C_\epsilon=-4
& \epsilon Q_{\epsilon}\beta_{\epsilon}
\label{nnn_11}
\end{align}
As $Q_\epsilon \propto \beta^2_{\epsilon}$, the condition that $C_{\infty}$ is finite yields
$\beta_\epsilon  \sim (1/\epsilon)^{1/3}$
 (hence, $Q_\epsilon \propto 1/\epsilon^{2/3} \to 0$).

  Third, as an independent check,  we set $\epsilon$ to be large but finite and evaluated $\Delta_\epsilon (\omega_m)$ near each nodal point $\omega_p$, for which
$\psi = \pi/2 + \pi p$. Setting $\omega_m = \omega_p + \delta \omega$ and expanding in $\delta \omega$, we obtain
\beq
\Delta_\epsilon (\omega_p + \delta \omega) \approx 4 (-1)^p \epsilon Q_{\epsilon}\beta_{\epsilon} \delta \omega, \label{eq:nodal}
\eeq
If  $\epsilon Q_{\epsilon}\beta_{\epsilon}$ tends to a constant at $\epsilon \to \infty$,  as we anticipated, $\Delta_\epsilon (\omega_p + \delta \omega) \sim (-1)^p \delta \omega$.
 This is consistent with the behavior of $\Delta_n (\omega_m)$ with finite $n$ at small but finite $2-\gamma$ in the frequency range, where $\Delta_n (\omega_m)$ changes sign $n$ times.

Substituting $\beta_\epsilon \propto 1/\epsilon^{1/3}$ into (\ref{nnn_15}), we find the relation between  $\epsilon$ and $n$ at $\gamma =2-0$:
\beq
\frac{n}{n^*} \sim \frac{4 a}{\epsilon^{1/3}}
\eeq
 This relation holds when $\epsilon \gg 1$, i.e., when $n \ll n^*$.
A more general expression, valid also for  $\epsilon \leq 1$,
 is $\beta_\epsilon = (\pi/4a) n/n^*$.
 A more sophisticated analysis is required to relate $\epsilon$ and $n$ at smaller $\epsilon$.

\subsection{Continuum spectrum of the condensation energy}

The appearance of a continuum of solutions of the Eliashberg gap equation at $\gamma =2-0$ and $\omega_D =0$ poses the question about the spectrum of the condensation energy. At a first glance,
the condensation energy should be flat as a function of $\epsilon$
 because each $\Delta_\epsilon (\omega_m)$ satisfies the stationary condition $\delta F_{sc}/\delta \Delta = \delta E_c /\delta \Delta =0$, hence
 $d E_c/d\epsilon = \delta E_c /\delta \Delta_\epsilon \times d \Delta_\epsilon/d\epsilon =0$,
We argue that
in
our case the situation is more tricky because the condensation energy formally diverges at $\gamma = 2-0$ for all $\epsilon$,  which creates
"zero times infinity" conundrum.
We argue that after proper regularization, the condensation energy becomes a regular non-flat function of $\epsilon$.


 To understand how to reconcile a non-flat $E_c (\epsilon)$ with the fact that each $\Delta_\epsilon$ is a stationary solution,
 consider $\gamma$ slightly below $2$.  For any $\gamma <2$, $\epsilon = \epsilon_n$ are discrete, and the condensation energy, given by
 Eq.~(\ref{eq:n2}),
 is also discrete $E_{c,n}$.  Let's start with $n=0$. At small frequencies the gap function $\Delta_0 (\omega_m)$
 tends to a finite value $\Delta_0 \sim {\bar g}$.  The first term in
 Eq.~(\ref{eq:n2})
 is regular and of order ${\bar g}^2 N_F$. The leading contribution to the second term in Eq.~(\ref{eq:n2}) comes from small frequencies and has the form
 \beq
 - N_F {\bar g}^\gamma \int_0^{\omega^*} d \omega_m \int_0^{\omega^*} d \omega_{m'} \frac{1}{|\omega_m + \omega_{m'}|^\gamma},
 \label{nnn_11_1}
 \eeq
 where the upper limit $\omega^* \sim {\bar g}$.  Evaluating the integral we find that the condensation energy diverges at $\gamma  \to 2$ as
\begin{equation}
E_{c,0} \approx -  {\bar g}^2 N_F \frac{1}{2-\gamma}.\label{eq:Ec1}
\end{equation}

Consider next the condensation energy for a finite $n$.
  Because all $\Delta_n (\omega)$ tend to  finite $\Delta_n (0)$ at vanishing $\omega_m$, the divergent $1/(2-\gamma)$ term is the same as in (\ref{eq:Ec1}). The difference  $\delta E_n = E_{c,n} - E_{c,0}$ comes from the range where $\Delta_n (\omega_m)$ changes sign
$n$
 times
  between the largest $\omega_m \sim {\bar g}e^{-\pi/2\beta_{\epsilon_n}}$ and the smallest $\omega_m \sim {\bar g}e^{-\pi n/2\beta_{\epsilon_n}}$.  We obtained (see Appendix~\ref{sec:ecn} for details)
  $\delta E_c=c {\bar g}^2 N_F n$, where $c = O(1)$. Together with
   Eq. (\ref{eq:Ec1}) this yields
\beq
E_{c,n}= - {\bar g}^2 N_F \left ({1\over 2-\gamma} -c n + ...\right ) =
E_{c,0} \left(1 - c n (2-\gamma) + ...\right).
\label{nnn_14}
\eeq

Eq. (\ref{nnn_14}) is valid as long as $n \ll n^*$,  more accurately,  when
$n \leq n^*/((2-\gamma)^{a/3} |\log{(2-\gamma)}|$ ($\epsilon \approx \epsilon_{max} \sim 1/(2-\gamma)^a$).
 For larger $n$,   the factor $n (2-\gamma)$ in the r.h.s. of (\ref{nnn_14}) is replaced by
 \beq
 J_n
= \frac{2\beta_\epsilon}{\pi} \left(1 - e^{- \frac{\pi n (2-\gamma)}{2 \beta_\epsilon}}\right)
= \frac{2\beta_\epsilon}{\pi} \left(1 - (\epsilon (2-\gamma)^{a})^2 \right) \approx \frac{2\beta_\epsilon}{\pi}
\label{nnn_17}
\eeq
(see again Appendix \ref{app:E} for details).  This result holds for $n \sim n^*$, i.e., $\epsilon \sim 1$. For numerically large $\epsilon$ (but $\epsilon \ll \epsilon_{max} \sim 1/(2-\gamma)^a$), $\beta_\epsilon \sim 1/\epsilon^3$, hence
\beq
E_{c} (\epsilon) \approx  E_{c,0}  \left(1 - \frac{{\bar c}}{\epsilon^{1/3}} \right)
\label{nnn_18}
\eeq
where ${\bar c} = O(1)$.
In the opposite limit $n >> n^*$, i.e. at $\epsilon \to 0$,  the gap function $\Delta_{\infty} (\omega_m) \propto \epsilon$   does not saturate at a finite value at $\omega_m =0$.  Yet we found in explicit calculation in Appendix \ref{app:E} that
 $E_c (\epsilon) \propto \epsilon^2/(2-\gamma)$  still diverges as $1/(2-\gamma)$.
  For a generic $\epsilon$, we expect
  \beq
  E_c = E_{c,0} g (\epsilon),
  \label{new_new}
  \eeq
   where
   $g (\epsilon) \approx 1 -{\bar c}/\epsilon^{1/3}$
  for large $\epsilon$ and $g (\epsilon) \sim \epsilon^2$ for  small $\epsilon$ (Ref.
  \footnote{The  emergence of the overall factor $1/(2-\gamma)$ in $E_c$ was not noted in Ref.\cite{paper_5}.  Eq. [40] in that paper is incorrect} ).

The singular $ 1/(2-\gamma)$ dependence in $E_c$
 is the reason why $E_{c} (\epsilon)$  disperses with $\epsilon$  despite that
  for each $\epsilon$, $\Delta_\epsilon (\omega_m)$ is a stationary point of $F_{sc}$.  To see this explicitly, we substituted $\Delta_\epsilon (\omega_m)$ from (\ref{eq:delta_2}) into the expression for $E_c$,
  Eq. (\ref{eq:n2}),
   and expanded around some representative $\epsilon = \epsilon_0$. The linear term in the expansion of $E_c (\epsilon)$ in  $\epsilon - \epsilon_0$ vanishes, as it should, but the prefactors for all higher derivatives diverge. In this special case Taylor expansion in $\epsilon - \epsilon_0$ must be kept to an infinite order to obtain the correct functional form of $E_c (\epsilon)$ near $\epsilon_0$.

\begin{figure}
\centering
\includegraphics{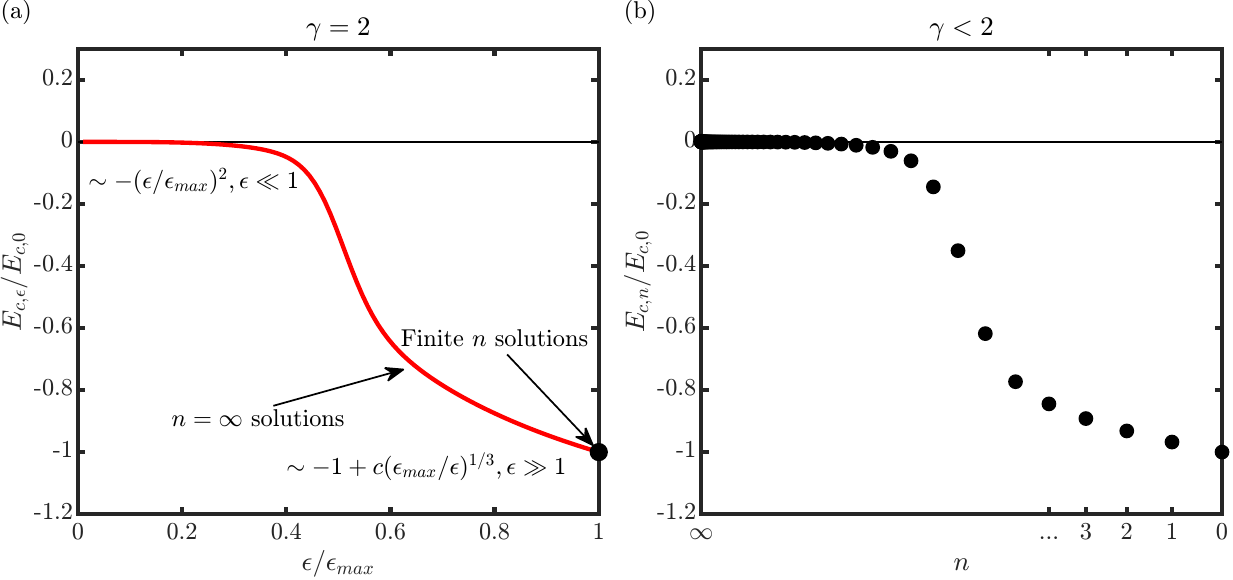}\caption{(Schematic) condensation energy $E_{c,n}$ for (a) $\gamma=2$ and (b) a generic $\gamma<2$.}
\label{fig:Ec}
\end{figure}

We see from (\ref{new_new}) that if we measure $E_c (\epsilon)$ in units of $E_{c} (\infty)$, we obtain the dispersing $E_c/E_{c} (\infty) = g(\epsilon)$. This, however, makes sense if $E_{c} (\infty)$ remains finite at $\gamma =2-0$, i.e., if $1/(2-\gamma)$ divergence  is regularized.  The way to do this is to take the double limit $\gamma \to 2$ and $N_F/N \to 0$ (i.e., $E_F \sim \Lambda \to \infty)$, such that the product $N_F/(N(2-\gamma))$ remains finite at $\gamma =2-0$.
  We illustrate this in Fig.~\ref{fig:Ec}
Once $E_c (\infty)$ is finite, the spectrum of the condensation energies becomes a regular gapless function of $\epsilon$.  In particular,  for a finite $n$, the difference $(E_{c,n} - E_{c,0})/N \sim N_F n/N$ vanishes.

The emergence  of a continuum spectrum of $E_c (\epsilon)$ at $\gamma =2+0$ due to a collapse of all $E_{c,n}$ with finite $n$ onto $E_{c,0}$ has a profound effect on the behavior of the superfluid stiffness $\rho_s$ at
$T= \Omega_D =0$
as it opens up a gapless channel of "longitudinal" gap fluctuations.  In \cite{paper_5} we argued semi-phenomenologically that these fluctuations give rise to singular downward renormalization of the stiffness and
 give rise to vanishing $T_c$ at $\gamma =2-0$. We leave the detailed analysis of $\rho_s$, based on the formula for $E_c (\epsilon)$, obtained in this paper,  to a separate study.

Note that the  $1/(2-\gamma)$
divergence
 can also be regularized by a finite $T$ or a finite $\omega_D$.
 In both cases, the would be  divergent term in $E_c$  becomes
\begin{equation}
 - {\bar g}^2 N_F \log {{\bar g}\over T} \text{ or }  -   {\bar g}^2 N_F \log {{\bar g}\over \omega_D}.\label{eq:Ec2}
\end{equation}
The spectrum of $E_{c,n}$ then remains discrete at $\gamma =2$, but becomes continuous in the properly defined double limit $T, \omega_D \to 0$ and $N_F \to 0$.  Incidentally, the same  double limit is also required to keep vertex corrections to Eliashberg equations small. Vertex corrections at $T=0$  are controlled by
    $\lambda_E =  (N_F/N) {\bar g}^2/\omega_D)$, and to keep them small one has to set $N_F/N$ to zero along with
    $\omega_D\to 0$.

\section{Conclusions}
\label{sec:sumary}

In this work we extended our earlier analysis of pairing in a metal tuned to a $T=0$ quantum critical point, at which it develops a spontaneous order in the particle-hole channel. At this point bosonic  fluctuations of a ctitical order parameter become massless.
    In earlier works~\cite{paper_1,paper_2,paper_3,paper_4,paper_5,paper_6} we analyzed the pairing
     mediated by a massless boson and argued that this problem is qualitatively different from BCS/Eliashberg pairing in a metal away from a critical point. Specifically, a massless boson  gives rise not only to the pairing, but also to a NFL behavior in the normal state.   The competition between NFL and pairing leads to new physics, not seen in non-critical metals.    We argued that under certain conditions low-energy properties of a critical metal are described by a
     $0+1$ dimensional dynamical model with an effective 4-fermion interaction $V(\Omega) \propto 1/|\Omega|^\gamma$. This model has been nicknamed the $\gamma$ model.  We argued that
       for $\gamma <2$, the ground state is a superconductor with a regular, sign-preserving
       gap
       function $\Delta_0 (\omega_m)$ on the Matsubara axis. Yet, the
      gap equation at $T=0$ also allows an infinite, discrete set of topologically distinct stationary solutions
      $\Delta_n (\omega_m)$ with $n$ nodal points at positive $\omega_m$ ($n=1,2...$).   We argued that the set evolves with increasing $\gamma$ and eventually becomes a continuous one at $\gamma =2-0$. At this $\gamma$,
     the  system undergoes a  topological transition into a state with
    novel measurable features.

     Here we performed  an in-depth analysis of the free energy
      of the $\gamma$-model for a generic $\gamma <2$ and for $\gamma$ infinitesimally close to $2$. One goal of our study was to understand the profile of the free energy near each stationary $\Delta_n (\omega_m)$, another was to understand in more detail how a discrete set of $\Delta_n (\omega_m)$ becomes continuous at $\gamma =2-0$ and how the spectrum of condensation energies evolves near this $\gamma$.

      For the free energy analysis we first compared the two forms of the free energy -- the variational one, obtained by extending Luttinger-Ward-Eliashberg
       variational approach to arbitrary $\gamma$, and the actual one, which appears in the partition function after
       Hubbard-Stratonovich transformation.  The variational free energy has a simple form and yields  the stationary  gap equation,  which agrees with the one obtained diagrammatically by summing up ladder diagrams with the fully dressed Green's functions.  Yet, for a massive pairing boson
         it does not adequately describe fluctuations around a stationary point as it neglects variations of the quasiparticle residue $Z (\omega_m)$.  We  showed that for a massless boson fluctuations of $Z$ around its stationary value  become irrelevant,  and the variational free energy can be used to obtain not only stationary solutions, but also fluctuations around them.
       For $\gamma =2$ this has been earlier found in Ref.  \cite{yuz}.

       To obtain stationary $\Delta_n (\omega_m)$ at $T=0$ we used as a point of departure the exact solution
         for infinitesimally small $\Delta_{\infty} (\omega_m)$, which we found in previous works, expanded in the gap amplitude $\epsilon$, and explicitly demonstrated  that $\epsilon$ is  discretized into  $\epsilon_n$.
         Using this reasoning, we obtained $\Delta_n (\omega_m)$ with $n=0,1, 2$, which satisfy the gap equation with high numerical accuracy.

         We then expanded the free energy near stationary $\Delta_n (\omega_m)$ with these $n$ to second order in deviations and analyzed the stability of each solution. We found that the $n=0$ solution is a global minimum, while the expansion around $\Delta_1 (\omega_m)$ yields one negative eigenvalue for deviations towards $\Delta_0 (\omega_m)$, and the expansion around $\Delta_2 (\omega_m)$ yields two negative eigenvalues for deviations towards $\Delta_0 (\omega_m)$ and $\Delta_1 (\omega_m)$.  By continuity, we expect $n$ negative eigenvalues for $\Delta_n (\omega_m)$.
          This  result implies that $\Delta_0$ is a true minimum, while $\Delta_n$ with $n >0$  is a saddle point with $n$ unstable directions.

      As a bi-product of this analysis, we  obtained the free energy and the specific heat $C(T)$ in the normal state  and compared the results with recent studies by other groups.

   For our second goal we analyzed in more detail than before the transformation from a discrete set of $\Delta_n (\omega_m)$  into a continuous one-parameter set $\Delta (\epsilon, \omega_m)$ at $\gamma =2-0$. We demonstrated that when the exponent $\gamma$ increases towards $2$, the set of discretized gap amplitudes  $\epsilon_n$ progressively splits into two subsets: the ones with
 $n < n^* \sim |\log(2-\gamma)|/(2-\gamma)$
            all approach $\epsilon_0$, while the ones with $n > n^*$ form a continuum set $\Delta (\epsilon, \omega_m)$, where $\epsilon$ is a function of the ratio $n/n^*$ when both tend to infinity.
             At $\gamma =2-0$, all functions $\Delta (\epsilon, \omega_m)$  oscillate infinite number of times down to $\omega_m =0$, i.e. in the count of nodal points they all are  $n = \infty$  solutions (the solution of the linearized gap equation is the one at infinitesimally small $\epsilon$).
               This holds for all finite $\epsilon$. However, the end point at $\epsilon = \infty$ is a sign-preserving
               $\Delta_0 (\omega_m)$, which at $\gamma =2-0$ incorporates all gap functions $\Delta_n (\omega_m)$ with  finite $n$. We showed that the range, where $\Delta_n (\omega_m)$ changes sign $n$ times, shrinks to progressively lower frequencies  as $\gamma \to 2$.

We used these results to obtain the condensation energy  $E_{c,n}$ - the difference between free energy of a superconductor and of a would be normal state at the same $T$.  We found
            $E_c (\epsilon) = E_c (\infty)
            g(\epsilon)$,
            where
            $E_{c}(\infty) = {\bar g}^2 N_F (1/(2-\gamma)$, and
 $
 g
 (\epsilon \gg 1) =1 -O(1/\epsilon^{1/3})$
and $
g(\epsilon \ll 1) \sim \epsilon^2$.  We argued that this result makes sense if $E_{c} (\infty)$ remains finite at $\gamma =2-0$, i.e., if $1/(2-\gamma)$ divergence  is regularized.  The way to do this is to take the double limit $\gamma \to 2$ and $N_F/N \to 0$ (i.e., $E_F \sim \Lambda \to \infty)$, such that the product $N_F/(N(2-\gamma))$ remains finite at $\gamma =2-0$.   At the same time, we argued that $1/(2-\gamma)$ divergence in $E_c (\infty)$ is
 is the reason why $E_{c} (\epsilon)$  disperses with $\epsilon$  despite that
  all $\Delta_\epsilon (\omega_m)$ are stationary points of the free energy.
    We argued that if we expand $E_c (\epsilon)$  around some  $\epsilon_0$, the
     linear term in the expansion  vanishes, as it should, but the prefactors for all higher derivatives diverge.
   In this singular case, Taylor expansion in $\epsilon - \epsilon_0$ must be kept to an infinite order to obtain the correct functional form of $E_c (\epsilon)$.

 We argued in \cite{paper_6} that $\gamma =2$ is a critical point of a topological phase transition.  The argument is that the gap function $\Delta_0 (\omega_m)$, for which the condensation energy is the largest by magnitude for $\gamma <2$ and $\gamma >2$,   lives on different Riemann surfaces  at $\gamma <2$ and $\gamma >2$.  A related argument that $\gamma =2$ is special is that in the upper frequency half-plane, $\Delta_0 (z)$, where $z = \omega' + i \omega^{''}$, possesses dynamical vortices, and their number becomes infinite when $\gamma$ approaches $2$ from either side.  It is natural to expect a gapless branch of excitations at the critical point, and the emergence of a gapless continuum of $E_{c} (\epsilon)$ is in line with this reasoning.

The emergence  of a continuum spectrum of $E_c (\epsilon)$ at $\gamma =2+0$ due to a collapse of all $E_{c,n}$ with finite $n$ onto $E_{c,0}$ has a profound effect on the behavior of the superfluid stiffness $\rho_s$ at
$T= \Omega_D =0$
as it opens up a gapless channel of "longitudinal" gap fluctuations.  In \cite{paper_5} we argued semi-phenomenologically that these fluctuations give rise to singular downward renormalization of the stiffness and
 give rise to vanishing $T_c$ at $\gamma =2-0$.
 We leave the detailed analysis of $\rho_s$, based on the formula for $E_c (\epsilon)$, obtained in this paper,  to a separate study.  This analysis has to be made at small, but finite $\omega_D$. The limit $\omega_D \to 0$ also has to be taken together with $E_F \to \infty$ to keep vertex corrections to Eliashberg gap equation small.

\acknowledgements We thank everyone with whom we discussed  this and previous works.
 S.-S.Z and A.V. C.  were supported by the NSF DMR-1834856. Y.-M.W is grateful to the support of Shuimu Fellow Foundation at Tsinghua.

\appendix

\section{Expansion of free energy and equivalence of $F_{sc}$ and $F^{\text{var}}_{sc}$ at $\omega_D \to 0$. }
\label{app:B}

In this Appendix we show that for a finite $\omega_D$, the expansion of the actual free energy of an Eliashberg  superconductor, $F_{sc}$, to second order in variations around a stationary solution $\Delta_n (\omega_m)$ is non equivalent to the expansion of the variational free energy $F^{\text{var}}_{sc}$. However, at $\omega_D \to 0$, the expansions become equivalent for any value of $\gamma$ at a finite $T$ and for $\gamma >1$ at $T=0$.
We verified that $F^{\text{var}}_{sc}$ and  $F_{sc}$ remain equivalent also beyond the second order.

\subsection{Computation of $V^{-1} (\Omega_m)$}

  The free energy  $F_{sc}$ is given by  Eq. (\ref{eq:var_energy_actual}).
  It contains the
  ``inverse''
  interaction $V^{-1} (\Omega_m)$, which is the Fourier transform of $1/V(\tau)$.
 We will need the explicit form of  $V^{-1} (\Omega_m)$ at a finite $T$ and $\omega_D \ll T$.
 We
  compute
  it in this subsection and use the result in the next subsection.

  The interaction $V(\Omega_m)$ is given by Eq. (\ref{eq:1}):
    \beq
  V(\Omega_m) = \frac{{\bar g}^\gamma}{(\Omega^2_m + \omega^2_D)^{\gamma/2}}
 \label{eq:1_1}
  \eeq
  Its inverse Fourier transform is
  \bea
  &&V(\tau) = T \left(\frac{\bar g}{\omega_D}\right)^\gamma \times \nonumber \\
  &&\left[1 + \left(\frac{\omega_D}{2\pi T}\right)^\gamma \left(Li_\gamma (e^{2\pi T \tau}) + Li_\gamma (e^{-2\pi T \tau})\right)\right]
  \label{f:1_a}
  \eea
   where $Li_\gamma (x)$ is a polylogarithmic function.
   For $\omega_D \ll T$, the second term in the last line in (\ref{f:1}) is a small correction. Then
    \bea
  &&V^{-1}(\tau) =
  \frac{1}{T} \left(\frac{\omega_D}{\bar g}\right)^\gamma \times \nonumber \\
  &&\left[1 - \left(\frac{\omega_D}{2\pi T}\right)^\gamma \left(Li_\gamma (e^{2\pi T \tau}) + Li_\gamma (e^{-2\pi T \tau})\right)\right]
  \label{f:1}
  \eea
   The Fourier transform of (\ref{f:1}) is
   \bea
   V^{-1} (\Omega_m) &=& \int_0^{1/T} V^{-1}(\tau) \cos{(\Omega_m \tau)} \nonumber \\
    &=&
    \frac{1}{T^2} \left(\frac{\omega_D}{\bar g}\right)^\gamma \nonumber \\
   && \left[\delta_m - \left(\frac{\omega_D}{2\pi T}\right)^\gamma I_m\right]
    \label{f:2}
  \eea
  where  $\delta_{m,0}$ is a Kronecker symbol, and
  \beq
  I_m = \frac{1}{2\pi} \int_0^{2\pi} dx \cos{(mx)} \left(Li_\gamma (e^x) + Li_\gamma (e^{-x})\right)
 \eeq
 We verified that $I_0 =0$ and $I_{m>0} = 1/|m|^\gamma$.  As a result,
 \beq
 V^{-1} (\Omega_m) =
 \frac{1}{T^2}
  \left(\frac{\omega_D}{\bar g}\right)^\gamma \left[\delta_{m,0} - (1-\delta_{m,0}) \left(\frac{\omega_D}{|\Omega_m|}\right)^\gamma\right]
  \label{eq:17}
 \eeq
 The original $V (\Omega_m)$ can be equally represented as
 \beq
 V (\Omega_m) = \left(\frac{\bar g}{\omega_D}\right)^\gamma \left[\delta_{m,0} + (1-\delta_{m,0}) \left(\frac{\omega_D}{|\Omega_m|}\right)^\gamma\right]
  \label{eq:17_a}
 \eeq
 We will use Eqs. (\ref{eq:17}) and (\ref{eq:17_a}) below.

 We note in passing that at $T=0$ the computation of $V^{-1} (\Omega_m)$ for a continuous $\Omega_m$ is more involved and requites one to regularize the integral over $\tau$ and then take the proper double limit.  To be brief, we show how it works for $\gamma =2$.  We have
 \beq
 V(\tau) = \frac{1}{2\pi}
 \int_{-\infty}^{\infty}
 d \Omega_m e^{i\Omega_m \tau} V(\Omega_m) = \frac{{\bar g}^2}{2\omega_D} e^{-\omega_D |\tau|}
 \eeq
 Then
 \beq
 V^{-1}(\tau) = \frac{2\omega_D}{{\bar g}^2} e^{\omega_D |\tau|}
 \eeq
 The inverse interaction $V^{-1} (\Omega_m) = \int_{-\infty}^{\infty}
 d\Omega_m
  e^{-i\Omega_m \tau} V^{-1} (\tau)$.
 To make this integral convergent, we add to the integrand $e^{-\delta \tau^2}$ and set $\delta \to 0$ at the end of calculations.  Integrating over $\tau$ we obtain
 \beq
 V^{-1} (\Omega_m) = \frac{2\omega_D}{{\bar g}} \left(S(\Omega_m) + S(-\Omega_m)\right),
 \eeq
  where at large $|\Omega_m| \gg (\delta)^{1/2}$,
  \beq
  S(\Omega_m) = -\frac{1}{\omega_D + i \Omega_m} + \sqrt{\frac{\pi}{\delta}} e^{\frac{(\omega_D + i \Omega_m)^2}{4\delta}}
  \eeq
  Then
  \bea
&& V^{-1} (\Omega_m) = \frac{4}{{\bar g}^2} \times \nonumber \\
&& \left[-\pi \omega_D \delta (\Omega_m) - \frac{\omega^2_D}{\Omega^2_m} + \frac{\sqrt{\pi}}{\sqrt{\delta}}\omega_D \cos{(\frac{\omega_D \Omega_m}{2\delta})}
e^{\frac{(\omega_D + i \Omega_m)^2}{4\delta}}\right]
 \label{f:3}
 \eea
 A simple experimentation shows that in the double limit, in which we set $\omega_D \to 0$ first and set $\delta \to 0$ after that, the last term in (\ref{f:3})  reduces to $2\pi \omega_D \delta (\Omega_m)$ (To see this one can just integrate the last term in (\ref{f:3}) over $\Omega_m$. The integration yields $2\pi \omega_D$.)
 The total $ V^{-1} (\Omega_m)$ is then
 \beq
 V^{-1} (\Omega_m) = \frac{4 \omega_D}{{\bar g}^2} \left[\pi \delta (\Omega_m) - \frac{\omega_D}{\Omega^2_m}\right]
 \eeq
 In the same limit $\omega_D \to 0$, we can  express $V(\Omega_m)$ as
  \beq
 V (\Omega_m) = \frac{{\bar g}^2}{\omega_D} \left[\pi \delta (\Omega_m) + \frac{\omega_D}{\Omega^2_m}\right]
 \eeq
 This analysis  can be easily extended to $\gamma <2$.

 \subsection{Expansion of Free energy $F_{sc,n}$ in $\omega_D$}

  The free energy, obtained in Hubbard-Stratonovich formalism, is given by
  Ref. 
  (\ref{eq:var_energy_actual}).
  We
   select the stationary solution  for the gap $\Delta_n (\omega_m)$
  and the quasiparticle residue $Z_n(\omega_m) =  Z (\Delta_n (\omega_m))$  with some $n$ and expand around them to second order in  $\delta \Delta (\omega_m) = \Delta (\omega_m) - \Delta_n (\omega_m)$ and $\delta Z (\omega_m) = Z (\omega_m) - Z_n (\omega_m)$. The zeroth order term $F^{(0)}_{sc,n}$ is
   \bea
   && F^{0}_{sc,n} = - N_F 2\pi T \sum_{m} Z_n (\omega_m) \sqrt{\omega^2_m + \Delta^2_n (\omega_m)} \nonumber \\
   &&+ N_F T^2 \sum_{m,m'} V^{-1} (\omega_m -\omega_{m'}) \nonumber \\
    && \left[Z_n (\omega_m) \Delta_n (\omega_m) Z_n (\omega_{m'}) \Delta_n (\omega_{m'}) + \omega_m \omega_{m'} (Z_n (\omega_m)-1) (Z_n (\omega_{m'})-1)\right].
   \label{app_1}
   \eea
   Using the Eliashberg equations
   \bea
   Z_n(\omega_m) \Delta_n (\omega_m) &=& \pi T \sum_{m'} V(\omega_m - \omega_{m'})  \frac{\Delta_n (\omega_{m'})}{\sqrt{\omega^2_{m'} + \Delta^2_n (\omega_{m'})}} \label{app_2} \\
   \omega_m (Z_n(\omega_m)-1)  &=& \pi T \sum_{m'} V(\omega_m - \omega_{m'})  \frac{\omega_{m'}}{\sqrt{\omega^2_{m'} + \Delta^2_n (\omega_{m'})}}
   \label{app_2_a}
   \eea
   we re-express $ F^{0}_{sc,n}$ as
  \bea
   F^{0}_{sc,n} &=& - N_F 2\pi T \sum_{m} Z_n (\omega_m) \sqrt{\omega^2_m + \Delta^2_n (\omega_m)} \nonumber \\
   &&+ N_F \pi^2 T^2 \sum_{m,m'} \frac{\Delta_n (\omega_{m}) \Delta_n (\omega_{m'}) + \omega_{m} \omega_{m'}}{\sqrt{\omega^2_{m} + \Delta^2_n (\omega_{m})} \sqrt{\omega^2_{m'} + \Delta^2_n (\omega_{m'})}} S(m,m')
   \label{app_3}
   \eea
   where
   \beq
   S(m,m') = T^2 \sum_{m^{''}, m^{'''}} V^{-1} (\omega_{m} - \omega_{m^{''}}) V(\omega_{m^{''}} - \omega_{m^{'''}}) V(\omega_{m^{'''}} - \omega_{m'})
   \eeq
  The sum is evaluated using the relation
   \beq
   T^2 \sum_{m^{''}} V^{-1} (\omega_m - \omega_{m^{''}}) V(\omega_{m^{''}} - \omega_{m'}) = \delta_{m,m'}
  \label{qq}
   \eeq
  Using (\ref{qq}),  we obtain $S(m,m') = V(\omega_m - \omega_{m^{'}})$ and
   \bea
   F^{0}_{sc,n} &=& - N_F 2\pi T \sum_{m} Z_n (\omega_m) \sqrt{\omega^2_m + \Delta^2_n (\omega_m)} \nonumber \\
   &&+ N_F \pi^2 T^2 \sum_{m,m'} \frac{\Delta_n (\omega_{m}) \Delta_n (\omega_{m'}) + \omega_{m} \omega_{m'}}{\sqrt{\omega^2_{m} + \Delta^2_n (\omega_{m})} \sqrt{\omega^2_{m'} + \Delta^2_n (\omega_{m'})}} V(\omega_m - \omega_{m'})
   \label{app_3_1}
   \eea
   Using the Eliashberg equations once again, we re-express  $F^{0}_{sc,n}$ as
   \bea
   F^{0}_{sc,n} &=& - N_F 2\pi T \sum_{m}
   \frac{\omega^2_m}{\sqrt{\omega^2_m + \Delta^2_n (\omega_m)}}\nonumber \\
  && -  \pi^2 T^2 N_F \sum_{m, m'} V(\omega_m-\omega_{m'}) \frac{\omega_m \omega_{m'} + \Delta_n (\omega_m) \Delta_n (\omega_{m'})}{\sqrt{\omega^2_m + \Delta^2_n (\omega_m)} \sqrt{(\omega_{m'})^2 + \Delta^2_n (\omega_{m'})}}
  \label{eq:10}
  \eea
  This is the same expression as the variational free energy
  $F^{\text{var}}_{sc}$  for stationary $\Delta_n (\omega_m)$,  Eq. (\ref{eq:var_energy_2}).

   First order terms in the expansion of $F_{sc,n}$ in $\delta \Delta$ and $\delta Z$ vanish  because  $F_{sc}$ is stationary with respect to variations around $\Delta_n (\omega_m)$ and $Z (\Delta_n (\omega_m))$. The expansion to second order in variations yields
   \bea
   &&F^{(2)}_{sc,n} = - N_F \pi T \sum_{m} \frac{(\delta \Delta_m)^2 Z_n (\omega_m) \omega^2_m}{(\omega^2_m + \Delta^2_n (\omega_m))^{3/2}} + N_F T^2  \sum_{m, m'} V^{-1} (\omega_m-\omega_{m'}) \times \label{eq:11} \\
  && \left[\delta \Delta_m \delta \Delta_{m'} Z_n (\omega_m) Z_n (\omega_{m'}) + \delta Z_m \delta Z_{m'}
  \left(\omega_m \omega'_m + \Delta_n (\omega_m) \Delta_n (\omega'_m)\right) + 2 \delta Z_m \delta \Delta_{m'} Z_{n} (\omega_{m'}) \Delta_n (\omega_m)\right]. \nonumber
  \eea

 For a generic $\omega_D$, $\delta Z_m$ and $\delta \Delta_m$ are two independent variations.
 For small $\omega_D$, the largest contribution to $Z_n (\omega_m)$ comes from the term with
 $m=m'$ 
 in the r.h.s. of (\ref{app_2}).  If we keep only this term, we find
 \beq
 Z_n (\omega_m) =  \left(\frac{{\bar g}}{\omega_D}\right)^\gamma \frac{\pi T}{\sqrt{\omega^2_m + \Delta^2_n (\omega_m)}}.
 \label{eq:12}
 \eeq

Based on this observation, we split $\delta Z_m$ into two components, one of which satisfies Eq. (\ref{eq:12}):
 \beq
 \delta Z_m = - \frac{\Delta_n (\omega_m) Z_n (\omega_m) \delta \Delta_m}{\omega^2_m + \Delta^2_n (\omega_m)} + {\bar \delta} Z_m
 \label{eq:15}
 \eeq
 The first term in (\ref{eq:15}) would be the variation of $Z_n (\omega_m)$  if  Eq. (\ref{eq:12}) was valid for arbitrary $\Delta (\omega_m)$, not just a stationary $\Delta_n (\omega_m)$

 Substituting $\delta Z_m$ from (\ref{eq:15}) into (\ref{eq:11}), we find $F^{(2)}_{sc,n}$  as the sum of three terms. The first one contains only $\delta \Delta$, the second one contains only ${\bar \delta}Z$, and the third is the mixed term.
  The term with ${\bar \delta}Z$ is
  \beq
  N_F T^2 \sum_{m,m'} V^{-1} (\omega_m-\omega_{m'})  {\bar \delta} Z_m {\bar \delta} Z_{m'} \left(\omega_m \omega'_m + \Delta_n (\omega_m) \Delta_n (\omega'_m)\right)
  \label{eq:16}
 \eeq
We now use Eq. (\ref{eq:17}) for $V^{-1} (\Omega_m)$.
 Substituting into (\ref{eq:16}), we find that this component of $F^{(2)}_{sc,n}$ contains $\omega^{\gamma}_D$ as the overall factor and hence vanishes as $\omega_D \to 0$.  The same happens with the mixed term.   It is
 \beq
  -2N_F T^2 \sum_{m,m'} V^{-1} (\omega_m-\omega_{m'})  {\bar \delta} Z_m \delta \Delta_{m'}
   \frac{Z_n (\omega_{m'}) \omega_{m'} \left(\Delta_n (\omega_{m'}) \omega_m - \Delta_n (\omega_m) \omega_{m'}\right)}{\omega^2_{m'} + \Delta^2_n (\omega_{m'})}
  \label{eq:18}
 \eeq
Substituting  $V^{-1} (\omega_m-\omega_{m'})$ from (\ref{eq:17}), we find that the leading term  with $m=m'$
 does not contribute because the summand in (\ref{eq:18}) vanishes at $m=m'$.   Combining the subleading term in $V^{-1}$ with $Z_n (\omega_m) \propto ({\bar g}/\omega_D)^\gamma$, we find that the overall factor in (\ref{eq:18}) again scales as $\omega^\gamma_D$ and vanishes at $\omega_D \to 0$.   This implies that variations ${\bar \delta} Z$ do not affect the free energy, at least for quadratic deviations from a stationary point.

The remaining term in $F^{(2)}_{sc,n}$ is quadratic in $\delta \Delta$.  Combing  all contributions of this kind, we obtain
 \bea
 && F^{(2)}_{sc,n} = - N_F \sum_{m} \frac{(\delta \Delta_m)^2 Z_n (\omega_m) \omega^2_m}{(\omega^2_m + \Delta^2_n (\omega_m))^{3/2}}  + N_F T^2  \sum_{m, m'} V^{-1} (\omega_m-\omega_{m'})
  \delta \Delta_m \delta \Delta_{m'} \times \nonumber \\
 &&\frac{Z_n (\omega_m) Z_n (\omega_{m'}) \omega_m \omega'_m
  \left(\omega_m \omega'_m + \Delta_n (\omega_m) \Delta_n (\omega'_m)\right)}{(\omega^2_{m} + \Delta^2_n (\omega_{m}))
  (\omega^2_{m'} + \Delta^2_n (\omega_{m'}))}
  \label{eq:19}
 \eea
Substituting $V^{-1}$ from (\ref{eq:17}), we find after  simple algebra that potentially divergent terms at $\omega_D \to 0$  cancel out, and the remaining piece is
\bea
 && F^{(2)}_{sc,n} =  -N_F \pi T \sum_{m} \frac{(\delta \Delta_m)^2 Z_n (\omega_m) \omega^2_m}{(\omega^2_{m} + \Delta^2_n (\omega_{m}))^{3/2}}  + N_F T^2  \sum_{m, m'} \delta \Delta_m \delta \Delta_{m'} V^{-1} (\omega_m -\omega_{m'})  \times \nonumber \\
 && \frac{Z_n (\omega_m) Z_n (\omega_{m'}) \omega_m \omega'_m
  \left(\omega_m \omega'_m + \Delta_n (\omega_m) \Delta_n (\omega'_m)\right)}{(\omega^2_{m} + \Delta^2_n (\omega_{m}))
  (\omega^2_{m'} + \Delta^2_n (\omega_{m'}))}
  \label{eq:20}
 \eea
Using again Eq. (\ref{eq:17}) for $V^{-1} (\omega_m-\omega_{m'})$ and re-expressing it via $V(\omega_m-\omega_{m'})$ from
(\ref{eq:17_a}),
 we obtain $F^{(2)}_{sc,n}$ as the sum of the two terms:
$F^{(2)}_{sc,n} = F^{(2,a)}_{sc,n} + F^{(2,b)}_{sc,n}$, where
\bea
 && F^{(2,a)}_{sc,n} =  N_F \pi T \sum_{m} \frac{(\delta \Delta_m)^2 \omega^2_m}{(\omega^2_{m} + \Delta^2_n (\omega_{m}))^{3/2}}  + \nonumber \\
  && N_F \frac{\pi^2}{2} T^2  \sum_{m \neq m'} V (\omega_m -\omega_{m'}) \frac{\omega_m \omega'_m}{ (\omega^2_{m} + \Delta^2_n (\omega_{m}))^{3/2}
  (\omega^2_{m'} + \Delta^2_n (\omega_{m'}))^{3/2}} \times \nonumber \\
  && \left[\left(\delta \Delta_m \omega_{m'} - \delta \Delta_{m'} \omega_{m}\right)^2 + \left(\delta \Delta_m \Delta_n (\omega_{m'}) - \delta \Delta_{m'} \Delta_n (\omega_m)\right)^2\right]
  \label{eq:21}
 \eea
 and
 \beq
  F^{(2,b)}_{sc,n}=  N_F \left(\frac{\omega_D}{\bar g}\right)^{\gamma} \pi T \sum_{m} \frac{(\delta \Delta_m)^2 \omega^2_m}{(\omega^2_{m} + \Delta^2_n (\omega_{m}))}  \left(Z_n (\omega_m) - \frac{\pi T}{\sqrt{\omega^2_{m} + \Delta^2_n (\omega_{m})}} \left(\frac{\bar g}{\omega_D}\right)^{\gamma} \right)^2
  \label{eq:22}
 \eeq
Using Eq. (\ref{ss_11_01}) we find that the  last bracket is the effective  $Z_n (\omega_m)$, without the
$m'=m$ term in the r.h.s. of  (\ref{ss_11_01}).   This expression is non-singular at $\omega_D \to 0$.  Then
 $F^{(2,b)}_{sc,n}$ vanishes at vanishing $\omega_D$ because of the overall factor. Hence,  in this limit,
 $F^{(2)}_{sc,n} = F^{(2,a)}_{sc,n}$.

The expression for $F^{(2)}_{sc,n}$ can be further simplified by using the Eliashberg equation  for $\Delta_n (\omega_m)$ and re-expressing  the first term in (\ref{eq:21})  as the double sum over $m$ and $m'$. The result looks more compact when expressed in terms of $D_n (\omega_m) = \Delta_n (\omega_m)/\omega_m$ and $\delta D_m = \delta \Delta_m/\omega_m$.  After simple algebra, we obtain
\beq
F^{(2)}_{sc,n} =  N_F \pi^2 T^2 \bar g^\gamma \sum_{m, m' \geq 0} \frac{1}{D_n (\omega_m)D_n (\omega_{m'})}
   \frac{1}{(1 + D^2_n (\omega_{m}))^{3/2}
  (1 + D^2_n (\omega_{m'}))^{3/2}} Q_{m,m'}
\label{eq:23}
\eeq
where
\bea
&&Q_{m,m'} = {\bar g}^\gamma \times  \nonumber \\
  && \left(\delta D_m D_n (\omega_m) - \delta D_{m'} D_n (\omega_{m'})^2\right) \left(\frac{1}{|\omega_m-\omega_{m'}|^\gamma} + \frac{1}{|\omega_m+\omega_{m'}|^\gamma}\right) \nonumber \\
  && +
\frac{\left(\delta D_m D^2_n (\omega_m) - \delta D_{m'} D^2_n (\omega_{m'})\right)^2}{|\omega_m-\omega_{m'}|^\gamma} +
\frac{\left(\delta D_m D^2_n (\omega_m) + \delta D_{m'} D^2_n (\omega_{m'})\right)^2}{|\omega_m+\omega_{m'}|^\gamma}
\label{eq:23_a1}
 \eea
Note that the term with $m=m'$ vanishes because of vanishing numerator (to see this clearly one should keep infinitesimally small $\omega_D$ in the denominator).

We now perform the same calculation staring from the variational free energy $F^{\text{var}}_{sc}$, Eq.~(\ref{eq:var_energy_2}). This free energy depends only on $\Delta (\omega_m)$.  Expanding Eq.~(\ref{eq:var_energy_2}) to second order in  $\delta \Delta_m = \Delta (\omega_m) - \Delta_n (\omega_m)$ and using the Eliashberg equation for stationary $\Delta_n (\omega_m)$, we obtain after long but straightforward algebra that $F^{\text{var}}_{sc,n} = F^{\text{var},(0)}_{sc,n} + F^{\text{var},(2)}_{sc,n}$, where $F^{\text{var},(0)}_{sc,n}$ and $F^{\text{var},(2)}_{sc,n}$ are given by the same Eqs.~(\ref{eq:10}) and (\ref{eq:23}) as $F^{(0)}_{sc,n}$ and $F^{(2)}_{sc,n}$, respectively. This implies that at vanishing $\omega_D$, the expansion of the actual free energy around a stationary point coincides with the expansion of the variational free energy.
 For $\gamma =2$ this has been established in Ref. \cite{yuz}.

More accurately, at a non-zero $T$,  $F^{\text{var},(0)}_{sc,n}$ coincides with $F^{(0)}_{sc,n}$ for arbitrary $\omega_D$, but $F^{\text{var},(2)}_{sc,n} = F^{(2)}_{sc,n}$  only up to corrections of order  $(\omega_D/{\bar g})^\gamma$.
This holds for all $n$, including $n=0$.
 At $T=0$, $F^{\text{var},(0)}_{sc,n}$ and  $F^{(0)}_{sc,n}$ again coincide, and $F^{\text{var},(2)}_{sc,n}$ and  $F^{(2)}_{sc,n}$  differ by terms  of order $(\omega_D/{\bar g})^{\gamma-1}$. This difference is small for $\gamma >1$, when $Z_n (\omega_m)$ diverges at vanishing $\omega_m$. For these $\gamma$, $F^{(2)}_{sc,n}$ is given by Eq. (\ref{eq:23}) with $\pi^2 T^2 \sum_{m,m' \geq 0}$ replaced by $(1/4) \int_{0}^\infty \int_0^\infty d\omega_m d\omega_{m'}$.   For $\gamma <1$, $Z_n (\omega_m)$ remains finite at $\omega_D =0$, and $F^{{\text{var}},(2)}_{sc,n}$ and $F^{(2)}_{sc,n}$ do not coincide, even at $\omega_D =0$. Specifically,
   $F^{{\text{var}},(2)}_{sc,n}$ is still given by Eq.~(\ref{eq:23}) with the sum replaced by the integral, but $F^{(2)}_{sc,n}$ has an additional contribution from fluctuations of ${\bar \delta} Z_m$.

\subsection{Beyond second order in $\delta \Delta_m$ }

  We now go further and verify that at $T >0$, $F_{sc}^{\text{var}}$ and $F_{sc}$ remain equivalent to all
   orders in $\delta \Delta_m$, up to terms  $O(\omega_D/{\bar g})^{\gamma}$.
    To see this, we use as inputs the facts that at vanishing $\omega_D$ and a finite $T$
      (i) $V^{-1}(\Omega_m)$ is represented by  series of $(\omega_D/{\bar g})^{\gamma}$ as in Eq.~(\ref{eq:17}), (ii) $\Delta_m$ remains a regular function of $\omega_m$, and (iii) $Z_m$ contains a divergent thermal piece. We single out the divergent term and redefine $Z_m$  as
 \bea
 Z_m = {\pi T\over \sqrt{\omega_m^2 + \Delta_m^2}} \left( {{\bar g} \over \omega_D } \right)^\gamma + {\bar Z}_m.
 \eea
 We then expand $F_{sc}$ in series of $(\omega_D/{\bar g})^{\gamma}$. After straightforward algebra, done without using the Eliashberg gap equations, we obtain:
 \bea
 F_{sc} = F_{sc}^{\text{var}} + N_F  \left( {\omega_D \over {\bar g}  } \right)^\gamma \sum_m \left[ \bar{Z}_m^2 (\omega_m^2 + \Delta_m^2) -2 u_m {\bar Z}_m \omega_m^2  + v_m \omega_m^2  \right] +  O\left( {\omega_D \over {\bar g} } \right)^{2\gamma},
\label{ss_e}
 \eea
where
\bea
u_m &=& 1 + 2 \pi T \sum_{m\neq m'} V(m-m') {\omega_m'/ \omega_m + \Delta_{m'} \Delta_{m}/\omega_m^2 \over \sqrt{\omega_{m'}^2 + \Delta_{m'}^2} } , \nonumber \\
v_m &=& 1+ 2 \pi T \sum_{m\neq m'} V(m-m') {\omega_m' / \omega_m \over \sqrt{\omega_{m'}^2 + \Delta_{m'}^2} }.
\eea
One can check that the prefactor for $\left(\omega_D/{\bar g}\right)^\gamma$ is non-singular. Eq/ (\ref{ss_e}) then shows that  $F_{sc}$ and $F_{sc}^{\text{var}}$ differ by terms $O\left( {\omega_D/{\bar g} } \right)^\gamma$ the difference between   $F_{sc}$ and $F_{sc}^{\text{var}}$ is of order  $O\left( {\omega_D/{\bar g} } \right)^\gamma$
for arbitrary $\Delta (\omega_m)$  and $Z(\omega_m)$,  $F_{sc}$ and $F_{sc}^{\text{var}}$ differ by terms $O\left( {\omega_D/{\bar g} } \right)^\gamma$. This difference vanishes in the limit $\omega_D \to 0$.

 \section{Expansion of free energy around stationary $F_{sc,n}$
  for the subclass of analytic gap functions. }
 \label{app:C}

In this Appendix we discuss the details of our calculation of the matrix of quadratic deviations from $\Delta_n (\omega_m)$, assuming that the deviations $\delta \Delta (\omega_{m})$ are analytic functions of complex $z = \omega'+i\omega^{''}$, when extended  into the upper half-plane of frequency.

 An analytic  $\delta \Delta (\omega_m)$ can by expressed via the Cauchy relation
  in terms of $\psi (\omega) =$ Im $\delta \Delta (\omega)$ along the real axis.  Substituting this relation into the expression for the free energy, we obtain $F^{(2)}_{sc,n}$ in the form
 \beq
 F^{(2)}_{sc,n} = N_F \int_0^\infty \int_0^\infty d\omega d \omega' \psi_n(\omega) \psi_n (\omega') \bar{{\cal K}}_{\omega, \omega'},
  \eeq
 where
 \beq
 \bar{{\cal K}}_{\omega, \omega'} = \sum_p \lambda^{p}_n S^{p}_n (\omega) S^{p}_n (\omega')
 \eeq
  and
  \beq
  S^{p}_n (\omega) = 2 T \sum_m \frac{\Delta^{p}_n (\omega_m) \omega}{\omega^2 + \omega^2_m}
  \eeq
   The functions $S^{p}_n (\omega)$ are not orthogonal
   and their set is over-complete,
   as one can explicitly verify.
    We now introduce a complete set of orthogonal functions ${\bar S}^{k}_n (x)$, which satisfy
    \beq
    \int_0^\infty dx {\bar S}^{k}_n (x) {\bar S}^{k'}_{n} (x) = \delta_{k,k'}
\eeq
 Expressing $Q_{\omega, \omega'}$ and  $ S^{p}_n (x)$ in terms of these functions, we obtain
   \bea
   Q_{\omega, \omega'} &=& \sum_p {\bar \lambda}^{p}_n {\bar S}^{p}_n (\omega) {\bar S}^{p}_n (\omega') \nonumber \\
    S^{p}_n (x) &=& \sum_{k} g_{p,k} {\bar S}^{k}_n (x)
    \eea
   A simple manipulation then shows that
   \beq
   {\bar \lambda}^{p}_n  = \sum_k \lambda^{k}_n g^2_{p,k}
  \eeq

\begin{figure}
\centering
\includegraphics[scale=1]{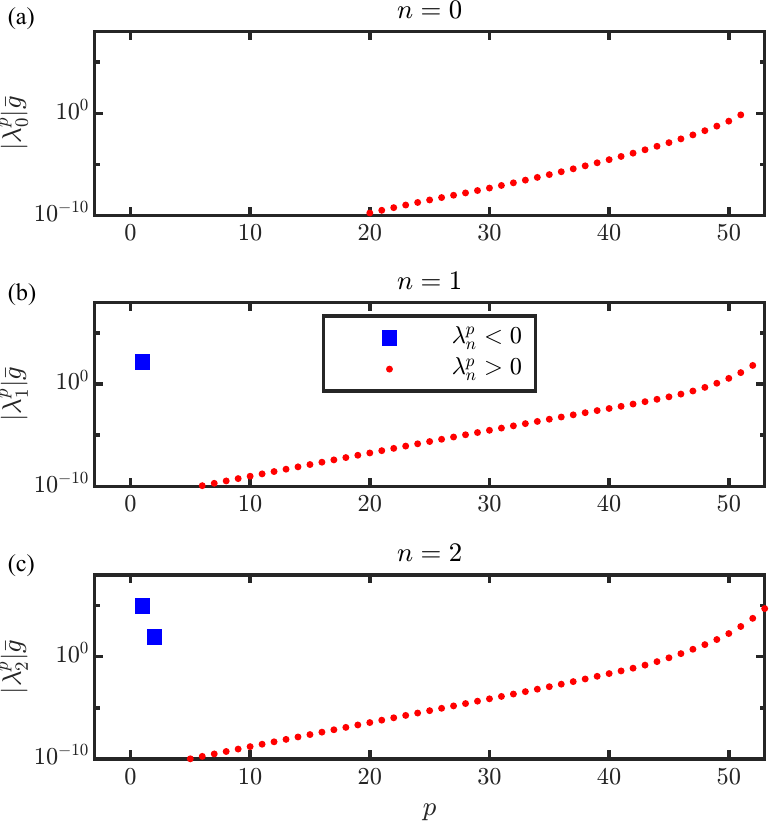}\caption{Eigenvalues of the kernel matrix of
$F^{(2)}_{sc,n}$ 
in a restricted subspace analytic in the upper half-plane of frequency, where (a) $n=0$, (b) $n=1$, and (c) $n=2$ solutions. Here we take $\gamma=0.8$.}
\label{fig:eigenval_phys}
\end{figure}

\begin{figure}
\centering
\includegraphics[scale=1]{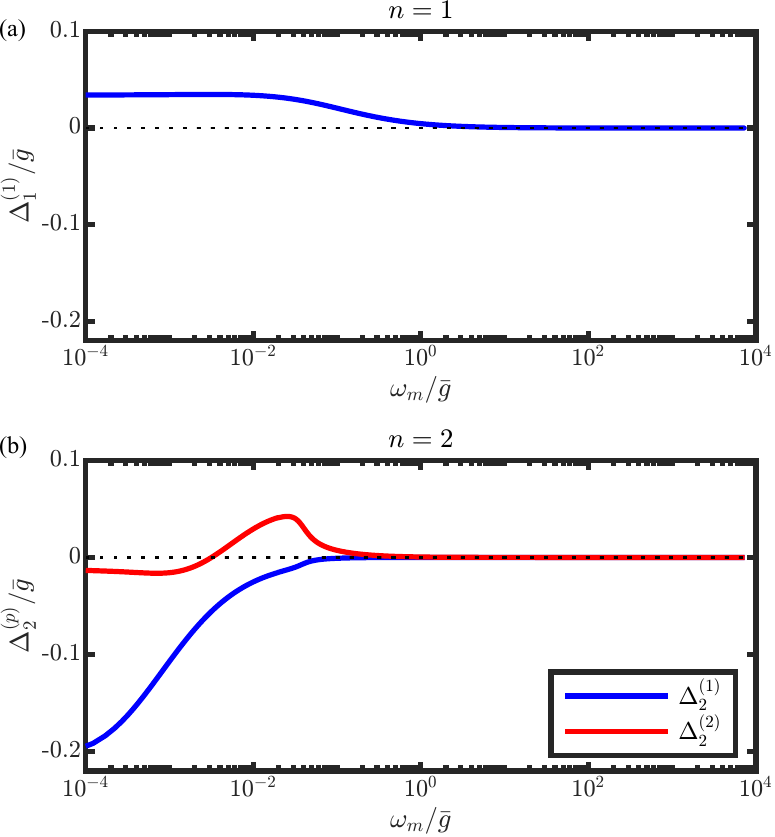}\caption{Eigenfunctions corresponding to the negative eigenvalues shown in Fig.~\ref{fig:eigenval_phys} (b) $n=1$ and (c) $n=2$, which are analytic in the upper half-plane of frequency. Here we take $\gamma=0.8$.  }
\label{fig:eigenfunc_phys}
\end{figure}

In Fig.~\ref{fig:eigenval_phys} we compare the eigenvalues ${\bar \lambda}^{p}_n$ ($n=0,1,2$)
 with $\lambda^{p}_n$, which we obtained by diagonalizing $F^{(2)}_{sc,n}$ from Eq. (\ref{eq:n4}) from the main text without imposing the condition of analyticity  on $\delta \Delta (\omega_m)$.
We see that the two sets are non-equivalent, yet in both cases there are $n$ negative eigenvalues for the expansion around $\Delta_n (\omega_m)$.
The eigenfunctions, which correspond to  negative  ${\bar \lambda}^{p}_n$, define the unstable directions in the restricted subspace towards $\Delta_{n'} (\omega_m)$ with $n' <n$, as shown in Fig.~\ref{fig:eigenfunc_phys}.
This  implies that all physically relevant perturbations are described by analytic $\delta \Delta (\omega_m)$. This result  is also consistent with the analysis in the main text of the free energy for hybrid trial functions, made of combinations of stationary $\Delta_n (\omega_m)$. All stationary gap functions are analytic in the upper half-plane of frequency, as we explicitly verified, hence the hybrid trial functions are also analytic.

 \section{Evolution of trial
 $\Delta_n^{\text{trial}}  
 (\omega_m)$  under iterations}

 Our trial solutions ``almost'' satisfy the gap equation, but are not the exact $\Delta_n (\omega_m)$ and hence evolve under iterations.  The analysis in the main text and in the previous Appendix shows that under iterations  all trial $\Delta_n^{\text{trial}} (\omega)$ should slowly move towards the actual $\Delta_0 (\omega_m)$. To verify this, we re-express the gap equation as
\beq
\label{eq:iteration}
\Delta(\omega_m) = \frac{\pi T {\bar g}^{\gamma} \sum_{m' \neq m} \frac{1}{\rvert \omega_m-\omega_m^{\prime} \rvert^{\gamma}} \frac{\Delta(\omega_m^{\prime})}{\sqrt{(\omega_m^{\prime})^2 + \Delta^2(\omega_m^{\prime})}}}
{ 1+ \pi T {{\bar g}^{\gamma} \over 2} \frac{\Delta (\omega_m)}{\omega_m} \sum_{m' \neq m} \frac{1}{\rvert \omega_m-\omega_m^{\prime} \rvert^{\gamma}}  { \omega_m^{\prime} \over \sqrt{(\omega_m^{\prime})^2 + \Delta^2(\omega_m^{\prime})} } }
\eeq
and initiate the iteration process by substituting the  trial $\Delta_n^{\text{trial}} (\omega_m)$ into the r.h.s.

\begin{figure}
\centering
\includegraphics[scale=1]{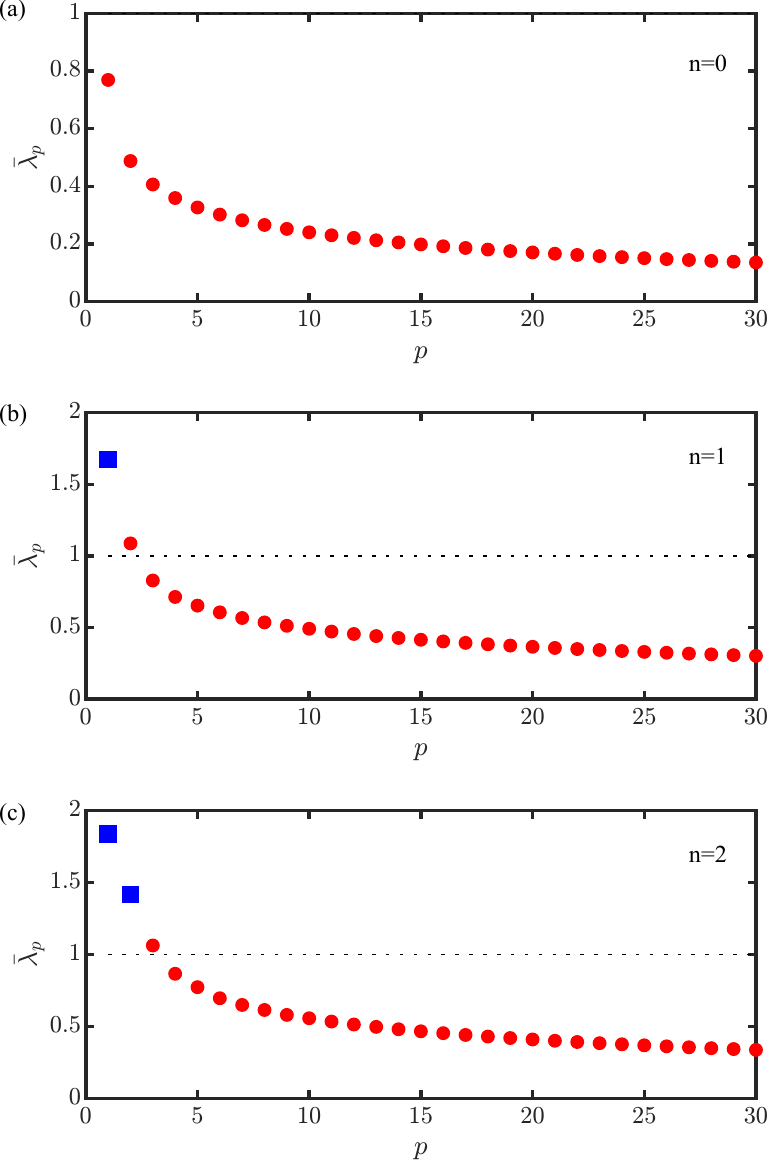}
\caption{Eigenvalues $\bar{\lambda}_p$ of the corresponding kernel of the iteration procedure following Eq.~(\ref{eq:iteration})
for the $n=0$, $n=1$, and $n=2$ gap functions, respectively, where $\gamma=0.8$. Solid symbols mark the eigenstates corresponding to
unstable ``directions'' of the iteration matrix. The eigenvalue $\bar{\lambda}_p$ slightly above $1$ for both $n=1,2$ solutions corresponds to stable ``direction''
of the iteration matrix, whose value becomes larger than one because the trivial gap function is close to not exactly located at the saddle point of the free-energy functional.}
\label{fig:iteration_lt1}
\end{figure}

For $n=0$, the iteration procedure converges to the exact $\Delta_0 (\omega_m)$, which we
already used in the calculation of eigenfunctions for the free energy variations. For $n =1$, the
trial $\Delta_1^{\text{trial}} (\omega_m)$  remains almost unchanged under first few iterations, but eventually gets modified and
moves towards $\Delta_0 (\omega_m)$ (see Fig.~\ref{fig:iteration}).   The same happens with the trial
 $\Delta_2^{\text{trial}} (\omega_m)$.

To quantify this analysis, we linearized the gap equation (\ref{eq:iteration}) in deviations from the trial gap functions  and found the eigenvalues of the corresponding kernel, ${\bar \lambda}_n^{(p)}$ . One can easily verify that the trial function is stable (i.e., the iteration procedure converges) if $|{\bar \lambda}_n^{(p)}| <1$ for all $p$ and unstable if  $|{\bar \lambda}_n^{(p)}| >1$ for at least for one value of $p$. We show the result in Fig.~\ref{fig:iteration_lt1}. We see that for $n=0$ all eigenvalues are below $1$, while for $n=1$ and $n=2$, there are one and two eigenvalues larger than $1$, respectively.

\section{Expansion in terms of the gap amplitude}
\label{sec:perturb}
In this Appendix we describe
 in some detail amplitude expansion for the function $D(\omega_m) = \Delta (\omega_m)/\omega_m$.
Expanding the r.h.s. of Eq.~(\ref{eq:8}) in powers of $D$, we obtain
\begin{equation}
\omega_{m}D(\omega_{m})=\frac{\bar{g}^{\gamma}}{2}\int_{-\infty}^{\infty}d\omega_{m}^{\prime}
\left(D(\omega_{m}^{\prime})-D(\omega_{m})\right)\frac{\text{sign}(\omega_{m}^{\prime}
)}{\rvert\omega_{m}^{\prime}-\omega_{m}\rvert^{\gamma}}\left(1-\frac{1}{2}D^{2}(\omega_{m}^{\prime})+
\frac{3}{8}D^{4}(\omega_{m}^{\prime})+...\right).\label{eq:expansion}
\end{equation}
We will be searching for the solution in the form
\begin{equation}
D(\omega_{m})=\sum_{j}  D^{(2j+1)}(\omega_{m}),
\label{app_4}
\end{equation}
where $j=0,1,2,...$. The first term $D^{(1)} (\omega_m)$ is the solution of the linearized gap equation (the one which we labeled as $D_\infty (\omega_m)$ in the main text).  We assign an overall factor $\epsilon$  to $D^{(1)} (\omega_m)$
and set $D^{(2j+1)} (\omega_m)$ to be proportional to $\epsilon^{(2j+1)}$.  The expansion in (\ref{app_4}) then holds in powers of $\epsilon^2$.
 At each order of the expansion in $\epsilon$ we obtain the equation on $D^{(2j+1)}(\omega_{m})$ with the source term expressed in terms of $D^{(2j+1)}(\omega_{m})$ with smaller $j$. Specifically,
\begin{align}
\omega_{m}D^{(2j+1)}(\omega_{m})-\frac{\bar{g}^{\gamma}}{2}\int_{-\infty}^{\infty}d\omega_{m}^{\prime}\left(D^{(2j+1)}(\omega_{m}^{\prime})-D^{(2j+1)}(\omega_{m})\right)\frac{\text{sign}(\omega_{m}^{\prime})}{\rvert\omega_{m}^{\prime}-\omega_{m}\rvert^{\gamma}} & =K^{(2j+1)}(\omega_{m}),
\label{ss_9}
\end{align}
where
\begin{align}
K^{(0)}(\omega_{m}) & =0,\\
K^{(3)}(\omega_{m}) & =- \frac{\bar{g}^{\gamma}}{4}\int_{-\infty}^{\infty} \frac{ d\omega_{m}^{\prime} \text{sign}(\omega_{m}^{\prime})}{\rvert\omega_{m}^{\prime}-\omega_{m}\rvert^{\gamma}} \left(D^{(1)}(\omega_{m}^{\prime})-D^{(1)}(\omega_{m})\right)(D^{(1)})^2(\omega_{m}^{\prime}), \label{eq:k3}\\
K^{(5)}(\omega_{m}) & =\frac{\bar{g}^{\gamma}}{2}\int_{-\infty}^{\infty} \frac{ d\omega_{m}^{\prime} \text{sign}(\omega_{m}^{\prime})}{\rvert\omega_{m}^{\prime}-\omega_{m}\rvert^{\gamma}} \Bigg[-\frac{1}{2}\left(D^{(3)}(\omega_{m}^{\prime})-D^{(3)}(\omega_{m})\right)(D^{(1)})^2(\omega_{m}^{\prime})\nonumber \\
 & +\left(D^{(1)}(\omega_{m}^{\prime})-D^{(1)}(\omega_{m})\right)\left(\frac{3}{8}(D^{(1)})^4(\omega_{m}^{\prime})-
 D^{(1)}(\omega_{m}^{\prime})D^{(3)}(\omega_{m}^{\prime})\right)\Bigg].
\end{align}
 and so on.

Below we
 focus on the solution for $D^{(2j+1)}(\omega_{m})$ at small $\omega \ll {\bar g}$. At these frequencies,
 \begin{equation}
 D^{(1)}(\omega_{m})=2\epsilon^{*}
 \cos\left(\psi_{0}(\omega_{m})\right),
\end{equation}
where $\epsilon^{*}=\epsilon\rvert\omega_{m}\rvert^{\gamma/2-1}$ and
\begin{equation}
\psi_{0}(\omega_{m})=\beta\log\rvert\omega_{m}/\bar{g}\rvert^{\gamma} + \phi
\end{equation}
  where $\beta =\beta_\gamma$ and $\phi = \phi (\gamma)$ are $\gamma-$dependent numbers (see the discussion around Eq. (\ref{eq:7}) in the main text and Refs. \cite{paper_1,paper_4,paper_5,paper_6}).
 Substituting this $D^{(1)}(\omega_{m})$ into (\ref{eq:k3}) we obtain the source term for $D^{(3)}(\omega_{m})$ in the form
 \begin{align}
K^{(3)}(\omega) & = (\epsilon^{*})^3 \bar{g}^{\gamma} \rvert\omega_{m}\rvert^{1-\gamma}(\epsilon^{*})^{3}\left(e^{i\psi_{\gamma}(\omega)}I_{1}^{(3)}+e^{3i\psi_{0}(\omega)}I_{3}^{(3)}\right)+c.c.,
\end{align}
where
\begin{align}
I_{1}^{(3)} & \equiv-\frac{1}{4}I\left(3\alpha+i\beta\gamma,2\alpha+2i\beta\gamma\right)-\frac{1}{2}I\left(3\alpha+i\beta\gamma,2\alpha\right),\\
I_{3}^{(3)} & \equiv-\frac{1}{4}I\left(3\alpha+3i\beta\gamma,2\alpha+2i\beta\gamma\right),
\end{align}
 $\alpha\equiv\gamma/2-1$, and
\begin{align}
I(a,b) & \equiv\int_{-\infty}^{\infty}\frac{dx}{\rvert x-1\rvert^{\gamma}}\left(\rvert x\rvert^{a}-\text{sign}(x)\rvert x\rvert^{b}\right).
\end{align}
This integral is infra-red convergent for
 $\gamma<3$. It  contains the "high-energy" part, which is determined by internal frequencies of order ${\bar g}$, and the universal part, which comes from $x = O(1)$ and determines the behavior of
 $D^{(3)}(\omega_{m})$  at $\omega_m \ll {\bar g}$. The universal part of $I(a,b)$ is
\bea
I(a,b) &=&B(\gamma-1-a,1+a)+B(\gamma-1-b,1+b)\nonumber \\
&&+B(1+a,1-\gamma)-B(1+b,1-\gamma)\nonumber \\
&&+B(\gamma-1-a,1-\gamma)-B(\gamma-1-b,1-\gamma).
\eea
Substituting this into the expression for $K^{(3)}(\omega_{m})$ and solving Eq. (\ref{ss_9}) for $D^{(3)}(\omega_{m})$, we obtain
\begin{equation}
D^{(3)}(\omega_{m}) =
\text{sign}(\omega_{m})(\epsilon^{*})^{3}
\left(Q_{1}^{(3)}e^{i\psi_{0}(\omega_{m})}+Q_{3}^{(3)}e^{3i\psi_{0}(\omega_{m})}\right)+c.c., \label{eq:Dw}
\end{equation}
where
\begin{align}
Q_{2r+1}^{(3)} & =-2\frac{I_{2r+1}^{(3)}}{I(3\alpha+i(2r+1)\beta\gamma,0)},\;r=0,1.
\end{align}
  Adding $D^{(1)}(\omega_{m})$  and $D^{(3)}(\omega_{m})$, we see that $D^{(3)}(\omega_{m})$
   affects the prefactor for $\cos{\psi_0 (\omega_m)}$ and also generates the higher harmonic $\cos{3\psi_0 (\omega_m)}$.

Further, we use the expressions for $D^{(1)}(\omega_{m})$  and $D^{(3)}(\omega_{m})$  and
compute the source term for $D^{(5)} (\omega_m)$:
\begin{equation}
K^{(5)}(\omega_{m})=\bar{g}^{\gamma}\rvert\omega_{m}\rvert^{\frac{3}{2}\gamma-4}(\epsilon^{*})^{5}\left(e^{i\psi_{0}(\omega_{m})}I_{1}^{(5)}+e^{3i\psi_{0}(\omega_{m})}I_{3}^{(5)}+e^{5i\psi_{0}(\omega_{m})}I_{5}^{(5)}\right)+c.c.,
\end{equation}
where
\begin{align}
I_{1}^{(5)} & =\frac{3}{16}\left[6I(5\alpha+i\beta\gamma,4\alpha)+4I(5\alpha+i\beta\gamma,4\alpha+2i\beta\gamma)\right]\nonumber \\
 & -\frac{1}{4}\left[2Q_{1}^{(3)}I(5\alpha+i\beta\gamma,2\alpha)+Q_{1}^{(3)*}I(5\alpha+i\beta\gamma,2\alpha+2i\beta\gamma)+Q_{3}^{(3)}I(5\alpha+i\beta\gamma,2\alpha-2i\beta\gamma)\right]\nonumber \\
 & -\frac{1}{2}\left[Q_{1}^{(3)}I(5\alpha+i\beta\gamma,4\alpha)+Q_{1}^{(3)*}I(5\alpha+i\beta\gamma,4\alpha)+\left(Q_{1}^{(3)}+Q_{3}^{(3)*}\right)I(5\alpha+i\beta\gamma,2\alpha+2i\beta\gamma)\right],\\
I_{3}^{(5)} & =\frac{3}{16}\left[4I(5\alpha+3i\beta\gamma,4\alpha+2i\beta\gamma)+I(5\alpha+3i\beta\gamma,4\alpha+4i\beta\gamma)\right]\nonumber \\
 & -\frac{1}{4}\left[2Q_{3}^{(3)}I(5\alpha+3i\beta\gamma,2\alpha)+Q_{1}^{(3)}I(5\alpha+3i\beta\gamma,2\alpha+2i\beta\gamma)\right]\nonumber \\
 & -\frac{1}{2}\left[\left(Q_{1}^{(3)}+Q_{3}^{(3)}\right)I(5\alpha+3i\beta\gamma,4\alpha+2i\beta\gamma)+Q_{3}^{(3)*}I(5\alpha+3i\beta\gamma,4\alpha+4i\beta\gamma)\right],\\
I_{5}^{(5)} & =\frac{3}{16}I(5\alpha+5i\beta\gamma,4\alpha+4i\beta\gamma)\nonumber \\
 & -\frac{1}{4}Q_{3}^{(3)}I(5\alpha+5i\beta\gamma,2\alpha+2i\beta\gamma)\nonumber \\
 & -\frac{1}{2}Q_{3}^{(3)}I(5\alpha+5i\beta\gamma,4\alpha+4i\beta\gamma),
\end{align}
The induced solution is
\begin{align}
D^{(5)}(\omega) &
 =
 \text{sign}(\omega_{m})(\epsilon^{*})^{5}\left(Q_{1}^{(5)}e^{i\psi_{0}(\omega_{m})}+Q_{3}^{(5)}e^{3i\psi_{0}(\omega_{m})}+Q_{5}^{(5)}e^{5i\psi_{0}(\omega_{m})}\right)+c.c.,
\end{align}
where
\begin{align}
Q_{2r+1}^{(5)} & =-2\frac{I_{2r+1}^{(5)}}{I(5\alpha+i(2r+1)\beta\gamma,0)}.
\end{align}

This expansion can be straightforwardly extended to higher orders. The generic form
of the gap function is
\begin{align}
D(\omega_{m}) & =\text{sign}(\omega_{m})\epsilon^{*}\Bigg[\left(1+(\epsilon^{*})^{2}Q_{1}^{(3)}+(\epsilon^{*})^{4}Q_{1}^{(5)}+...\right)e^{i\psi_{0}(\omega_{m})}\nonumber \\
 & +(\epsilon^{*})^{2}\left(Q_{3}^{(3)}+(\epsilon^{*})^{2}Q_{3}^{(5)}+...\right)e^{3i\psi_{0}(\omega_{m})}\nonumber \\
 & +(\epsilon^{*})^{4}\left(Q_{5}^{(5)}+...\right)e^{3i\psi_{0}(\omega_{m})}\nonumber \\
 & +...\Bigg]+c.c..
\label{app_5}
\end{align}

The expression for $D(\omega_m)$ in the main text is obtained by further expressing $\sum_{j=k}^{\infty}Q_{2k+1}^{(2j+1)}(\epsilon^{*})^{2j+1}$
 in each line in (\ref{app_5}) as
\begin{equation}
\sum_{j=k}^{\infty}Q_{2k+1}^{(2j+1)}(\epsilon^{*})^{2j+1} = (\epsilon^{*})^{2k+1}Q_{2k+1}e^{i(\phi_{2k+1}+(2k+1)\psi)}
\label{eq:reformulate}
\end{equation}
where  $Q_{2k+1}$, $\phi_{2k+1}$, and $\psi$ are given by series in $\epsilon^{*}$. These series are presented in
 Eqs.~(\ref{eq:8a})-(\ref{eq:8c}) in the main text.

\section{Condensation energy near $\gamma=2$}
\label{sec:ecn}

The condensation energy for a stationary $\Delta_n (\omega_m)$ is given by  Eq.~(\ref{eq:n2}). We reproduce this formula here for convenience of a reader. At $T=0$, which we consider here, we have
\bea
E_{c,n} &=& N_F \int d \omega_m \left(|\omega_m|-\frac{\omega^2_m}{\sqrt{\omega^2_m + \Delta^2_n (\omega_m)}}\right) -  \frac{1}{4} N_F \int \int d\omega_m d \omega'_{m} V(\omega_m-\omega'_m)   \nonumber \\
 && \times \left(  \frac{\omega_m \omega'_m + \Delta_n (\omega_m) \Delta_n (\omega'_m) }{\sqrt{\omega^2_m + \Delta^2_n (\omega_m)} \sqrt{(\omega'_m)^2 + \Delta^2_n (\omega'_m)}}  - {\text{sign}} (\omega_m \omega'_{m}) \right).
\label{eq:n2_1_1}
\eea
The first term in (\ref{eq:n2_1_1}) is infra-red convergent  and is determined by fermions with $\omega_m \sim {\bar g}$. This term is  not singular at $\gamma =2$ and below we just skip it.
We argue in this Appendix that the  second term in (\ref{eq:n2_1_1}) becomes singular in the limit $\gamma \to 2$  To see this we first consider $E_{c,0}$, which is the largest by magnitude.

\subsection{Condensation energy for $\Delta_0 (\omega_m)$}

The gap function $\Delta_0 (\omega_m)$ is sign-preserving and tends to $\Delta_0 (0) \sim {\bar g}$ at $\omega_m \ll {\bar g}$.  Using this, we express the
condensation energy as
\beq
E_{c,0}
 \approx
-\frac{1}{4}\bar{g}^{\gamma}N_{F} \left[\int_{-O(\bar g)}^{O(\bar g)}
\frac{d\omega_{m}d\omega_{m}^{\prime}}{\rvert\omega_{m}-\omega_{m}^{\prime}\rvert^{\gamma}}
\left(1-{\text{sign}}(\omega_{m}){\text{sign}}(\omega_{m}^{\prime})\right) +
 \text{regular terms}
\right]
\label{app_9}
\eeq
 This integral can be easily evaluated and for $\gamma \lesssim 2$ yields
\begin{equation}\label{eq:ec1}
E_{c,0}\simeq-\bar{g}^{2}N_{F} [ \frac{1}{2-\gamma} + \text{regular terms}].
\end{equation}
At a finite $\omega_D$, the demominator in (\ref{app_9}) is replaced by
$((\omega_m -\omega'_m)^2 + \omega^2_D)^{\gamma/2}$. The integral then remains finite even at $\gamma =2$,  and $E_{c,0}$ becomes
\begin{equation}\label{eq:ec3}
E_{c,0}\simeq - \bar{g}^{2}N_{F} [\ln {{\bar g}\over \omega_D} + \text{regular terms}].
\end{equation}
The same happens when $\omega_D =0$, but temperture $T$ is finite. At $\gamma =2$  we have
\begin{equation}\label{eq:ec2}
E_{c,0}\simeq-\bar{g}^{2}N_{F} [\ln {{\bar g}\over T} + \text{regular terms}],
\end{equation}

\subsection{Condensation energy for $\Delta_n (\omega_m)$ with $n >0$}
\label{app:E}

For any finite $n$, the gap function $\Delta_n (\omega_m)$  still saturates at a finite value $\Delta_n (0)$ at $\omega_m =0$.  At low enough frequencies,
 $\Delta_n (0) \gg \omega_m$, and the integrand of the condensation energy $E_{c,n}$ has the same as for $n=0$.  Then $E_{c,n}$ contains the same divergent piece as in (\ref{eq:ec1}).

We will also need the expression for the difference $E_{c,n} - E_{c,0}$.  We show below that it primarily
comes from the regions near the nodal points in
$\Delta_n (\omega_m)$, where $\rvert \Delta_n(\omega_m) \rvert \leq  \rvert \omega_m \rvert$. To demonstrate this, we divide the frequency range into two subranges -- $\Omega_{N}$ around
 the nodal points and $\Omega_{A}$ away from nodal points.
We use the convention
\begin{equation}
\Omega_{N}:\; \cup_p \{ \omega^{(p)}-\delta\omega^{(p)}<\rvert\omega_{m}\rvert<\omega^{(p)}+\delta\omega^{(p)} \},
\end{equation}
where $p=1,...,n$ and $\rvert \Delta_{n}(\omega^{(p)}\pm\delta\omega^{(p)})\rvert=|\omega^{(p)} \pm \delta^{(p)}|$.
  Then $\rvert \Delta_{n}(\omega_{m})\rvert<\rvert \omega_m \rvert $ for $\omega_{m}\in\Omega_{N}$  and
  $\rvert \Delta_{n}(\omega_{m})\rvert>\rvert \omega_m \rvert $ for $\omega_{m}\in\Omega_{A}$.
  Because the positions of nodal points of $\Delta_n (\omega_m)$ depend on the logarithm of frequency, it is convenient to introduce the logarithmic variable   $x=\log (\rvert \omega_{m}\rvert/{\bar g})^{\gamma}$.
The neighborhood of the nodal point $\omega_{p}$ transforms to $x_{p}-\delta_{p}<x<x_{p}+\delta_{p}$, where $x_{p}\simeq-\pi p/\beta_{\epsilon}$.  We argued in the main text that for finite $n$ and $\gamma \to 2$, $\epsilon_n \approx 1/(2-\gamma)^a$ ($a >0$), $\beta_\epsilon \sim \epsilon^{-1/3} \propto (2-\gamma)^{(a/3)}$, and
 $\Delta_n(\omega^{(p)}\pm\delta\omega_m)/|\omega_p| \simeq (-1)^{p}
 (x-x_p)$.
The condition $\rvert \Delta_{n}(\omega_{m})\rvert \sim \rvert \omega_m \rvert $ corresponds to $|x-x_p|\sim1$.

Below we consider contributions to $\delta E_{c,n}$ from different
combinations of $\omega_{m},\omega_{m}^{\prime}$, each within either
 $\Omega_{N}$ or $\Omega_{A}$.

(1) $\omega_{m},\omega_{m}^{\prime}\in\Omega_{N}$

In this frequency domain,
$\rvert \Delta(\omega_m) \rvert < \rvert \omega_m \rvert \  $ and
$\rvert \Delta(\omega_m^{\prime}) \rvert  < \rvert \omega_m^{\prime} \rvert$. The
  the corresponding contribution to $\delta E_{c,n}$ is
\begin{equation}
\delta E_{c,n}^{(1)} \simeq -\frac{1}{4}\bar{g}^{\gamma}N_{F}\iint_{\omega_{m},\omega_{m}^{\prime}\in\Omega_{N}}
\frac{d\omega_{m}d\omega_{m}^{\prime}}{\rvert\omega_{m}-\omega_{m}^{\prime}\rvert^{\gamma}}
\left(\text{sign}(\omega_{m})\text{sign}(\omega_{m}^{\prime})-1\right),
\label{ss_11}
\end{equation}
where the superscript $(1)$ refers to the integration domain.
Converting the integral in (\ref{ss_11}) to positive $\omega_m$ and negative $\omega_{m}^{\prime}$ or vice versa
and changing the frequencies to $x = x_p + y$ and $x^{\prime}=x_{p^{\prime}}+y^{\prime}$, we
obtain to leading order in $2-\gamma$
\begin{equation}
\delta E_{c,n}^{(1)}
\simeq
\frac{1}{4} \bar{g}^{\gamma}
N_{F}\sum_{p,p^{\prime}=1}^{n}
\iint_0^
 {O(1)}
dy dy^{\prime}
\left[2\cosh\frac{x_{p}-x_{p^{\prime}}+y-y^{\prime}}{4}\right]^{-
2}e^{\frac{2-\gamma}{4}(x_{p}+x_{p^{\prime}})}.
\end{equation}
The integrand decays exponentially as a function of $\rvert p - p^{\prime} \rvert $, hence the
dominant contribution to the integral comes from
$p=p^{\prime}$. Keeping only this term, we obtain
\bea
\delta E_{c,n}^{(1)} &\simeq& \frac{1}{4}\bar{g}^{\gamma}N_{F}\iint_{0}^{O(1)}dydy^{\prime}\left[2\cosh\frac{y-y^{\prime}}{4}\right]^{-
2}\sum_{p=1}^{n}e^{\frac{2-\gamma}{2}x_{p}} \nonumber \\
&=& {{\bar g}^2 N_F} {c_1 \over 2-\gamma} J_{n},
\eea
where $J_{n}= {2 \beta_{\epsilon_n} \over \pi} \left( 1- e^{-{\pi n (2-\gamma) \over 2 \beta_{\epsilon_n}}} \right)$ and $c_1 = O(1)$.
At
 $\gamma \to 2$ and finite $n$, the exponential factor in $J_n$
   can be expanded in $n$, yielding
    $\delta E_{c,n}^{(1)} ={\bar g}^2 N_F c_1 n$.

(2) $\omega_{m},\omega_{m}^{\prime}\in\Omega_{A}$

In this frequency
range,
$\rvert \Delta(\omega_m) \rvert  > \rvert \omega_m \rvert$ and $\rvert \Delta(\omega_m^{\prime}) \rvert  > \rvert \omega_m^{\prime} \rvert$. The contribution to the
 condensation energy is
\begin{equation}
\delta E_{c,n}^{(2)}=-\frac{1}{4}\bar{g}^{\gamma}N_{F}\iint_{\omega_{m},\omega_{m}^{\prime}\in\Omega_{A}}\frac{d\omega_{m}d\omega_{m}^{\prime}}{\rvert\omega_{m}-\omega_{m}^{\prime}\rvert^{\gamma}}\left(\text{sign}(\Delta_{n}(\omega_{m}))\text{sign}(\Delta_{n}(\omega_{m}^{\prime}))-1\right).
\end{equation}
 We further divide $\Omega_{A}$
into sub-intervals $\Omega_{A}^{p}$ between  two adjacent nodal points
$\omega_{p}$ and $\omega_{p+1}$, where $p=1,2,...,n$ and $\omega_{n+1}\equiv0$.
The sign structure of $\Delta_n (\omega_m)$ i such that
 $\omega_{m}$
and $\omega_{m}^{\prime}$ must come from intervals $\Omega_{A}^{p}$
and $\Omega_{A}^{p^{\prime}}$ with odd $p+p^{\prime}
$.
Introducing again logarithmic variables
$x=x_{p}-y$ and $x^{\prime}=x_{p^{\prime}+1}+y^{\prime}$, we obtain after simple algebra
\bea
\delta E_{c,n}^{(2)} & \simeq &
 \frac{\bar{g}^{2}N_{F}}{4} \sum_{p+p^{\prime}\in\text{odd}}\int_{O(1)}^{O(\pi/\beta_{\epsilon})}dy
 \int_{O(1)}^{O(\pi/\beta_{\epsilon})}dy^{\prime}\nonumber \\
 && \left[\frac{1}{\rvert2\sinh\frac{x_{p}-x_{p^{\prime}+1}-y-y^{\prime}}{2}\rvert^{2}}+
 \frac{1}{(2\cosh\frac{x_{p}-x_{p^{\prime}+1}-y+y^{\prime}}{2})^{2}}\right]e^{(x_{p}+x_{p^{\prime}+1})
 \left(\frac{2-\gamma}{4}\right)}
\eea
Given that $x_{p}-x_{p^{\prime}+1}\propto(p^{\prime}-p+1)/\beta_{\epsilon}$
 and $\beta_\epsilon \ll 1$,
this integral is exponentially small unless
$p^{\prime}+1=p$. Keeping only such term, we obtain
\bea
\delta E_{c,n}^{(2)} && \simeq \frac{\bar{g}^{\gamma}N_{F}}{4} \int_{O(1)}^{O(\pi/\beta_{\epsilon})}dy\int_{O(1)}^{O(\pi/\beta_{\epsilon})}dy^{\prime}\left[
\frac{1}{\rvert2\sinh\frac{y+y^{\prime}}{2}\rvert^{2}}+\frac{1}{(2\cosh\frac{y+y^{\prime}}{2})^{2}}\right]  \sum_{p=1}^{n}e^{\frac{2-\gamma}{2}x_{p}} \nonumber \\
&&= {\bar g}^2 N_F  {c_2 \over 2-\gamma} J_{n},
\eea
where $c_2 = O(1)$. The integral over $y,y^{\prime}$ is confined to $y,y' \sim 1$, hence it is not
surprising that $\delta E_{c,n}^{(2)}$ is of the sam order as $\delta E_{c,n}^{(1)}$.

(3) $\omega_{m}\in\Omega_{N},\omega_{m}^{\prime}\in\Omega_{A}$

 On general grounds we expect that the contribution from (3) is the same as from (1) and (2), but
 it is instructive to explicitly verify this.
  We compute the contribution from $\rvert \Delta(\omega_m) \rvert <\rvert \omega_m \rvert$ and $\rvert \Delta(\omega_m^{\prime}) \rvert > \rvert \omega_m^{\prime} \rvert$ and multiply the result by 2.
   The contribution to
   the condensation energy is
\begin{equation}
\delta E_{c,n}^{(3)}\simeq
\frac{1}{2}\bar{g}^{\gamma}N_{F}\iint_{\omega_{m}\in\Omega_{N},\omega_{m}^{\prime}\in\Omega_{A}}\frac{d\omega_{m}d\omega_{m}^{\prime}}{\rvert\omega_{m}-\omega_{m}^{\prime}\rvert^{\gamma}}.
\end{equation}
In terms of  logarithmic variables, it becomes
\begin{align}
\delta E_{c,n}^{(3)}
&\simeq \frac{1}{4}\bar{g}^{\gamma}N_{F}\sum_{p=1}^{n}\int_{x_{p}-\delta_{p}}^{x_{p}+\delta_{p}}dx
\sum_{p^{\prime}=1}^{n}\int_{x_{p^{\prime}+1}+
\delta_{p^{\prime}+1}}^{x_{p^{\prime}}-\delta_{p^{\prime}}}dx^{\prime}\left(
\frac{1}{\rvert2\sinh\frac{x-x^{\prime}}{2}\rvert^{2}}+\frac{1}{(2\cosh\frac{x-x^{\prime}}{2})^{2}}\right) \nonumber \\
& \times e^{(x+x^{\prime})\left(\frac{2-\gamma}{4}\right)}.
\end{align}
The dominant contribution again comes from two neighboring
patches. Keeping only this contribution, we obtain
\begin{align}
\delta E_{c,n}^{(3)} & \simeq\frac{1}{4}\bar{g}^{\gamma}N_{F}\int_{-O(1)}^{O(1)}dy\int_{O(1)}^{O(\pi/\beta_{\epsilon})}dy^{\prime}
\left(\frac{1}{\rvert2\sinh\frac{y-y^{\prime}}{2}\rvert^{2}}+\frac{1}{(2\cosh\frac{y-y^{\prime}}{2})^{2}}\right)\nonumber \\
 & \times\left(\sum_{p=2}^{n+1}+\sum_{p=1}^{n}\right)e^{\frac{2-\gamma}{2} x_p } = {{\bar g}^2 N_F} {c_3 \over 2-\gamma} J_{n},
\end{align}
The integral over $y,y'$ is free from divergences as $y$ and $y$ belong to different ranges. It is again  confined to $y' \geq 1$, $y \leq 1$.

In total,  $\delta E_{c,n} = {{\bar g}^2 N_F} {c \over 2-\gamma} J_{n}$, where $c=c_1+c_2+c_3$.
 This is what we used in the main text.

\subsection{Condensation energy for infinitesimally small gap function}

The reasoning in the main text for the dispersion of $E_c (\epsilon)$ at   $\gamma =2-0$ implies that the  overall factor $1/(2-\gamma)$ is present in $E_c (\epsilon)$ for all non-zero $\epsilon$, including the smallest $\epsilon \to 0$.  At vanishing $\epsilon$,
 $\Delta(\omega_m)= \epsilon \Delta_{\infty} (\omega_m)$ is the solution of the linearized gap equation.  This gap function does not saturate at $\omega_m =0$ and instead keep oscillating down to  the smallest frequencies.
  In this sutuation, the argument about the overall factor $1/(2-\gamma)$, which we presented in the previous two subsections, does not hold, and it is a'priori unclear whether $E_c (\epsilon)$ at $\epsilon \to 0$ does contain $1/(2-\gamma)$. We show that it does.

It is convenient to use the condensation energy in the form given by Eq. (\ref{eq:n3}).
 The first term in (\ref{eq:n3}) is not singular for any $\gamma$, so we focus on the second term. For small $D (\omega_m)$ we have
 \bea \label{eq:ec}
E_c \simeq - N_F \frac{{\bar g}^\gamma}{16} \iint d \omega_m d\omega_m' \left( {1\over \rvert \omega_m - \omega_m' \rvert^\gamma} - {1\over \rvert \omega_m + \omega_m' \rvert^\gamma}  \right) \left( D^2(\omega_m) - D^2(\omega_m') \right)^2,
\eea
 where $D(\omega_m)$
 \beq
 |D(\omega_m)| = 2 \epsilon^* \cos{\left[\left(\beta(\gamma)
 \log{({\bar g}/|\omega_m|)^\gamma}  + \phi (\gamma)\right)\right]},
 \label{eq:8_f}
 \eeq
 and $\epsilon^* = \epsilon ({\bar g}/|\omega_m|)^{1-\gamma/2}$ (see Eq. (\ref{eq:8}). For simplicity, we absorbed the numerical factor $Q_1$  into $\epsilon$).  For $\gamma \leq 2$, $\epsilon^*$ increases
  for decreasing $\omega_m$, and  becomes of order one at $\omega_m = \omega_* \sim {\bar g} \epsilon^{2/(2-\gamma)}$.  At smaller $\omega_m$, $|D(\omega_m)|$ becomes larger than one.  Eq. (\ref{eq:ec}) is  valid at $\omega_m$ and $\omega_m'$ larger than $\omega_*$, which then
 sets the lower limit of integration in
(\ref{eq:ec}).  At infinitesimally small $\epsilon$, $\omega_*$ is also infinitesimally small and further decreases at $\gamma \to 2$. Yet, for any finite $\epsilon$ and finite $2-\gamma$, $\omega_*$ remains finite.

We now evaluate the 2D integral in (\ref{eq:ec}).  It is convenient to
  introduce logarithmic variables $x = \log({\bar g}/\omega_m)^\gamma$ and  $x'= \log({\bar g}/\omega_m')^\gamma$.
   The integration over $x$ and over $x'$ is between $O(1)$ and
   $x_* = \gamma \log({\bar g}/\omega_*)  = 2 \gamma/(2-\gamma)\log(C/\epsilon)$, where $C = O(1)$.
    We further introduce $a=x+x'$ and $b=x-x'$.  The condensation energy is re-expressed in terms of $a$ and $b$ as
\bea
&& E_c =  -N_F {\bar g}^2 {\epsilon^4 \over 16 \gamma^2 2^\gamma } \int_ {O(1)}^{2x_*} d a e^{{ 2-\gamma \over 2 \gamma } a} \int _{0} ^{2x_*-a}  d b
\left( {1\over (\sinh {b\over 2\gamma})^\gamma} -  {1\over (\cosh {b\over 2\gamma})}  \right) \times   \nonumber \\
&&
\left[ \sin(\beta (\gamma) a + 2 \phi (\gamma)) \sin {\beta (\gamma) b} \cosh ({\gamma-2 \over 2\gamma}b) + (1+\cos(\beta (\gamma) a + 2\phi (\gamma)) \cos(\beta (\gamma) b)) \sinh( {\gamma-2 \over 2 \gamma} b )  \right]^2.
\label{f:4}
\eea
We see that the integral over $a$ is confined to the upper limit $a \sim x_*$, but the one over $b$  is confined to $b = O(1)$.  We can then safely extend the upper limit of the integration over $b$ to infinity.
Taking the limit $\gamma \to 2$ in the last factor in (\ref{f:4}) and averaging over rapidly oscillating factor
$\sin^2 (\beta (\gamma) a + 2 \phi (\gamma))$, we obtain
\bea
E_c  -N_F {\bar g}^2 {\epsilon^4\over 512 }  I \int_ {O(1)}^{2x_*} d a
e^{{ 2-\gamma \over 4} a}
\eea
where
\bea
I = \int_0^\infty d b \left( {1\over \sinh^2 {b\over 4}}-{1\over \cosh^2 {b\over 4}} \right) \sin^2 (\beta (2) b) \simeq 5.75683.
\eea
Evaluating the remaining integral over $a$, we obtain
\bea
E_c = -N_F {\bar g}^2 {\epsilon^4\over 512 }  I    \frac{4}{2-\gamma} e^{(2-\gamma) x_*/2}
\label{f:5}
\eea
Substituting  the expression for $x_*$, we obtain
\bea
E_c = -N_F {\bar g}^2 {C^2 I \over 128 }    \frac{\epsilon^2}{2-\gamma}
\label{f:6}
\eea
The computation of the prefactor $C$ requires more sophisticated analysis.

We see that $E_c$ scales as $1/(2-\gamma)$, even for the smallest $\epsilon$. We cited this result in the main text.

The $1/(2-\gamma)$ divergence at small $\epsilon$ can be regularized by a finite bosonic mass $\omega_D$, like we found for larger $\epsilon$.  We found that at a finite $\omega_D$  the factor $1/(2-\gamma)$ is replaced by $\epsilon^2 \log{{\bar g}/\omega_D}$. As a result, the condensation energy scales as $E_c \propto \epsilon^4 \log{{\bar g}/\omega_D}$.

\section{The solutions of the gap equation, found in Ref. \cite{yuz} and their causality}\label{sec:causality}

In the Ref.~\cite{yuz} the authors found the new class of solutions of Eliashberg equations for $\gamma =2$  both in the normal and superconducting state.
In the spin-chain language, introduced in \cite{yuz}, these solutions correspond to ``spin-flip'' configurations, in which spins on some lattice sites are antiparallel to the local field, generated by other spins.
In terms of the Green's function  they correspond to
  sign-changing $G(k_F,\omega_{m})$ along $\omega_m >0$.

In this Appendix we show that such solutions cannot be analytically continued to the upper half plane (UHP) of complex frequency without singularities, i.e., they violate
causality
principle.  We also show that these solutions do not exist at $T=0$, in the limit of vanishing bosonic mass.

The analysis of the analytic properties
 can be done most easily the normal state. We focus on
$k = k_F$ and  define $G(\omega_m) = G(k_F,\omega_{m})$.
We recall that the solution on the Matsubara axis is casual
if there exists
a function $G(z)$ of complex $z = \omega' + i \omega^{''}$, which is analytic and has no singularities in the UHP and for which $G(z=i\omega_{m})=G(\omega_{m})$. On the real axis, the function $G(\omega)$ is a conventional retarded Green's function; at large $|z|$ in the UHP, $G(z\rightarrow \infty )\approx 1/z$.

Like in~\cite{yuz}, we assume particle-hole symmetry, in which case $G(z=i \omega^{''})$ must be real.
Then if $G(\omega_{m})$ changes sign
once
between Matsubara points $\omega_{m} $ and $\omega_{m+1}$,
then the real $G( z=i\omega^{''})$ must either have a pole or have a zero in the interval $z\in (i\omega_{m},i\omega_{m+1})$.
The former possibility immediately indicates that $G(z)$ has a pole in the UHP and thus violate the causality principle.
Now we discuss the latter possibility and demonstrate that $G(z)$ still must have a pole somewhere in the UHP.

\begin{figure}
\includegraphics[width=\columnwidth]{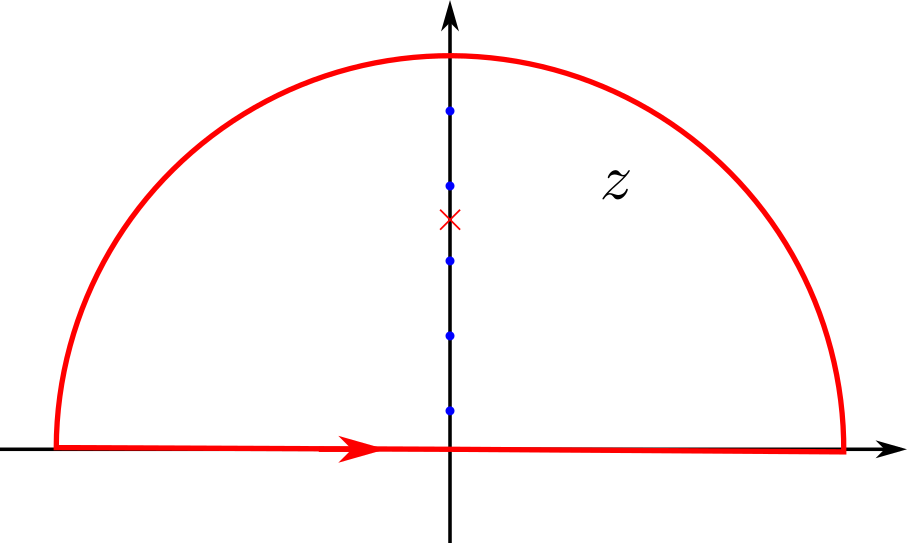}
\caption{\label{fig:AppF:contour} The contour $\Gamma $ used in the definition (\ref{eq:AppF:winding}). The radius of the arc is very large. The blue dots represent the Matsubara points. The red cross represents a zero of the function $G(z)$.
}
\end{figure}

For the proof, consider the following integral
\begin{equation}\label{eq:AppF:winding}
\frac{1}{2\pi }\oint_{\Gamma }\partial_{z}\Imm \log G(z) dz,
\end{equation}
where the closed contour $\Gamma $ is shown in
 Fig. \ref{fig:AppF:contour}. The contour $\Gamma $ consists of two parts: the large arc of radius
 $R \to \infty$ and the line along the real axis. The integral along the large arc can be evaluated using
  the asymptotic large $z$ form
$G(z) \approx 1/z$.
    Substituting this form into \eqref{eq:AppF:winding} and using $z=Re^{i\phi }$, we find $\oint_{\text{arc} }\partial_{z}\Imm \log G(z) dz=-\pi $.

To evaluate
 the integral along the interval $(-R,R)$ on the real axis, we note that
 i) the Eliashberg self energy
 depends only on $\omega$, and ii) the density of states, $\sim-\int dk \Imm G(\omega ,k)$, must be positive at all real $\omega $. One can then make sure
 that
   $\Imm G$ is negative and does not change sign on the real axis, while
  $\Ree G$ is negative at $\omega \rightarrow -\infty $ and positive at $\omega \rightarrow +\infty $. It then immediately follows that $\oint_{-R}^{R}\partial_{\omega }\Imm \log G(\omega ) d\omega =\pi$. Hence
\begin{equation}\label{eq:AppF:largeContour}
 \frac{1}{2\pi }\oint_{\Gamma }\partial_{z}\Imm \log G(z) dz=0
\end{equation}

On the other hand,
the function $\partial_{z} \log G(z)$ is an analytic function in the UHP with a simple pole of residue $+1$ at the point where $G(z)=0$.
 Using the residue theorem we then immediately find that
\begin{equation}\label{eq:AppF:residue}
 \frac{1}{2\pi }\oint_{\Gamma }\partial_{z}\Imm \log G(z) dz=1
\end{equation}
We see that \eqref{eq:AppF:residue} is  incomparable with \eqref{eq:AppF:largeContour}.
 This can only be avoided if  $G(z)$ has a pole in the UHP as at this point $\partial_{z} \log G(z)$
 has a pole with residue $-1$, and the r.h.s of \eqref{eq:AppF:residue}
 is $1-1=0$.
 In general, the r.h.s of \eqref{eq:AppF:residue} is $n-p$, where $n$ is the number of zeros and $p$ is the number of poles, counted with their multiplicity. Then if $G(\omega_{m})$ changes sign on the Matsubara axis more than once, there must be multiple poles of $G(z)$  in the UHP.
The requirement that $G(z)$ must have a poles in the UHP
contradicts the
 causality principle.

\subsection{T=0}\label{sec:add}

Now we return to Eliashberg equations and show that $Z_{\omega }$ cannot change sign on the positive Matsubara axis at $T=0$ in the limit of vanishing bosonic mass.  At $T=0$ the system is in the superconducting state, and we
 analyze Eliashberg equations for a non-zero $\Delta (\omega_m)$.

First,  we notice that at $T=0$,  $Z_{\omega }$ is a real function of $\omega_m$. It also must be a continuous function, as the Green's function in the superconducting state must be analytic in the UHP.

As the authors of \cite{yuz} noticed, the r.h.s. of the Eliashberg equation  \eqref{ss_11_01}
$\text{sign}(Z_{\omega_{m'}})$ in the numerator under the sum. In the main text we set this term to 1 as we were only interested in the sign-preserving solutions for $Z(\omega_m)$. Now we keep this term.
At $T=0$ the equation for $Z(\omega_m)$ becomes
\begin{equation}\label{eq:SCZ}
 Z_{\omega_m }=1+\frac{g^{\gamma }}{2\omega }\int_{-\infty }^{\infty }\frac{d\omega '_m}{\sqrt{{\omega '_m }^{2}+\Delta_{\omega '_m}^{2}}}\text{sign}(Z_{\omega '_m})\frac{1}{((\omega_m -\omega '_m)^{2}+\omega^{2}_D)^{\gamma/2 }}
\end{equation}
 We will be interested in the limit $\omega_D\rightarrow 0$.

Both functions $Z_{\omega_m }$ and $\Delta_{\omega_m }$ are real and $Z_{\omega_m \rightarrow \infty }\rightarrow 1$, $\Delta_{\omega_m \rightarrow \infty }\rightarrow 0$.

We want to check, if it is possible that
\begin{itemize}
\item both $Z_{\omega_m }$ and $\Delta_{\omega_m }$ are smooth functions.
\item $Z_{\omega_m }$ changes sign $N$ times at frequencies $\omega_n$, where $n=0\dots N-1$.
\end{itemize}

We take the derivative of $Z_{\omega_m }$ over $\omega_m $ and evaluate the integral  by parts
\begin{eqnarray}
&&
\partial_{\omega_m } Z_{\omega_m }=-\frac{1}{\omega_m }(Z_{\omega_m }-1)
-\frac{g^{\gamma }}{2\omega_m }\int_{-\infty }^{\infty }\frac{d\omega '_m}{\sqrt{{\omega '_m}^{2}+\Delta_{\omega '_m}^{2}}}\text{sign}(Z_{\omega '_m})\partial_{\omega '_m}
\frac{1}{((\omega_m -\omega '_m)^{2}+\omega^{2}_D)^{\gamma /2}}=\nonumber\\
&&-\frac{1}{\omega_m }(Z_{\omega_m }-1)-\frac{g^{\gamma }}{2\omega_m}
\left[
\left.
\frac{\text{sign}(Z_{\omega '_m})}{\sqrt{{\omega '_m}^{2}+\Delta_{\omega '_m}^{2}}}\frac{1}{((\omega_m -\omega '_m)^{2}+\omega^{2}_D)^{\gamma/2 }}
\right|_{-\infty }^{\infty }
-
\int_{-\infty }^{\infty }d\omega '_m\frac{1}{((\omega_m -\omega '_m)^{2}+\omega^{2}_D)^{\gamma /2}}\partial_{\omega '_m}\frac{\text{sign}(Z_{\omega '_m})}{\sqrt{{\omega '_m}^{2}+\Delta_{\omega '_m}^{2}}}
 \right]=\nonumber\\
&&
-\frac{1}{\omega }(Z_{\omega }-1)+\frac{g^{\gamma }}{2\omega }
\left[
2(-1)^{N+1}\sum_{n=0}^{N-1}\frac{(-1)^{n}}{((\omega -\omega_{n})^{2}+\omega^{2}_D)^{\gamma/2 }}\frac{1}{\sqrt{{\omega_{n}}^{2}+\Delta_{\omega_{n}}^{2}}}
+\int_{-\infty }^{\infty }d\omega '_m\frac{\text{sign}(Z_{\omega '_m})}{((\omega -\omega '_m)^{2}+\omega^{2}_D)^{\gamma/2 }}\partial_{\omega '_m}\frac{1}{\sqrt{{\omega '_m}^{2}+\Delta_{\omega '_m}^{2}}}
 \right], \nonumber \\
\label{nn}
\end{eqnarray}
where we used $\partial_{\omega '_m}\text{sign} (Z_{\omega '_m})=2(-1)^{N+1}\sum_{n=0}^{N-1}(-1)^{n}\delta (\omega_m -\omega_{n})$.
By assumption, $Z_{\omega_m }$ is a smooth function, so $Z_{\omega_{n}}=0$, {\it for all} $n$. Setting
 $\omega_m=\omega_{k}$ in (\ref{nn}),
 we obtain
\begin{eqnarray}
&&\left. \partial_{\omega_m } Z_{\omega_m }\right|_{\omega_m =\omega_{k}}=
\frac{1}{\omega_{k} }
+\frac{g^{\gamma }}{2\omega_{k} }\int_{-\infty }^{\infty }d\omega '_m\frac{\text{sign}(Z_{\omega '_m})}{((\omega_{k} -\omega '_m)^{2}+\omega^{2}_D)^{\gamma/2 }}\partial_{\omega '_m}\frac{1}{\sqrt{{\omega '_m}^{2}+\Delta_{\omega '_m}^{2}}}\nonumber\\
&&
+\frac{g^{\gamma }}{\omega_{k} }(-1)^{N+1}\sum_{n=0}^{N-1}\frac{(-1)^{n}}{((\omega_{k} -\omega_{n})^{2}+\omega^{2}_D)^{\gamma/2 }}\frac{1}{\sqrt{{\omega_{n}}^{2}+\Delta_{\omega_{n}}^{2}}}.
\label{eq:SCderZomegaK}
\end{eqnarray}
We see that in the limit $\omega_D \rightarrow 0$, the most singular term is the term in the sum with $n=k$. Keeping only this term we get in this limit
\begin{equation}\label{eq:SCderZomegaKlim}
\lim \limits_{m\rightarrow 0}\left. \partial_{\omega_m } Z_{\omega_m }\right|_{\omega_m =\omega_{k}}\sim
\frac{g^{\gamma }}{\omega_{k} }\frac{(-1)^{k+N+1}}{\omega^{\gamma/2 }_D}\frac{1}{\sqrt{{\omega_{k}}^{2}+\Delta_{\omega_{k}}^{2}}}\rightarrow \pm \infty .
\end{equation}
This shows that  in the limit $\omega_D \rightarrow 0$, the sign-changing function $Z_{\omega_m}$ is not smooth function  as its derivative over $\omega_m$ diverges at any finite $\omega_m$.

\bibliography{free_energy_bib}

\end{document}